\documentclass[sigconf]{acmart}

\makeatletter
\def\@ACM@checkaffil{% Only warnings
    \if@ACM@instpresent\else
    \ClassWarningNoLine{\@classname}{No institution present for an affiliation}%
    \fi
    \if@ACM@citypresent\else
    \ClassWarningNoLine{\@classname}{No city present for an affiliation}%
    \fi
    \if@ACM@countrypresent\else
        \ClassWarningNoLine{\@classname}{No country present for an affiliation}%
    \fi
}
\makeatother

\makeatletter
\let\@authorsaddresses\@empty
\makeatother

\AtBeginDocument{%
  }

\copyrightyear{2024}
\acmYear{2024}
\setcopyright{acmcopyright}
\acmConference[FAccT '24]{The 2024 ACM Conference on Fairness, Accountability, and Transparency}{June 3--6, 2024}{Rio de Janeiro, Brazil}
\acmBooktitle{The 2024 ACM Conference on Fairness, Accountability, and Transparency (FAccT '24), June 3--6, 2024, Rio de Janeiro, Brazil}\acmDOI{10.1145/3630106.3659037}
\acmISBN{979-8-4007-0450-5/24/06}
\acmDOI{3630106.3659037}

\acmConference[FAccT '24]{}{June 03--06,
  2024}{Rio de Janeiro, Brazil} % for submission, remove this for arxiv
\usepackage{cleveref}
\usepackage{enumitem}
\usepackage{pifont}
\usepackage{hyperref}
\usepackage{multirow}
\usepackage{xcolor,pifont}
\usepackage{tabularx}
\usepackage{booktabs}
\usepackage{soul}
\begin{document}

\title{Black-Box Access is Insufficient\\for Rigorous AI Audits\\ $\;$
}

\author{Stephen Casper}
\authornote{Equal Contribution.}
\email{scasper@mit.edu}
\affiliation{\institution{MIT CSAIL}}

\author{Carson Ezell}
\authornotemark[1]
\email{cezell@college.harvard.edu}
\affiliation{\institution{Harvard University}}

% \author{$\;$}
% \affiliation{\institution{$\;$}}

\author{Charlotte Siegmann}
\affiliation{\institution{MIT}}

\author{Noam Kolt}
\affiliation{\institution{University of Toronto}}

\author{Taylor Lynn Curtis}
\affiliation{\institution{MIT CSAIL}}

\author{Benjamin Bucknall}
\affiliation{\institution{Centre for the Governance of AI}}

\author{Andreas Haupt}
\affiliation{\institution{MIT}}

\author{Kevin Wei}
\affiliation{\institution{Harvard Law School}}

\author{Jérémy Scheurer}
\affiliation{\institution{Apollo Research}}

\author{Marius Hobbhahn}
\affiliation{\institution{Apollo Research}}

\author{Lee Sharkey}
\affiliation{\institution{Apollo Research}}

\author{Satyapriya Krishna}
\affiliation{\institution{Harvard University}}

\author{Marvin Von Hagen}
\affiliation{\institution{MIT}}

\author{Silas Alberti}
\affiliation{\institution{Stanford University}}

\author{Alan Chan}
\affiliation{\institution{\hspace{-6pt}Mila, Centre for the Governance of AI\hspace{-6pt}}}

\author{Qinyi Sun}
\affiliation{\institution{MIT}}

\author{Michael Gerovitch}
\affiliation{\institution{MIT}}

\author{David Bau}
\affiliation{\institution{Northeastern University}}

\author{Max Tegmark}
\affiliation{\institution{MIT}}

\author{David Krueger}
\affiliation{\institution{University of Cambridge}}

\author{Dylan Hadfield-Menell}
\affiliation{\institution{MIT CSAIL}}

\author{$\;$}
% \affiliation{$\;$}

\renewcommand{\shortauthors}{Casper, Ezell, et al.}

\begin{abstract}  
  External audits of AI systems are increasingly recognized as a key mechanism for AI governance. The effectiveness of an audit, however, depends on the degree of access granted to auditors. Recent audits of state-of-the-art AI systems have primarily relied on \emph{black-box} access, in which auditors can only query the system and observe its outputs. However, \emph{white-box} access to the system's inner workings (e.g., weights, activations, gradients) allows an auditor to perform stronger attacks, more thoroughly interpret models, and conduct fine-tuning. Meanwhile, \emph{outside-the-box} access to training and deployment information (e.g., methodology, code, documentation, data, deployment details, findings from internal evaluations) allows auditors to scrutinize the development process and design more targeted evaluations. In this paper, we examine the limitations of black-box audits and the advantages of white- and outside-the-box audits. We also discuss technical, physical, and legal safeguards for performing these audits with minimal security risks. Given that different forms of access can lead to very different levels of evaluation, we conclude that (1) transparency regarding the access and methods used by auditors is necessary to properly interpret audit results, and (2) white- and outside-the-box access allow for substantially more scrutiny than black-box access alone.
\end{abstract}  





\settopmatter{printfolios=true}  % adds page numbers
\maketitle

\newpage
\tableofcontents
% \newpage

\section{Introduction} \label{sec:intro}

External evaluations of AI systems are emerging as a key component of AI oversight \citep{brown2021algorithm, watkins2021governing, metcalf2021algorithmic, liu_medical_2022, metcalf2022relationship, raji2022outsider, raji2022anatomy, raji2022actionable, koshiyama2022algorithm, schuett2022three, schuett2023towards, M_kander_2023, shevlane2023model, seger2023open, landers2023auditing, anderljung2023frontier, solaiman2023gradient, solaiman2023evaluating, wilkinson2023accountability, mokander2023auditing, miotti2023taking, sharkey2023auditing, anderljung2023publicly, bengiomanaging, sharkey2024causal, birhane2024ai, metr} and governance frameworks \citep{european2021laying, airmf2023, tc260_2023, dsit2023, biden2023executive}.
There is a rich history in academic AI research of evaluating systems to explain their behaviors \citep{zhang2022explainable,agarwal2022openxai} and evaluate risks including those related to privacy \citep{yao2023survey, smith2023identifying, carlini2021extracting, singhal2022large, shi2023detecting}, intellectual property rights \citep{karamolegkou2023copyright, casper2023measuring, carlini2022quantifying}, fairness and discrimination \citep{bolukbasi2016man, buolamwini2018gender, eckhouse2019layers, coe2021evaluating, mehrabi2021survey, weidinger2021ethical, angwin2022machine, gichoya2022ai, chen2022fairness, tao2023auditing, dhamala2021bold,krishna2022measuring}, harmful content, \citep{birhane2021multimodal, birhane2023into, thiel2023identifying, thiel2023generative, qu2023unsafe, rando2022red}, circumvention of safeguards (\emph{jailbreaks}) \citep{shayegani2023survey, liu2023jailbreaking, rao2023tricking, wei2023jailbroken, zou2023universal, shah2023scalable, carlini2023aligned}, misinformation and deception \citep{ji2023survey, huang2023survey, scheurer2023technical, park2023ai, hubinger2024sleeper}, dangerous capabilities \citep{charan2023text, shevlane2023model, kinniment2023evaluating, chan2023harms, openai2023gpt4}, and broader societal impacts \citep{solaiman2023evaluating, weidinger2023sociotechnical}.

Historically, most academic work on evaluating AI systems has been conducted on models where parameters, data, and methodology are openly available. 
AI systems that are not available to the public, including ones that are proprietary or in pre-deployment, pose challenges for oversight. 
AI audits are structured evaluations designed to identify risks and improve transparency by assessing how well models and methods meet specific desiderata \citep{costanza-chock2022who, mokander_auditing_2023}. Norms for AI audits are not yet well established, and their effectiveness can vary depending on the degree of system access granted to auditors \citep{brundage_toward_2020, raji2022outsider, anderljung2023publicly}. 
This is crucial because existing calls for audits are often agnostic to the form of access, and industry actors have previously lobbied for limiting access given to auditors.\footnote{For example, Google made the unsubstantiated claim in 2021 that ``There will always be better methods for verifying the performance of an AI system (ex: input/output auditing) than direct access to source code,'' while lobbying the European Union on a draft of the EU AI Act \citep{google_consultation_2021}.}

% This is particularly crucial as existing calls for audits are often agnostic to form of access, and industry actors have supported limiting external auditors to black-box access in policy debates about audits \citep{google_consultation_2021}. % in current proposed legislation, access requirements in audits are not high per default. For example, in the Council of the European Union's Proposal for the AI Act, Article 63 (9) b) states that \emph{[Regulators] shall be granted access to the source code of the high-risk
% AI system upon a reasoned request [if] testing/auditing procedures and verifications based on the data and documentation provided by the provider have been exhausted or proved insufficient.} This paper clarifies that access to source code may not be a relevant distinction for the oversight over models, but whether access is black- or white-box.

Recently, some developers of prominent state-of-the-art AI systems have kept most details of their models private \citep{bommasani2023foundation}. 
To public knowledge, voluntary external audits of these systems have primarily involved analysis of the input/output behavior of models \citep{metr, openai2023gpt, anthropic2023challenges, touvron2023llama}. This form of access, in which auditors are only able to see outputs for given inputs, is known as \emph{black-box}.
Unfortunately, black-box access is very limiting for auditors.
Some problems, such as anomalous failures, are difficult to find with black-box access \citep{kolt2023algorithmic}, and others, such as dataset biases, can be actively reinforced by testing data \citep{shahbazi2023representation}.

The ability to query a black-box system is useful, but many of today's evaluation techniques require access to weights, activations, gradients, or the ability to fine-tune the model \citep{bucknall2023structured}.
\emph{White-box} access refers to the unrestricted ability to observe a system's internal workings. 
It enables evaluators to apply more powerful attacks to automatically identify weaknesses \citep{goodfellow2014explaining, papernot2016technical}, study internal mechanisms responsible for undesirable model behaviors \citep{jain2023mechanistically, lee2024mechanistic,DBLP:journals/corr/abs-2402-06625}, and identify harmful dormant capabilities through fine-tuning \citep{qi2023fine, zhan2023removing}.
Meanwhile, \emph{outside-the-box} access involves additional contextual information about a system's development or deployment such as methodology, code, documentation, hyperparameters, data, deployment details, and findings from internal evaluations. 
It allows auditors to study risks that stem from methodology or data \citep{mitchell2019model, birhane2021multimodal, shahbazi2023representation, casper_open_2023} and makes it easier to design useful tests.
This paper makes four contributions:
\begin{enumerate}
    \item We present shortcomings of black-box methods for evaluating AI systems (\Cref{sec:limitations}).
    \item We overview the ways in which white-box methods involving attacks, model interpretability, and fine-tuning substantially expand the capabilities of evaluators (\Cref{sec:white_box}). 
    \item Similarly, we examine how outside-the-box access, including methodology, code, documentation, hyperparameters, data, deployment details, and findings from internal evaluations, allow for more thorough evaluations (\Cref{sec:outside}).
    \item Finally, we describe methods to conduct white- and outside-the-box audits securely to avoid leaks of sensitive information. These include \textit{technical} solutions involving application programming interfaces, \textit{physical} solutions involving secure research environments, and \textit{legal} mechanisms that have precedent in other industries with audits (\Cref{sec:security}).
\end{enumerate}

Given the growing evidence that different forms of access can facilitate very different levels of evaluation, we draw two conclusions. 
First, transparency regarding model access and evaluation methods is necessary to properly interpret the results of an AI audit. 
Second, white- and outside-the-box access allow for substantially more scrutiny than black-box access alone.
When higher levels of scrutiny are desired, audits should be conducted with higher levels of access.\footnote{Jointly with this paper, we submitted a \href{https://www.regulations.gov/comment/NIST-2023-0009-0018}{\textcolor{blue}{\ul{comment on policy questions related to white- and outside-the-box audits}}} to a request for information from the US National Institute of Standards and Technology \citep{nist2023request}.}

\begin{figure*}[t!]
    \centering
    \includegraphics[width=0.75\linewidth]{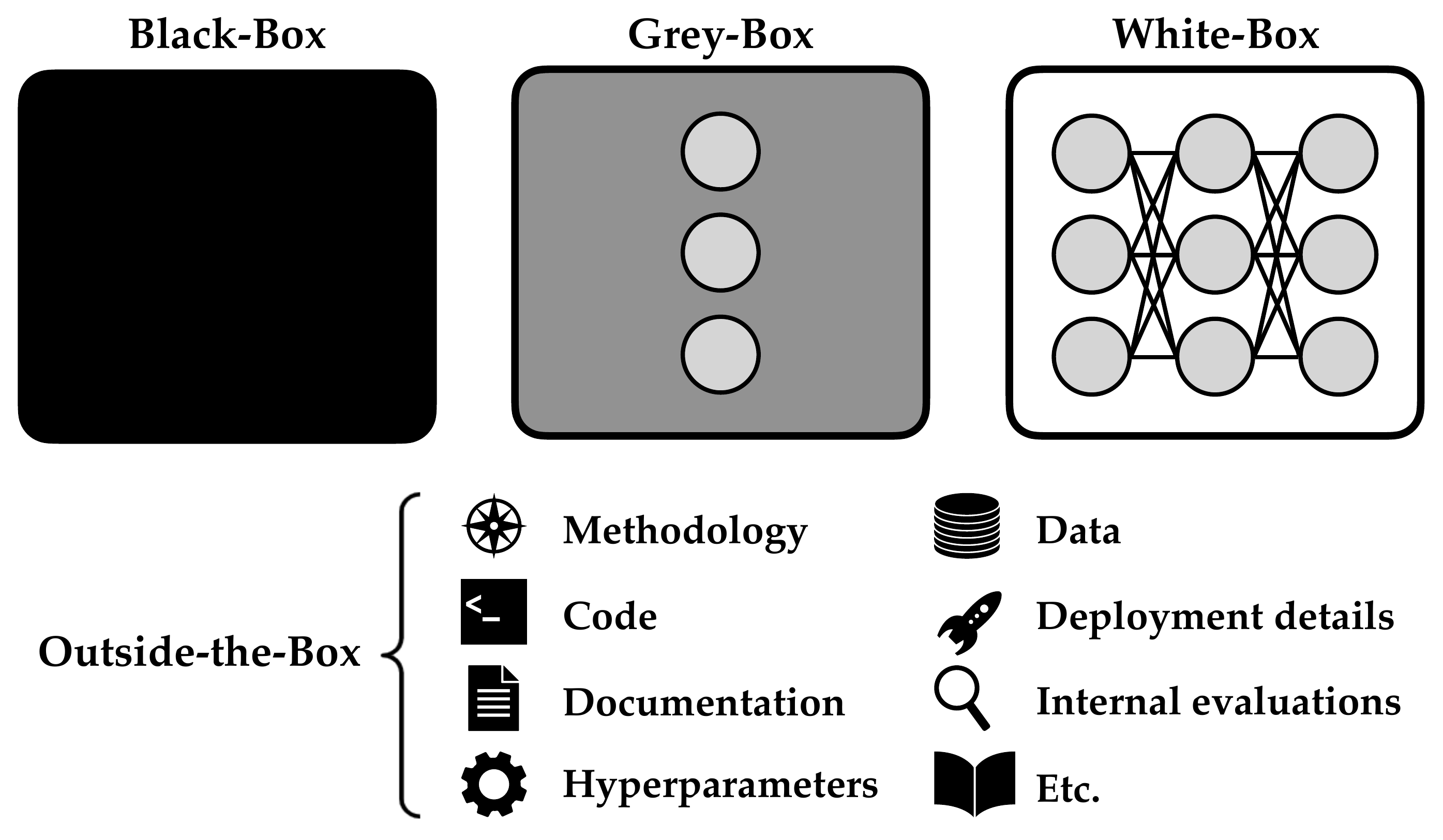}
    \caption{\emph{Black-box} access lets auditors query the system and analyze the resulting outputs. \emph{Grey-box} access lets auditors access limited internal information. \emph{White-box} access lets users access the full system. \emph{Outside-the-box} access gives auditors contextual information. In this paper, we argue that white- and outside-the-box access are key for rigorous AI audits.}
    \label{fig:taxonomy_diagram}
\end{figure*}

\section{Background} \label{sec:background}
% We start with background access types, and the legal and technical \emph{status quo} of AI audits.

\subsection{Black, Grey, White, and Outside-the-Box Access} \label{sec:black_white_grey}

\begin{table*}[t!]
\centering

\normalsize
\begin{tabular}{|l|c|c|c|c|c|c|}
\hline

& \textbf{Access} & \textbf{Black-} & \textbf{Grey-} & \textbf{De facto} & \textbf{White-} & \textbf{Outside-}  \\
& \textbf{Level} & \textbf{Box} & \textbf{Box} & \textbf{White-box} & \textbf{Box} & \textbf{the-box} \\ \hline

\textbf{Test sets} (\Cref{sec:limitations}) & \multirow{4}{*}{Queries} & \textcolor{green}{\ding{52}} & \textcolor{green}{\ding{52}} & \textcolor{green}{\ding{52}} & \textcolor{green}{\ding{52}} & \textcolor{red}{\ding{55}} \\ \cline{1-1} \cline{3-7}
\textbf{Manual attacks} (\Cref{sec:limitations}) & & \textcolor{green}{\ding{52}} & \textcolor{green}{\ding{52}} & \textcolor{green}{\ding{52}} & \textcolor{green}{\ding{52}} & \textcolor{red}{\ding{55}} \\ \cline{1-1} \cline{3-7}
\textbf{Transfer-based attacks} (\Cref{sec:attacks}) & & \textcolor{green}{\ding{52}} & \textcolor{green}{\ding{52}} & \textcolor{green}{\ding{52}} & \textcolor{green}{\ding{52}} & \textcolor{red}{\ding{55}} \\ \cline{1-1} \cline{3-7} 
\textbf{Gradient-free attacks} (\Cref{sec:attacks}) & & \textcolor{green}{\ding{52}} & \textcolor{green}{\ding{52}} & \textcolor{green}{\ding{52}} & \textcolor{green}{\ding{52}} & \textcolor{red}{\ding{55}} \\ \hline
\textbf{Sampling-probability-guided attacks} (\Cref{sec:attacks}) & Probabilities & \textcolor{red}{\ding{55}} & \textcolor{green}{\ding{52}} & \textcolor{green}{\ding{52}} & \textcolor{green}{\ding{52}} & \textcolor{red}{\ding{55}} \\ \hline
\textbf{Gradient-based attacks} (\Cref{sec:attacks}) & \multirow{2}{*}{Gradients} & \textcolor{red}{\ding{55}} & \textcolor{red}{\ding{55}} & \textcolor{green}{\ding{52}} & \textcolor{green}{\ding{52}} & \textcolor{red}{\ding{55}} \\  \cline{1-1} \cline{3-7}
\textbf{Hybrid attacks} (\Cref{sec:attacks}) & & \textcolor{red}{\ding{55}} & \textcolor{red}{\ding{55}} & \textcolor{green}{\ding{52}} & \textcolor{green}{\ding{52}} & \textcolor{red}{\ding{55}} \\ \cline{1-1} \hline
\textbf{Latent space attacks} (\Cref{sec:attacks}) & \multirow{2}{*}{\parbox{.1\textwidth}{\centering Weights/ \\ Activations}} & \textcolor{red}{\ding{55}} & \textcolor{red}{\ding{55}} & \textcolor{green}{\ding{52}} & \textcolor{green}{\ding{52}} & \textcolor{red}{\ding{55}} \\ \cline{1-1} \cline{3-7}
\textbf{Mechanistic interpretability} (\Cref{sec:interp}) & & \textcolor{red}{\ding{55}} & \textcolor{red}{\ding{55}} & \textcolor{green}{\ding{52}} & \textcolor{green}{\ding{52}} & \textcolor{red}{\ding{55}}  \\ \hline
\textbf{Fine-tuning} (\Cref{sec:fine-tuning}) & Fine-tuning & \textcolor{red}{\ding{55}} & \textcolor{red}{\ding{55}} & \textcolor{green}{\ding{52}} & \textcolor{green}{\ding{52}} & \textcolor{red}{\ding{55}} \\ \hline
\textbf{Methodological evaluations} (\Cref{sec:outside}) & \multirow{4}{*}{\parbox{.1\textwidth}{\centering Outside-\\the-Box}} & \textcolor{red}{\ding{55}} & \textcolor{red}{\ding{55}} & \textcolor{red}{\ding{55}} & \textcolor{red}{\ding{55}} & \textcolor{green}{\ding{52}} \\ \cline{1-1} \cline{3-7}
\textbf{Data evaluations} (\Cref{sec:outside}) & & \textcolor{red}{\ding{55}} & \textcolor{red}{\ding{55}} & \textcolor{red}{\ding{55}} & \textcolor{red}{\ding{55}} & \textcolor{green}{\ding{52}} \\ \cline{1-1} \cline{3-7}
\textbf{Complementary evaluations} (\Cref{sec:outside}) & & \textcolor{red}{\ding{55}} & \textcolor{red}{\ding{55}} & \textcolor{red}{\ding{55}} & \textcolor{red}{\ding{55}} & \textcolor{green}{\ding{52}} \\ \cline{1-1} \cline{3-7}
\textbf{Using source code} (\Cref{sec:outside}) & & \textcolor{red}{\ding{55}} & \textcolor{red}{\ding{55}} & \textcolor{red}{\ding{55}} & \textcolor{red}{\ding{55}} & \textcolor{green}{\ding{52}} \\ \hline
\textbf{Copying system parameters} (\Cref{sec:security}) & Unrestricted & \textcolor{red}{\ding{55}} & \textcolor{red}{\ding{55}} & \textcolor{red}{\ding{55}} & \textcolor{green}{\ding{52}} & \textcolor{red}{\ding{55}} \\ \hline

\end{tabular}

\vspace{8pt}
\caption{A summary of what evaluation techniques are possible with which types of access. A \textcolor{green}{\ding{52}} \hspace{1pt} means that a technique is possible while an \textcolor{red}{\ding{55}} \hspace{1pt} means it is not. Many levels of grey-box access are possible, but we highlight sampling-probability-attacks because they are a common example.
}
\label{tbl:methods} 

\large
\begin{tabularx}{\textwidth}{|p{.13\textwidth}|p{.15\textwidth}|X|}
\hline
\textbf{Goal} & \textbf{Technique} & \textbf{Advantage} \\
\hline
\multirow{9}{*}{\parbox{.19\textwidth}{Identifying\\ problems}} & \multirow{4}{*}{Attacks} & White-box methods are more reliable for detecting anomalous failures. \\ \cline{3-3} & & White-box attacks are more efficient. \\ \cline{3-3} & & White-box methods provide auditors with more attack options. \\ \cline{3-3} & & Robustness to white-box attacks confers greater robustness assurances. \\ \cline{2-3} & Fine-tuning & Fine-tuning enables searching for harmful dormant capabilities. \\ \cline{2-3} & \multirow{2}{*}{Interpretability} & Interpretability tools enable stronger assurances of what knowledge and capabilities a system has. \\ \cline{2-3} & \multirow{2}{*}{\parbox{.19\textwidth}{Outside-the-box \\ assessment}} & \multirow{2}{*}{Information helps auditors target complementary evaluations.} \\ & & \\ \hline 
\multirow{4}{*}{\parbox{.19\textwidth}{Incentivizing\\ responsible\\ development}} & \multirow{4}{*}{\parbox{.19\textwidth}{Outside-the-box \\ assessment}} & Information about deployment assists with assessing societal impacts. \\ \cline{3-3} & & Auditors can trace problems to datasets and methodological decisions made by developers. \\ \cline{3-3} & & Auditors can assess risk-mitigation strategies. \\ \hline 
\multirow{3}{*}{\parbox{.19\textwidth}{Increasing \\ transparency}} & \multirow{3}{*}{Interpretability} & White-box methods can address misconceptions about model outputs, characteristics, and risks. \\ \cline{3-3} & & White-box methods assist with providing more accurate and meaningful explanations. \\ \hline 
\multirow{3}{*}{\parbox{.19\textwidth}{Enabling \\ debugging}} & \multirow{2}{*}{Attacks} & Stronger attacks produce stronger instances of failures to incorporate into training data. \\ \cline{2-3} & Interpretability & Interpretability techniques enable more precise debugging. \\ \hline
\end{tabularx}
\caption{A summary of the advantages that various white-box and outside-the-box auditing techniques provide over black-box methods that are discussed in \Cref{sec:white_box} and \Cref{sec:outside}. In \Cref{app:goals}, we expand on the variety of motivations for AI audits.}
\label{tbl:advantages}

\end{table*}

In accordance with literature on security and software testing \citep{khan2012comparative}, we differentiate black-, grey-, and white-box access. We also introduce two new concepts: ``de facto white-box'' access and ``outside-the-box'' access.
\Cref{fig:taxonomy_diagram} illustrates these categories, and \Cref{tbl:methods} summarizes which techniques each type of access allows.

\begin{enumerate}
    \item \textbf{Black-box} access allows users to design inputs for a system, query it, and analyze the resulting outputs. 
    \item \textbf{Grey-box} access offers users limited access to a system's inner workings. For neural networks, this can include information such as input embeddings, inner neuron activations, or sampling probabilities.
    There are many ways that users can be given information about a system's inner workings and many corresponding shades of grey. \textbf{De facto white-box} access is a very light grey form of access that allows users to run arbitrary processes on a system indirectly with the constraint that the system's parameters cannot be copied. We discuss this and other methods to minimize the possibility of leaks in \Cref{sec:security}.
    \item \textbf{White-box} access allows users full access to the system. This includes access to weights, activations, gradients, and the ability to fine-tune the model.
    \item \textbf{Outside-the-box} access grants users access to additional information about the system's development and deployment. There are many types, which can include methodological details, source code, documentation, hyperparameters, training data, deployment details, and findings from internal evaluations. 
    Different forms of outside-the-box access can vary greatly in their comprehensiveness. For example, possessing high-level details (such as a ``model card,'' \citep{mitchell2019model}) is less informative compared to having comprehensive documentation from training and testing. 
\end{enumerate}

\subsection{Regulatory Frameworks' Reliance on Audits} \label{sec:regulatory}
Emerging frameworks for AI governance have been designed to rely on high-quality audits. 
Audits have been called for in the White House Executive Order on AI \citep{biden2023executive}, European Union policy \citep{gdpr2016, european2021laying, aiact2021}, other policy initiatives \citep{tc260_2023, dsit2023, chinaaiml2023, nyclaw2021}, general AI principles \citep{usaibor2020, un2022, nngai2023, caict2021, hiroshima2023}, voluntary standards \citep{oecd2019, airmf2023, sg2020}, multilateral commitments \citep{bletchley2023}, and position papers \citep{nngai2021, nistxai2021}.
In particular, audits in these proposals are intended to provide trustworthy \emph{assessments} of potential harm and \emph{explanations} of system behaviors.

\textbf{Regulatory frameworks have called for evaluations to accurately assess risks.}
Some jurisdictions may require risk assessment evaluations for AI systems used in certain contexts. 
These can include tests to ensure non-discrimination, such as New York City's requirement for bias audits of automated employment decision tools \citep{nyclaw2021}, or quality and performance evaluations \citep{hacker_varieties_2022}.
The draft EU AI Act \citep{aiact2021} has more recently harmonized quality assurance standards across several high-risk use cases, with provisions for external oversight. Regulators are also increasingly interested in external oversight of AI systems with potentially harmful capabilities. Recently, U.S. Executive Order 14110 \citep{biden2023executive} required developers of certain foundation models to share test results with the Federal Government. It also instructed the National Institute of Standards and Technology (NIST) to develop evaluation guidelines for harmful AI capabilities, and it tasked the Department of Energy with developing tools and testbeds to evaluate threats from AI systems to security and critical infrastructure. Companies may also voluntarily subject their systems to external evaluations beyond regulatory requirements. For example, the NIST Risk Management Framework provides recommendations for audits related to system design and reliable operation \citep{airmf2023}. In \Cref{sec:limitations}, \Cref{sec:white_box}, and \Cref{sec:outside}, we overview advantages of white- and outside-the-box access over black-box access for rigorous assessments.

\textbf{Assessing model explanations enables scrutiny of automated decisions.} 
Regulatory frameworks also use audits to provide those affected by automated decision-making with explanations of the decisions \citep{gryz2021black, ploug2021right}.
In some jurisdictions, such as the European Union, when individuals are harmed by automated decision-making systems, they may have the right to an explanation, and the results of these explanations may entitle them
to remediation \citep{hacker_varieties_2022, buiten_vision_2023}. 
Further, explanation requirements may exist for particularly high-risk systems, such as EU platform regulations (the Digital Markets Act and the Digital Services Act) that require transparency from large online platforms using AI systems (e.g., ranking algorithms) to protect against discrimination and abuse of market power \citep{dma2022,hacker_regulating_2023-1}. Finally, disclosure of evidence may be required under liability rules, such as the Product and AI Liability Directives in the EU, to enable potential claimants to adequately defend damage claims \citep{buiten_eu_2021, hacker_european_2023}. When producing explanations, a report from NIST emphasizes the importance of explanation accuracy, or \emph{[a]n explanation [that] correctly reflects the reason for generating the output and/or accurately reflects the system's process} \citep{phillips_four_2021}. In \Cref{sec:limitations} and \Cref{sec:interp}, we overview advantages of white-box access over black-box access for generating reliable explanations.

\subsection{Audits in the Status Quo} \label{sec:squo}
Recent advancements in AI capabilities -- especially from large generative models -- have increased public attention on AI audits.
As of January 2024, there are no widely adopted norms for conducting AI audits \citep{birhane2024ai}. 
Details of AI audits can vary because they depend upon the system, how it will be used, and what risks it poses. Auditing frameworks and metrics have been proposed for specific use cases, including hiring \citep{kazim_systematizing_2021,raghavan_mitigating_2020}, facial recognition \citep{khalil_investigating_2020,raji_saving_2020}, healthcare \citep{mahajan_algorithmic_2020,liu_medical_2022}, recommender systems \citep{robertson_auditing_2018, chen_bias_2023}, and general purpose language models \citep{mokander_auditing_2023}. 
However, \citet{raji2022outsider} identifies five \textit{general} limitations for algorithmic audits: scope, independence, level of access, professionalism, and public disclosure of methods and results.
% While greater access can increase the rigor of any high-stakes AI audit, this work pays particular attention to large language models. % because they are increasingly deployed in high-stakes settings with significant impacts.

Currently, evaluations of proprietary or pre-deployment AI systems are predominantly performed in-house by developers with selective disclosure of methods and outcomes. 
Some developers have voluntarily partnered with external auditors and provided them with black-box access to state-of-the-art systems \citep{metr, openai2023gpt4, anthropic2023challenges, touvron2023llama, kinniment2023evaluating}.
Additionally, some developers run programs for external researchers to support their internal evaluation process (e.g., OpenAI’s Preparedness Challenge \citep{oai_prep_challenge} and Red-Teaming Network \citep{oai_red_teaming_network}). % but these programs have consisted of black-box evaluations. % with limited input into how contributions are used thus far. 
%External stakeholders also conduct red-teaming, including academics (e.g., \citep{rando2022red, ma2023red, bhardwaj2023redteaming, zou2023universal, shah2023scalable}) and public stakeholders through challenges. 
% External groups, including academics \citep{rando2022red, ma2023red, bhardwaj2023redteaming, zou2023universal, shah2023scalable} and public stakeholders \citep{defcon}, have also conducted challenges for finding flaws in systems. 
% For example,  DEF CON 31, a US-based event for red-teaming language models, was supported by public entities including the White House Office of Science, Technology, and Policy \citep{defcon}. 
However, to public knowledge, these industry-based efforts have involved black-box and limited outside-the-box access such as \emph{model cards} \citep{mitchell2019model}. % stakeholders often have limited time and resources, and they are often conducted after models are released. 

\section{Limitations of Black-Box Access} \label{sec:limitations}

Black-box evaluations of AI systems are based on analysis of their inputs and outputs only. Such evaluations often involve assessing performance on test sets \citep{liang2022holistic, sarlin2020superglue, wang2018glue, srivastava2022beyond, hendrycks2020measuring, sun2024trustllm} or searching for inputs that elicit harmful outputs \citep{perez2022discovering, perez2022red, casper2023explore, kinniment2023evaluating, scheurer2023technical, wei2023jailbroken}. Generative AI audits often attempt to elicit undesirable capabilities or behaviors (e.g., \citep{mouton2023operational, sandbrink2023artificial, tann2023using, bran2023chemcrow, kinniment2023evaluating, soice2023can, shevlane2023model, hazell2023large, karanjai2022targeted, nasr_scalable_2023}).
However, black-box methods are inherently limited in their ability to identify harms or provide meaningful explanations.
Readers with a computer science background can consider the analogy of attempting to evaluate the performance of software without reading or modifying its source code.

\textbf{Black-box methods are not well suited to develop a generalizable understanding.} 
Black-box access limits evaluators to analyzing a system using only inputs and outputs. 
However, the vast number of possible inputs to AI systems makes it intractable to develop a complete understanding from this alone.  
This forces evaluators to rely on heuristics to produce `relevant' inputs for evaluation. For this reason, black-box methods have been shown to be unreliable for detecting failures that elude typical test sets including jailbreaks, adversarial inputs, or backdoors \citep{chen2017targeted, wei2023jailbroken, ziegler2022adversarial}. 

\textbf{Black-box access prevents system components from being studied separately.} 
Analyzing components of a system separately is ubiquitous in science and engineering. It enables engineers to trace problems to support more targeted interventions.
However, black-box access obscures what subsystems the AI system is composed of. For example, black-box access does not allow input or output filters to be studied separately from the rest of the system. 
Other issues can arise from a lack of outside-the-box access to data. 
Datasets can inform evaluations related to privacy and copyright \citep{roth_new_2023}, and can help to avoid problems from data contamination \citep{deng2023investigating, jacovi2023stop, golchin2023time}. 
Having a lack of outside-the-box knowledge about how the system is deployed also prevents evaluators from making a more practical assessment of broader societal impacts \citep{weidinger2023sociotechnical}.

\textbf{Black-box evaluations can produce misleading results.}
Since black-box evaluations rely entirely on the queries made to the system, they are biased by how evaluators design inputs \citep{wei2022chain, kapoor2023leakage,  narayanan2023evaluating}. 
This can lead to misconstrued conclusions about the system’s characteristics. For example, systems may satisfy simple statistical tests for non-discrimination, but they may still have undesirable biases in their underlying reasoning \citep{raghavan_limitations_2023}. Developers with information about the black-box tests can exacerbate this problem by modifying the model's output behavior on test cases despite unresolved flaws in its internal reasoning.
In addition, \citet{schaeffer2023emergent} provide examples of black-box prompt-based evaluation methods for language models that can lead to misunderstandings of their emergent capabilities.

\textbf{Black-box explanation methods are often unreliable.} 
Using black-box methods alone to produce explanations for an AI system's decisions is difficult \citep{hamon_bridging_2022, rudin2018please}. 
Many black-box techniques to provide counterfactual explanations for model decisions are misleading because they fail to reliably identify causal relationships between the system's input features and outputs \citep{chou_counterfactuals_2021}. 
Explanation methods for black-box systems can also be exploited by adversaries to produce misleading explanations for harmful decisions \citep{aivodji_fairwashing_2019, slack_fooling_2020}. 
Furthermore, when generative language models are asked to explain their decisions, their justifications do not tend to be faithful to their actual reasoning \citep{turpin2023language}.

\textbf{Black-box evaluations offer limited insights to help address failures.}
Black-box evaluations offer little insight into ways to address problems they discover. The main technique they enable is to train on problematic examples, but this can fail to address the underlying problem \citep{geirhos2020shortcut, du2023shortcut}, be sample-inefficient \citep{wang2023knowledge}, and may introduce new issues. 
Corrective actions are not robust when they fail to address a problem at its root. 
For example, some recent works have shown that safety measures built into large language models can be almost entirely undone by fine-tuning on a small number of harmful examples \citep{yang2023shadow, qi2023fine, lermen2023lora, zhan2023removing}. 
In contrast, white-box methods reveal more about the nature of flaws, facilitating more precise debugging methods \citep{wang2023knowledge}.

\section{Advantages of White-Box Access} \label{sec:white_box}

White-box offers a wider range of techniques to detect symptoms, understand causes, and mitigate harms in a targeted manner \citep{bucknall2023structured}.  
% White-box audits are especially appropriate when a system's weights will be released, which amounts to giving everyone white-box access that can be used to exploit or fine-tune the system.
Even for a system that will only be deployed as a black box, white-box audits are still more useful for finding problems.
% In this section, we survey research on why.
% See also \citet{bucknall2023structured} for a related taxonomy of system access and a survey of researchers' access needs. However, unlike \citet{bucknall2023structured}, we focus specifically on the access needs of auditors. 
Here, we survey techniques for white-box evaluations and their advantages over black-box ones. % We focus on current techniques and the pace of research in this area may improve many of the presented techniques. % In \Cref{app:supporting}, we discuss how supporting open research on evaluation techniques can help to equip future auditors with more powerful tools. 

\subsection{White-box attack algorithms are more effective and efficient.} \label{sec:attacks} 

In machine learning, \emph{adversarial attacks} refer to inputs that are designed specifically to make a system fail. AI systems have a long history of having unexpected failure modes that can be triggered by very subtle features in their inputs \citep{akhtar2018threat, zhang2020adversarial}. Attacks play a central role in evaluations because they help to assess a system's \emph{worst-case behavior}. 

\textbf{White-box algorithms produce stronger attacks.}
White-box algorithms allow for gradient-based optimization of adversarial inputs, which is powerful compared to simpler search methods. For example, white-box adversarial attack algorithms against vision systems typically use the gradient of the adversarial objective with respect to the input pixels to design adversarial inputs \citep{goodfellow2014explaining, papernot2016technical}. This is much more effective for finding vulnerabilities than unguided black-box search methods. Consequently, white-box attacks are dominant in vision applications.  
In reinforcement learning, white-box access to a target agent also helps develop stronger adversarial attacks against it \citep{casper2023red, wang2023adversarial}.
For language models, optimizing adversarial inputs with gradient-based methods is more challenging because text (unlike pixels) is discrete which prevents gradient propagation. 
Nonetheless, there are various state-of-the-art white-box techniques for attacking language models. These include using a differentiable approximation to the process of sampling text~\citep{wallace2019universal, song2020universal, guo2021gradient}, projecting adversarial embeddings onto text embeddings~\citep{wen2023hard}, and performing gradient-informed searches over modifications to textual changes~\citep{ebrahimi2017hotflip, li2018textbugger, ren2019generating, shin2020autoprompt, Liu2022CharacterlevelWA, jones2023automatically, zou2023universal}.

\textbf{Many black-box and grey-box attack algorithms are simply indirect or inefficient versions of white-box ones.}
Many black-box attacks against AI systems involve attacking a white-box model with a white-box algorithm and then testing the resulting attack on the target black-box model \citep{liu2016delving, zhou2018transferable}. For vision models, the main motivation behind studying black-box attacks is that white-box access is not always available to attackers \citep{bhambri2019survey}. 
Additionally, several of the most effective attacks against state-of-the-art models, such as GPT-4 and Claude-2, have simply been the result of transferring a white-box attack generated against an open-source model to the intended black-box target model \citep{zou2023universal}. 
Other types of black- and grey-box attack algorithms involve inefficiently estimating gradients by analyzing outputs or sampling probabilities across many queries \citep{grathwohl2017backpropagation, ilyas2018black} when more precise gradients could be obtained trivially with white-box access. 

\textbf{Latent space attacks help to make stronger assurances.}
Typically, AI systems are attacked by crafting \emph{inputs} meant to make them exhibit undesirable behavior. 
However, input space attacks are not well-suited to diagnose certain hard-to-find issues, including high-level misconceptions \citep{suri2023large}, anomalous failures \citep{ziegler2022adversarial}, backdoors \citep{chen2017targeted, wu2022backdoorbench}, and deception \citep{christiano2019worst, park2023ai}.
A complementary technique for attacking systems in the input space is to relax the problem and attack their internal \emph{latent representations}. 
The motivation of latent space attacks is that some failure modes are easier to find in the latent space than in the input space \citep{casper2024defending} because concepts important to the system's reasoning are represented at a higher level of abstraction inside the model \citep{athiwaratkun2015feature, johnston2023abstract, schwinn2023adversarial, turner2023activation, zou2023representation}.
Thus, robustness to latent attacks enables evaluators to make stronger assurances of safe worst-case performance. 
Latent space attacks are also more efficient to produce because they require less gradient propagation than input space attacks \citep{park2021reliably, qian2021towards}, allowing for more thorough debugging work to be conducted on a limited time and computing budget.
Latent space attacks are still an active area of research, but some works have emerged showing that robustness to latent space attacks effectively indicates robustness to input space attacks in vision models \citep{sankaranarayanan2018regularizing, kumari2019harnessing, zhou2019latent, osada2022latent, liu2023latent, zhang2023adversarial}. 
Since textual inputs to language models are discrete, only latent space attacks allow for the direct use of gradient-based optimization, rendering them especially useful for language models \citep{jiang2019smart, zhu2019freelb, liu2020adversarial, he2020deberta, kuangscale, li2021token, sae2022weighted, pan2022improved, kitada2023making}.

\textbf{White-box methods expand the attack toolbox.}
Some black-box attack methods are competitive, particularly against language models. These include methods based on local search \citep{prasad2022grips}, rejection sampling at scale~\citep{ganguli2022red}, Langevin dynamics~\citep{shi2022toward, kumar2022gradient}, evolutionary algorithms~\citep{lapid2023open}, and reinforcement learning~\citep{deng2022rlprompt, perez2022red, casper2023explore}. Additionally, some of the most effective methods for attacking language models involve human or human-guided generation of adversarial prompts \citep{liu2023jailbreaking, wei2023jailbroken, shah2023scalable}. However, even when black-box attacks are useful, white-box 
algorithms are complementary because they generate qualitatively different kinds of attacks. For example, many black-box techniques produce attacks that appear as natural language (e.g., \citep{shah2023scalable}) while white-box algorithms are state-of-the-art for synthesizing adversarial prompts which appear to humans as unintelligible text (e.g., \citep{zou2023universal}). 
Relying only on black-box methods would likely miss out on these types of failures entirely because finding them with black-box searches or human inductive biases would be highly improbable. 
Black- and white-box methods can also be combined to conduct \emph{hybrid} attacks by using the results of one method as an initialization for another. 
Combinations of attacks tend to be better at helping humans find vulnerabilities than a single method alone \citep{casper2023red1}.

\subsection{White-box interpretability tools aid in diagnostics.} \label{sec:interp}

While it is possible to infer properties of a system from studying inputs and outputs, understanding its internal processes allows evaluators to more thoroughly assess its trustworthiness \citep{gilpin2019explaining, rauker2023toward}. 
Interpreting the inner mechanisms of models has been recognized as a key part of agendas for reducing harms from AI systems \citep{hubinger2020overview, ngo2022alignment, jose2021fairness}, and explaining how models make specific decisions has also been recognized as a way to protect the rights of individuals affected by AI \citep{gryz2021black, ploug2021right}.

\textbf{White-box interpretability tools help evaluators discover novel failure modes.}
White-box algorithms and interpretability tools have aided researchers in finding vulnerabilities. 
Examples have involved identifying novel attacks \citep{ziegler2022adversarial}, internal representations of spurious features \citep{gandelsman2023interpreting, marks2024sparse}, brittle feature representations \citep{carter2019exploring, ghorbani2020neuron, mu2020compositional, hernandez2021natural, yuksekgonul2022post, casper2022robust, casper2022diagnostics, casper2023red1, wu2023depn}, and limitations of key-value memories in transformers \citep{geva2020transformer, meng2022locating, geva2023dissecting}. 
As an added benefit, attributing the problem to specific parts of the system's architecture or representations allows developers to address it in a more precise way \citep{wang2023knowledge}. 

\textbf{Studying internal representations can help to establish the presence or lack of specific capabilities.}
White-box methods allow for more precise identification of what knowledge and capabilities a system has \citep{alain2018understanding, belinkov2022probing}. Tools such as concept vectors \citep{kim_interpretability_2018, abid2022meaningfully} and probes \cite{conneau2018you} allow humans to assess the extent to which system internals can be understood in terms of familiar concepts. 
For example, these techniques have been used to 
study features related to fairness in visual classifiers \citep{goyal2022fairness}, provide evidence that language models internally represent space and time \citep{gurnee2023language}, and show that networks sometimes represent truth-like features along linear directions \citep{burns2022discovering, marks2023geometry}. 
Methods for this are imperfect and still an active area of research \citep{ravichander2020probing, elazar2021amnesic, antverg2021pitfalls}, but interpretations like these offer a potentially powerful way to identify whether a model represents specific concepts. 

Consider an example.
Suppose that an auditor wants to assess \emph{sycophancy}: a language model's tendency to pander to the biases of users who chat with it \citep{perez2022discovering, sharma2023understanding}.
For example, an evaluator might be concerned that the system will respond differently when the user says they are conservative or liberal in the chat. 
Black-box techniques could only be used to argue that the system is not sycophantic by producing examples and analyzing them for apparent sycophancy.
However, a white-box interpretability-based approach could offer much more information. 
For example, if it were not possible for a classifier to distinguish whether the user revealed themselves to be conservative or liberal from the model's internal representations, then this would offer stronger evidence that the system will reliably not exhibit this type of sycophancy. 

\textbf{Mechanistic understanding helps to make stronger assurances.} In general, it is impossible to make guarantees about black-box systems using a finite number of queries without additional assumptions. 
In contrast to black-box methods, which can only show the existence of failures by finding inputs that elicit them, thoroughly understanding the computations inside of a model gives auditors a complementary way to find evidence against the existence of failure modes.   
A mechanistic understanding can help researchers develop a predictive model of how the system would act for broad classes of inputs. 
Some works have aimed to provide thorough investigations of how networks perform simple tasks \citep{cammarata2020thread, nanda2023progress, zhong2023clock}.
Although scaling thorough analysis is an open challenge \citep{tegmark_provably_2023}, it offers a strategy for making strong assurances. 
Recent works have attempted to make progress on this problem by using sparse autoencoders to allow evaluators to more thoroughly study the features represented inside of large language models \citep{cunningham2023sparse, bricken2023monosemanticity, marks2024sparse}.

\textbf{White-box methods expand the toolbox for explaining specific AI system decisions.}
As discussed in \Cref{sec:background}, existing regulatory frameworks have been designed with specific desiderata for model explanations in order to determine accountability and protect individual rights. Many techniques are used to provide explanations of model behaviors during audits \citep{zhang2022explainable}. Black-box techniques can only attribute decisions to input features using techniques that involve modifying inputs and analyzing how model outputs change \citep{ribeiro2016model}.
However, these techniques are frequently misleading \citep{rudin2018please} and can fail to reliably identify causal relationships between the system’s input features and output \citep{chou_counterfactuals_2021}.
White-box access expands and strengthens the toolbox by allowing for gradient-based techniques \citep{das2020opportunities, linardatos2020explainable, zhao2023explainability, dwivedi2023explainable}. 
It also allows for explainability tools to be combined with interpretations of the model mechanisms to explain a model's behaviors in terms of more abstract concepts.

\subsection{Fine-tuning reveals risks from latent knowledge or post-deployment modifications.} \label{sec:fine-tuning}

State-of-the-art AI systems are typically trained on large amounts of internet data, often in multiple stages. 
This can cause them to learn undesirable capabilities, such as knowledge of how to perform illegal activities \citep{shah2023scalable, zou2023representation} or the ability to produce harmful content \citep{Geirhos2020ShortcutLI, luccioni2021s, birhane2021multimodal, rando2022red, birhane2023into, thiel2023identifying, thiel2023generative, shahbazi2023representation}. 
Developers attempt to remove harmful abilities through fine-tuning, but they can unexpectedly resurface through ``jailbreaks'' \citep{jailbreakChat2023, oneal2023, li2023multi, liu2023jailbreaking, rao2023tricking, wei2023jailbroken, shen2023anything, qi2023visual, yu2023gptfuzzer, deng2023jailbreaker, yong2023lowresource, zou2023universal, chao2023jailbreaking, shah2023scalable, tsai2023ring} or further fine-tuning models on a small number of new examples \citep{yang2023shadow, qi2023fine, lermen2023lora, zhan2023removing}.
The existence of harmful dormant capabilities in models thus poses risks from attacks and fine-tuning, especially if they are leaked (e.g. Stable Diffusion \citep{lanz2023stable}), open-sourced (e.g., Llama-2 \citep{touvron2023llama}), or deployed with fine-tuning access via API (e.g., GPT-3.5 \citep{openai2023gpt}). 
Consequently, being able to fine-tune the model offers another strategy to search for evidence of undesirable capabilities and assess the risks in deployment.

\section{Advantages of Outside-the-Box Access} \label{sec:outside}

In addition to having access to AI systems themselves, giving auditors outside-the-box access to contextual information also helps to identify risks. This can include methodological details, source code, documentation, hyperparameters, training data, deployment details, and the findings of internal evaluations. While it can come in many types, all outside-the-box information can be useful to auditors for three common reasons: (1) helping auditors more effectively design and implement tests, (2) offering clues about potential issues, and (3) helping auditors trace problems to their sources. See also \Cref{app:technical} where we discuss how outside-the-box access to technical assistance from developers can also be useful for auditors.

\textbf{Code, documentation, and hyperparameters help auditors work more efficiently.} As discussed in \Cref{sec:white_box}, audits can require a number of technical evaluations. Having code and documentation from developers can streamline the process of designing them. For example, consider fine-tuning evaluations. Fine-tuning a model typically requires precisely configured code and hyperparameters that have been carefully selected after extensive testing, often over the course of weeks or months. Using the developer's existing resources is a much more efficient option for auditors compared to re-implementing everything from scratch.

\textbf{Access to methodological details helps to identify risks.} 
Knowing methodological details can reveal shortcuts taken during development, which can guide evaluators toward discovering problems. For example, if a system was trained with human-generated data using a non-representative cohort of humans, this can suggest specific social biases that the system may have internalized \citep{santurkar2023whose, sun2023aligning, casper_open_2023}.
Knowing the findings of internal evaluations is especially useful for helping auditors target their efforts toward a set of complementary evaluations. Furthermore, when developers attempt to mitigate flaws, auditors can better assess the effectiveness of these efforts if they have detailed information about the attempted mitigation (e.g., fine-tuning datasets, both old and new versions of model weights, etc.) \citep{pozzobon2023challenges}. 

\textbf{Access to data helps auditors trace problems and assess fair use.} 
Recent work has highlighted the ability of dataset audits to identify harmful and biased content used to train models \citep{Geirhos2020ShortcutLI, luccioni2021s, birhane2021multimodal, birhane2023into, thiel2023identifying, thiel2023generative, shahbazi2023representation}. 
For example, \citet{birhane2021multimodal} and \citet{thiel2023identifying} were able to identify previously-overlooked examples of unintended hateful, sexual, and child-abuse-related content in widely-used training datasets. 
Access to training data also helps to investigate risks of data-poisoning attacks (which is especially important for systems trained on internet data) \citep{carlini2023poisoning, rando2023universal, wang2023exploitability, wan2023poisoning}.
Meanwhile, legal questions are currently being debated involving the extent to which training generative AI systems on copyrighted content constitutes fair use  
\citep{karamolegkou2023copyright,henderson2023foundation, rodriguez2023copyright, nytimes_v_openai_2023}.
Auditors may require access to training data to properly assess whether it was used in accordance with copyright law.

\textbf{Contextual information makes it easier to hold developers accountable.} 
Requirements to produce documentation place greater responsibility on developers to detail their methods, especially if subject to regulatory penalty defaults \citep{wieringa2020account, yew2022penalty}. 
Contextual information provides information about whether developers made decisions in a responsible manner \citep{mokander_ethics-based_2021}. 
For example, documentation can provide insights into why certain design choices were made over others \citep{metcalf2022relationship}.
Datasets and training details can help trace risks to intentional choices, and internal evaluation reports provide insights into how the developer responded to findings.
By increasing the scrutiny placed on decisions in the development process, requirements for greater methodological transparency to auditors can deter developers from taking risks in the first place \citep{casper_open_2023, lambert2023entangled}.

% \section{Security concerns are nonunique and addressable.} \label{sec:security}
\section{Methods to Address Security Risks} \label{sec:security}

A concern with white- and outside-the-box audits is an increased risk that a developer's models or intellectual property could be leaked \citep{bommasani2023foundation}.
In turn, leaks could compromise developers' trade secrets and pose risks to the public if they enable misuse \citep{nevo_securing_2023}. Widespread norms for secure audits in AI do not yet exist, but there is precedent in other fields for navigating similar challenges to enable secure oversight. %as there are in other industries. 
The risk of leaks can be minimized through several \textit{technical}, \textit{physical}, and \textit{legal} mechanisms. With these measures, developers can provide white- and outside-the-box access to auditors without the system's parameters leaving their servers. These can reduce leakage risks to a level comparable to ones posed by common existing practices.

\textbf{Technical: API access can offer remote auditors de facto white-box access.} 
Forms of structured access, particularly research application programming interfaces (APIs) \citep{bucknall2023structured, shevlane2022structured, bluemke_exploring_2023, nnsight}, could enable auditors to analyze systems using some white-box tools without giving auditors direct access to model parameters. 
We refer to this form of access as \emph{de facto} white-box access if it enables auditors to indirectly run arbitrary white-box processes on models while restricting direct access to model parameters.
One example of running an algorithm that accesses a model's parameters via API is an OpenAI GPT-3.5 API which allows for fine-tuning \citep{openai2023gpt}. 
However, more customizable APIs (e.g., \citep{nnsight}) would be needed to allow for more flexible access. 
Another proposed paradigm is a \emph{flexible query API} \citep{openmined_how_2023}, where auditors are given complete access to mock versions of a model and data. 
The auditors then develop evaluations using their complete access to these mock artifacts before submitting them to be run on the true model and dataset. 
This allows auditors to better customize their evaluations.

The goal of API access is to ensure that the system cannot easily be reconstructed. However, prohibiting the sharing of weights is neither necessary nor sufficient for this. 
For example, sharing a small subset of weights with auditors is unlikely to pose significant security risks \citep{martic2018scaling}, but sharing other information, such as the product of weights with their pre-synaptic neuron's activations, may allow for parameter reconstruction. 
This suggests the need for a process by which a developer can raise grievances about specific requests from auditors and have them adjudicated. 
Overall, while conceptually simple, designing APIs that simultaneously provide the comprehensiveness, flexibility, and security required for rigorous auditing is an open area of research \citep{bucknall2023structured}.
Greater clarity is required regarding how to balance different desiderata. For example, more comprehensive access may impact security by facilitating model reconstruction, as discussed above.
In \Cref{app:supporting}, we also discuss how investment, research, and development into secure auditing infrastructure can help with progress toward improved techniques.

\textbf{Physical: Secure research environments can be used for auditors given unrestricted white-box access.}
Auditing personnel could securely be given white-box access to a system by hosting them on-site at the developer’s facilities in a secure research environment. This is a common practice in other industries despite the costs of requiring auditors to be physically on-site \citep{Brom2022OnsiteIA, openmined_how_2023, trager2023international}. %, and there is precedent in other fields for solving very challenging oversight problems. 
For example, the International Atomic Energy Agency employs over 300 expert inspectors from approximately 80 countries who do on-site inspections of nuclear facilities \citep{IAEASafeguardsInspector2016, IAEA2023}. 
Compared to API access, secure research environments could allow auditors to access systems more flexibly and efficiently while minimizing the risk that the model is leaked or reconstructed. Safeguards for limiting information leakage through lab employees (such as NDAs) are already common practice and could be adapted for application to on-site auditors.
However, it is unclear whether, absent external legal structures, labs could incentivize adherence to protective measures to the same extent as with employees.

\textbf{Legal: Other industries have developed practices to address the risk of leaks from audits.} 
Across many industries, auditors require privileged access to systems and data in order to perform effective assessments.
There are established mechanisms from other fields, such as financial auditing, employed to reduce the risk of leaks. 
In the finance industry, this manifests in three main ways. 
First, policies for confidentiality and handling of sensitive information are enforced by formal training and non-disclosure clauses in contracts to hold auditors accountable for violations \citep{eycode}. 
Second, there are clear terms of engagement that govern the relationship between auditor and auditee. These typically include specific restrictions on confidentiality expectations tailored to a particular client \citep{asa2006auditing}. While some specifics of auditing are managed through contracting, the Public Company Accounting Oversight Board (PCAOB) requires all registered auditors to adhere to common standards in the US \citep{sarbanesoxley2002}. 
Finally, auditors can be legally required to avoid conflicts of interest. In the US, this is done through a regime specifying general provisions for auditor independence (such as reporting requirements) outlined in Title II of the Sarbanes Oxley Act \citep{sarbanesoxley2002}. Provisions are enforced through various agencies, including the Securities Exchange Commission (SEC), which prevents auditor manipulation or the use of financial information for personal gain \citep{sec-reg-m}. 
This type of enforcement could allow for auditors to be held accountable in a way that could reduce the risk of leakage to a level comparable to risks posed by the developer's employees. In fact, employees may pose greater risks of sharing tacit knowledge with competitors than auditors because developers regularly attempt to recruit AI researchers from competitors.

\section{Discussion} \label{sec:discussion}

\textbf{White- and outside-the-box audits offer several benefits to developers.}  
One potential benefit of more rigorous audits for developers is increased credibility by creating the perception that their systems are of higher quality.
Meanwhile, white- and outside-the-box evaluations also offer developers greater insight into addressing problems with systems they build \citep{raji2019actionable}. We discuss in \Cref{sec:limitations} how black-box evaluations can only establish when problems exist, while white- and outside-the-box methods can provide a clearer diagnosis to help address them. For example, if a flaw in a system can be attributed to a specific set of components, this enables more targeted interventions to fix it.

\textbf{However, absent legal requirements, developers have strong incentives to limit access granted to external auditors.} 
Thus far, external audits of state-of-the-art AI systems, when they have occurred, have been black-box (to public knowledge) \citep{metr, kinniment2023evaluating, openai2023gpt, anthropic2023challenges, touvron2023llama}, suggesting a failure of existing incentive structures to provide greater access to auditors.
Developers are typically reluctant to provide more permissive access to their models and related resources \citep{raji2022outsider}. 
This may stem from concerns that information collected through white- and outside-the-box audits could be leaked \citep{bommasani2023foundation} (though this is addressable--see \Cref{sec:security}). 
Furthermore, it could expose a system’s lackluster performance, vulnerabilities, or poor risk management processes by developers, which could lead to reputational harm and possibly legal liability \citep{henderson2023wheres}.

\textbf{Current black-box audits may set a precedent for future norms.} 
Established norms frequently become ``sticky'' and entrenched in regulatory regimes \citep{nielson2018sticky}.
Accordingly, the current norm for black-box audits \citep{metr, kinniment2023evaluating, openai2023gpt, anthropic2023challenges, touvron2023llama} may set the future standard.
Furthermore, in policy debates about audits, industry actors have also lobbied for limiting external auditors to black-box access \citep{google_consultation_2021}. 
Without sufficient access and resources, non-industry researchers will struggle to iterate upon methods for more thorough audits \citep{bucknall2023structured}. 
Over time, this could limit or bias public understanding of AI systems. 
A lack of open research on transparency tools and the view that there is little social benefit from greater transparency are mutually reinforcing.
\Cref{app:supporting} expands on how investment, research, and development into auditing techniques and infrastructure can facilitate further progress.

\textbf{Low-quality audits can be counterproductive.}
Poor (e.g., black-box) audits can have counterproductive effects: they can increase public or regulatory trust in systems on false grounds, preventing appropriate levels of external scrutiny \citep{fuerman2009bernard, linthicum2010social}. They also enable safety- or ethics-washing by developers \citep{farrell1996cheap, westra2021virtue, krawiec2003cosmetic} who make AI systems that contribute to risks without sufficiently investing in methods to address them.

\textbf{Conclusion:} \label{sec:conclusion}
We have argued that providing auditors with white-box and thorough outside-the-box access to systems is feasible and allows for more meaningful oversight from audits. 
% To gain more meaningful insight from audits, policies can be designed to (1) require transparency regarding the forms of access given to auditors and the methods auditors use and (2) ensure that systems deployed in high-stakes applications are subject to audits with greater levels of access.
We draw two conclusions. First transparency regarding model access and evaluation methods is necessary to properly interpret the results of an AI audit. Second, white- and outside-the-box access allow for substantially more scrutiny than black-box access alone. When higher levels of scrutiny are desired, audits should be conducted with higher levels of access.
Finally, we emphasize that white-box and thorough outside-the-box access are necessary but not sufficient for rigor. Audits can and do fail for many reasons. Without careful institutional design, the incentives of developers and auditors may result in audits that do not consistently align with public interest \citep{raji2020closing, raji2022outsider, costanza-chock2022who, anderljung2023publicly, birhane2024ai, ojewale2024towards}. In \Cref{app:beyond}, we examine additional ways in which the quality of audits can be compromised.

\section*{Ethics Statement}

Given the role of rigorous audits in improving accountability and representing public interests, we expect the foreseeable impacts of this paper to be positive. 
However, audits can fail to benefit the public for a variety of reasons. 
We discuss these challenges in \Cref{app:beyond}.

\section*{Contributions}

Carson Ezell and Stephen Casper were the central writers and organizers. Charlotte Siegmann, Kevin Wei, Andreas Haupt, and Taylor Curtis contributed primarily to \Cref{sec:background}. Noam Kolt contributed primarily to \Cref{sec:discussion} and \Cref{app:beyond}. Jérémy Scheurer, Marius Hobbhahn, and Lee Sharkey contributed primarily to \Cref{sec:limitations}, \Cref{sec:interp}, \Cref{sec:discussion}, \Cref{app:supporting}, and \Cref{app:beyond}. Satyapriya Krishna contributed primarily to \Cref{sec:attacks}, Marvin von Hagen and Silas Alberti contributed primarily to \Cref{app:goals} and \Cref{sec:squo}. Qinyi Sun, Michael Gerovitch, and Benjamin Bucknall contributed primarily to \Cref{sec:black_white_grey}, \Cref{sec:security}, and \Cref{sec:discussion}. Alan Chan, David Bau, Max Tegmark, David Krueger, and Dylan Hadfield-Menell offered high-level feedback and guidance. 

\begin{acks}
We are grateful for discussions and feedback from Markus Anderljung, Thomas Krendl Gilbert, Matthijs Maas, Javier Rando, Stewart Slocum, Luke Bailey, Adam Jermyn, Alexandra Bates, Miles Wang, Audrey Chang, and Erik Jenner.
\end{acks}

\newpage

\bibliographystyle{ACM-Reference-Format}  % use this for submission
\bibliography{bibliography}


\begin{thebibliography}{355}

%%% the Web address.

\ifx \showCODEN    \undefined \def \showCODEN     #1{\unskip}     \fi
\ifx \showDOI      \undefined \def \showDOI       #1{#1}\fi
\ifx \showISBNx    \undefined \def \showISBNx     #1{\unskip}     \fi
\ifx \showISBNxiii \undefined \def \showISBNxiii  #1{\unskip}     \fi
\ifx \showISSN     \undefined \def \showISSN      #1{\unskip}     \fi
\ifx \showLCCN     \undefined \def \showLCCN      #1{\unskip}     \fi
\ifx \shownote     \undefined \def \shownote      #1{#1}          \fi
\ifx \showarticletitle \undefined \def \showarticletitle #1{#1}   \fi
\ifx \showURL      \undefined \def \showURL       {\relax}        \fi
% invisible to TeX
\providecommand\bibfield[2]{#2}
\providecommand\bibinfo[2]{#2}
\providecommand\natexlab[1]{#1}
\providecommand\showeprint[2][]{arXiv:#2}

\bibitem[Abdalla and Abdalla(2021)]%
        {abdalla2021grey}
\bibfield{author}{\bibinfo{person}{Mohamed Abdalla} {and} \bibinfo{person}{Moustafa Abdalla}.} \bibinfo{year}{2021}\natexlab{}.
\newblock \showarticletitle{The Grey Hoodie Project: Big tobacco, big tech, and the threat on academic integrity}. In \bibinfo{booktitle}{\emph{Proceedings of the 2021 AAAI/ACM Conference on AI, Ethics, and Society}}. \bibinfo{pages}{287--297}.
\newblock


\bibitem[Abid et~al\mbox{.}(2022)]%
        {abid2022meaningfully}
\bibfield{author}{\bibinfo{person}{Abubakar Abid}, \bibinfo{person}{Mert Yuksekgonul}, {and} \bibinfo{person}{James Zou}.} \bibinfo{year}{2022}\natexlab{}.
\newblock \showarticletitle{Meaningfully debugging model mistakes using conceptual counterfactual explanations}. In \bibinfo{booktitle}{\emph{International Conference on Machine Learning}}. PMLR, \bibinfo{pages}{66--88}.
\newblock


\bibitem[Adebayo et~al\mbox{.}(2020)]%
        {adebayo2020debugging}
\bibfield{author}{\bibinfo{person}{Julius Adebayo}, \bibinfo{person}{Michael Muelly}, \bibinfo{person}{Ilaria Liccardi}, {and} \bibinfo{person}{Been Kim}.} \bibinfo{year}{2020}\natexlab{}.
\newblock \showarticletitle{Debugging tests for model explanations}.
\newblock \bibinfo{journal}{\emph{arXiv preprint arXiv:2011.05429}} (\bibinfo{year}{2020}).
\newblock


\bibitem[Agarwal et~al\mbox{.}(2022)]%
        {agarwal2022openxai}
\bibfield{author}{\bibinfo{person}{Chirag Agarwal}, \bibinfo{person}{Satyapriya Krishna}, \bibinfo{person}{Eshika Saxena}, \bibinfo{person}{Martin Pawelczyk}, \bibinfo{person}{Nari Johnson}, \bibinfo{person}{Isha Puri}, \bibinfo{person}{Marinka Zitnik}, {and} \bibinfo{person}{Himabindu Lakkaraju}.} \bibinfo{year}{2022}\natexlab{}.
\newblock \showarticletitle{Openxai: Towards a transparent evaluation of model explanations}.
\newblock \bibinfo{journal}{\emph{Advances in Neural Information Processing Systems}}  \bibinfo{volume}{35} (\bibinfo{year}{2022}), \bibinfo{pages}{15784--15799}.
\newblock


\bibitem[{AI Safety Summit}(2023)]%
        {bletchley2023}
\bibfield{author}{\bibinfo{person}{{AI Safety Summit}}.} \bibinfo{year}{2023}\natexlab{}.
\newblock \bibinfo{title}{The {Bletchley} {Declaration} by {Countries} {Attending} the {AI} {Safety} {Summit}}.
\newblock
\newblock
\urldef\tempurl%
\url{https://www.gov.uk/government/publications/ai-safety-summit-2023-the-bletchley-declaration/the-bletchley-declaration-by-countries-attending-the-ai-safety-summit-1-2-november-2023}
\showURL{%
\tempurl}


\bibitem[Aivodji et~al\mbox{.}(2019)]%
        {aivodji_fairwashing_2019}
\bibfield{author}{\bibinfo{person}{Ulrich Aivodji}, \bibinfo{person}{Hiromi Arai}, \bibinfo{person}{Olivier Fortineau}, \bibinfo{person}{Sébastien Gambs}, \bibinfo{person}{Satoshi Hara}, {and} \bibinfo{person}{Alain Tapp}.} \bibinfo{year}{2019}\natexlab{}.
\newblock \showarticletitle{Fairwashing: the risk of rationalization}. In \bibinfo{booktitle}{\emph{Proceedings of the 36th {International} {Conference} on {Machine} {Learning}}}. \bibinfo{publisher}{PMLR}, \bibinfo{pages}{161--170}.
\newblock
\urldef\tempurl%
\url{https://proceedings.mlr.press/v97/aivodji19a.html}
\showURL{%
\tempurl}
\newblock
\shownote{ISSN: 2640-3498}.


\bibitem[Akhtar and Mian(2018)]%
        {akhtar2018threat}
\bibfield{author}{\bibinfo{person}{Naveed Akhtar} {and} \bibinfo{person}{Ajmal Mian}.} \bibinfo{year}{2018}\natexlab{}.
\newblock \showarticletitle{Threat of adversarial attacks on deep learning in computer vision: A survey}.
\newblock \bibinfo{journal}{\emph{Ieee Access}}  \bibinfo{volume}{6} (\bibinfo{year}{2018}), \bibinfo{pages}{14410--14430}.
\newblock


\bibitem[Alain and Bengio(2018)]%
        {alain2018understanding}
\bibfield{author}{\bibinfo{person}{Guillaume Alain} {and} \bibinfo{person}{Yoshua Bengio}.} \bibinfo{year}{2018}\natexlab{}.
\newblock \showarticletitle{Understanding intermediate layers using linear classifier probes}.
\newblock  (\bibinfo{year}{2018}).
\newblock
\showeprint[arxiv]{1610.01644}~[stat.ML]


\bibitem[Albert(2023)]%
        {jailbreakChat2023}
\bibfield{author}{\bibinfo{person}{Alex Albert}.} \bibinfo{year}{2023}\natexlab{}.
\newblock \showarticletitle{Jailbreak Chat}.
\newblock  (\bibinfo{year}{2023}).
\newblock
\urldef\tempurl%
\url{https://www.jailbreakchat.com/}
\showURL{%
\tempurl}


\bibitem[Anderljung et~al\mbox{.}(2023a)]%
        {anderljung2023frontier}
\bibfield{author}{\bibinfo{person}{Markus Anderljung}, \bibinfo{person}{Joslyn Barnhart}, \bibinfo{person}{Jade Leung}, \bibinfo{person}{Anton Korinek}, \bibinfo{person}{Cullen O'Keefe}, \bibinfo{person}{Jess Whittlestone}, \bibinfo{person}{Shahar Avin}, \bibinfo{person}{Miles Brundage}, \bibinfo{person}{Justin Bullock}, \bibinfo{person}{Duncan Cass-Beggs}, {et~al\mbox{.}}} \bibinfo{year}{2023}\natexlab{a}.
\newblock \showarticletitle{Frontier AI regulation: Managing emerging risks to public safety}.
\newblock \bibinfo{journal}{\emph{arXiv preprint arXiv:2307.03718}} (\bibinfo{year}{2023}).
\newblock


\bibitem[Anderljung et~al\mbox{.}(2023b)]%
        {anderljung2023publicly}
\bibfield{author}{\bibinfo{person}{Markus Anderljung}, \bibinfo{person}{Everett~Thornton Smith}, \bibinfo{person}{Joe O'Brien}, \bibinfo{person}{Lisa Soder}, \bibinfo{person}{Benjamin Bucknall}, \bibinfo{person}{Emma Bluemke}, \bibinfo{person}{Jonas Schuett}, \bibinfo{person}{Robert Trager}, \bibinfo{person}{Lacey Strahm}, {and} \bibinfo{person}{Rumman Chowdhury}.} \bibinfo{year}{2023}\natexlab{b}.
\newblock \showarticletitle{Towards Publicly Accountable Frontier LLMs: Building an External Scrutiny Ecosystem under the ASPIRE Framework}.
\newblock  (\bibinfo{year}{2023}).
\newblock
\showeprint[arxiv]{2311.14711}~[cs.CY]


\bibitem[Angwin et~al\mbox{.}(2022)]%
        {angwin2022machine}
\bibfield{author}{\bibinfo{person}{Julia Angwin}, \bibinfo{person}{Jeff Larson}, \bibinfo{person}{Surya Mattu}, {and} \bibinfo{person}{Lauren Kirchner}.} \bibinfo{year}{2022}\natexlab{}.
\newblock \showarticletitle{Machine bias}.
\newblock In \bibinfo{booktitle}{\emph{Ethics of data and analytics}}. \bibinfo{publisher}{Auerbach Publications}, \bibinfo{pages}{254--264}.
\newblock


\bibitem[Anthropic(2023)]%
        {anthropic2023challenges}
\bibfield{author}{\bibinfo{person}{Anthropic}.} \bibinfo{year}{2023}\natexlab{}.
\newblock \showarticletitle{Challenges in evaluating AI systems}.
\newblock  (\bibinfo{year}{2023}).
\newblock
\urldef\tempurl%
\url{https://www.anthropic.com/index/evaluating-ai-systems}
\showURL{%
\tempurl}


\bibitem[Antverg and Belinkov(2021)]%
        {antverg2021pitfalls}
\bibfield{author}{\bibinfo{person}{Omer Antverg} {and} \bibinfo{person}{Yonatan Belinkov}.} \bibinfo{year}{2021}\natexlab{}.
\newblock \showarticletitle{On the pitfalls of analyzing individual neurons in language models}.
\newblock \bibinfo{journal}{\emph{arXiv preprint arXiv:2110.07483}} (\bibinfo{year}{2021}).
\newblock


\bibitem[ASA(2006)]%
        {asa2006auditing}
\bibfield{author}{\bibinfo{person}{Compiled Auditing~Standard ASA}.} \bibinfo{year}{2006}\natexlab{}.
\newblock \bibinfo{title}{Auditing standard ASA 210 terms of audit engagements}.
\newblock
\newblock


\bibitem[Athiwaratkun and Kang(2015)]%
        {athiwaratkun2015feature}
\bibfield{author}{\bibinfo{person}{Ben Athiwaratkun} {and} \bibinfo{person}{Keegan Kang}.} \bibinfo{year}{2015}\natexlab{}.
\newblock \showarticletitle{Feature representation in convolutional neural networks}.
\newblock \bibinfo{journal}{\emph{arXiv preprint arXiv:1507.02313}} (\bibinfo{year}{2015}).
\newblock


\bibitem[Baghai and Becker(2020)]%
        {baghai2020reputations}
\bibfield{author}{\bibinfo{person}{Ramin~P. Baghai} {and} \bibinfo{person}{Bo Becker}.} \bibinfo{year}{2020}\natexlab{}.
\newblock \showarticletitle{Reputations and credit ratings: {Evidence} from commercial mortgage-backed securities}.
\newblock \bibinfo{journal}{\emph{Journal of Financial Economics}} \bibinfo{volume}{135}, \bibinfo{number}{2} (\bibinfo{date}{Feb.} \bibinfo{year}{2020}), \bibinfo{pages}{425--444}.
\newblock
\showISSN{0304-405X}
\urldef\tempurl%
\url{https://doi.org/10.1016/j.jfineco.2019.06.001}
\showDOI{\tempurl}


\bibitem[Belinkov(2022)]%
        {belinkov2022probing}
\bibfield{author}{\bibinfo{person}{Yonatan Belinkov}.} \bibinfo{year}{2022}\natexlab{}.
\newblock \showarticletitle{Probing classifiers: Promises, shortcomings, and advances}.
\newblock \bibinfo{journal}{\emph{Computational Linguistics}} \bibinfo{volume}{48}, \bibinfo{number}{1} (\bibinfo{year}{2022}), \bibinfo{pages}{207--219}.
\newblock


\bibitem[Bengio et~al\mbox{.}({[n.\,d.]})]%
        {bengiomanaging}
\bibfield{author}{\bibinfo{person}{Yoshua Bengio}, \bibinfo{person}{Geoffrey Hinton}, \bibinfo{person}{Andrew Yao}, \bibinfo{person}{Dawn Song}, \bibinfo{person}{Pieter Abbeel}, \bibinfo{person}{Yuval~Noah Harari}, \bibinfo{person}{Ya-Qin Zhang}, \bibinfo{person}{Lan Xue}, \bibinfo{person}{Shai Shalev-Shwartz}, \bibinfo{person}{Gillian Hadfield}, {et~al\mbox{.}}} \bibinfo{year}{[n.\,d.]}\natexlab{}.
\newblock \showarticletitle{Managing AI Risks in an Era of Rapid Progress}.
\newblock  (\bibinfo{year}{[n.\,d.]}).
\newblock


\bibitem[Bengio et~al\mbox{.}(2023)]%
        {bengio2023managing}
\bibfield{author}{\bibinfo{person}{Yoshua Bengio}, \bibinfo{person}{Geoffrey Hinton}, \bibinfo{person}{Andrew Yao}, \bibinfo{person}{Dawn Song}, \bibinfo{person}{Pieter Abbeel}, \bibinfo{person}{Yuval~Noah Harari}, \bibinfo{person}{Ya-Qin Zhang}, \bibinfo{person}{Lan Xue}, \bibinfo{person}{Shai Shalev-Shwartz}, \bibinfo{person}{Gillian Hadfield}, {et~al\mbox{.}}} \bibinfo{year}{2023}\natexlab{}.
\newblock \showarticletitle{Managing AI Risks in an Era of Rapid Progress}.
\newblock \bibinfo{journal}{\emph{arXiv preprint arXiv:2310.17688}} (\bibinfo{year}{2023}).
\newblock


\bibitem[Bhambri et~al\mbox{.}(2019)]%
        {bhambri2019survey}
\bibfield{author}{\bibinfo{person}{Siddhant Bhambri}, \bibinfo{person}{Sumanyu Muku}, \bibinfo{person}{Avinash Tulasi}, {and} \bibinfo{person}{Arun~Balaji Buduru}.} \bibinfo{year}{2019}\natexlab{}.
\newblock \showarticletitle{A survey of black-box adversarial attacks on computer vision models}.
\newblock \bibinfo{journal}{\emph{arXiv preprint arXiv:1912.01667}} (\bibinfo{year}{2019}).
\newblock


\bibitem[Birhane et~al\mbox{.}(2022)]%
        {birhane2022values}
\bibfield{author}{\bibinfo{person}{Abeba Birhane}, \bibinfo{person}{Pratyusha Kalluri}, \bibinfo{person}{Dallas Card}, \bibinfo{person}{William Agnew}, \bibinfo{person}{Ravit Dotan}, {and} \bibinfo{person}{Michelle Bao}.} \bibinfo{year}{2022}\natexlab{}.
\newblock \showarticletitle{The values encoded in machine learning research}. In \bibinfo{booktitle}{\emph{Proceedings of the 2022 ACM Conference on Fairness, Accountability, and Transparency}}. \bibinfo{pages}{173--184}.
\newblock


\bibitem[Birhane et~al\mbox{.}(2023)]%
        {birhane2023into}
\bibfield{author}{\bibinfo{person}{Abeba Birhane}, \bibinfo{person}{Vinay Prabhu}, \bibinfo{person}{Sang Han}, \bibinfo{person}{Vishnu~Naresh Boddeti}, {and} \bibinfo{person}{Alexandra~Sasha Luccioni}.} \bibinfo{year}{2023}\natexlab{}.
\newblock \showarticletitle{Into the LAIONs Den: Investigating Hate in Multimodal Datasets}.
\newblock \bibinfo{journal}{\emph{arXiv preprint arXiv:2311.03449}} (\bibinfo{year}{2023}).
\newblock


\bibitem[Birhane et~al\mbox{.}(2021)]%
        {birhane2021multimodal}
\bibfield{author}{\bibinfo{person}{Abeba Birhane}, \bibinfo{person}{Vinay~Uday Prabhu}, {and} \bibinfo{person}{Emmanuel Kahembwe}.} \bibinfo{year}{2021}\natexlab{}.
\newblock \showarticletitle{Multimodal datasets: misogyny, pornography, and malignant stereotypes}.
\newblock \bibinfo{journal}{\emph{arXiv preprint arXiv:2110.01963}} (\bibinfo{year}{2021}).
\newblock


\bibitem[Birhane et~al\mbox{.}(2024)]%
        {birhane2024ai}
\bibfield{author}{\bibinfo{person}{Abeba Birhane}, \bibinfo{person}{Ryan Steed}, \bibinfo{person}{Victor Ojewale}, \bibinfo{person}{Briana Vecchione}, {and} \bibinfo{person}{Inioluwa~Deborah Raji}.} \bibinfo{year}{2024}\natexlab{}.
\newblock \bibinfo{title}{AI auditing: The Broken Bus on the Road to AI Accountability}.
\newblock
\newblock
\showeprint[arxiv]{2401.14462}~[cs.CY]


\bibitem[Bluemke et~al\mbox{.}(2023)]%
        {bluemke_exploring_2023}
\bibfield{author}{\bibinfo{person}{Emma Bluemke}, \bibinfo{person}{Tantum Collins}, \bibinfo{person}{Ben Garfinkel}, {and} \bibinfo{person}{Andrew Trask}.} \bibinfo{year}{2023}\natexlab{}.
\newblock \showarticletitle{Exploring the {Relevance} of {Data} {Privacy}-{Enhancing} {Technologies} for {AI} {Governance} {Use} {Cases}}.
\newblock  (\bibinfo{date}{March} \bibinfo{year}{2023}).
\newblock
\urldef\tempurl%
\url{https://arxiv.org/abs/2303.08956v2}
\showURL{%
\tempurl}


\bibitem[Bolton et~al\mbox{.}(2012)]%
        {bolton2012credit}
\bibfield{author}{\bibinfo{person}{Patrick Bolton}, \bibinfo{person}{Xavier Freixas}, {and} \bibinfo{person}{Joel Shapiro}.} \bibinfo{year}{2012}\natexlab{}.
\newblock \showarticletitle{The {Credit} {Ratings} {Game}}.
\newblock \bibinfo{journal}{\emph{The Journal of Finance}} \bibinfo{volume}{67}, \bibinfo{number}{1} (\bibinfo{year}{2012}), \bibinfo{pages}{85--111}.
\newblock
\showISSN{1540-6261}
\urldef\tempurl%
\url{https://doi.org/10.1111/j.1540-6261.2011.01708.x}
\showDOI{\tempurl}
\newblock
\shownote{\_eprint: https://onlinelibrary.wiley.com/doi/pdf/10.1111/j.1540-6261.2011.01708.x}.


\bibitem[Bolukbasi et~al\mbox{.}(2016)]%
        {bolukbasi2016man}
\bibfield{author}{\bibinfo{person}{Tolga Bolukbasi}, \bibinfo{person}{Kai-Wei Chang}, \bibinfo{person}{James~Y Zou}, \bibinfo{person}{Venkatesh Saligrama}, {and} \bibinfo{person}{Adam~T Kalai}.} \bibinfo{year}{2016}\natexlab{}.
\newblock \showarticletitle{Man is to computer programmer as woman is to homemaker? debiasing word embeddings}.
\newblock \bibinfo{journal}{\emph{Advances in neural information processing systems}}  \bibinfo{volume}{29} (\bibinfo{year}{2016}).
\newblock


\bibitem[Bommasani et~al\mbox{.}(2023)]%
        {bommasani2023foundation}
\bibfield{author}{\bibinfo{person}{Rishi Bommasani}, \bibinfo{person}{Kevin Klyman}, \bibinfo{person}{Shayne Longpre}, \bibinfo{person}{Sayash Kapoor}, \bibinfo{person}{Nestor Maslej}, \bibinfo{person}{Betty Xiong}, \bibinfo{person}{Daniel Zhang}, {and} \bibinfo{person}{Percy Liang}.} \bibinfo{year}{2023}\natexlab{}.
\newblock \showarticletitle{The {Foundation} {Model} {Transparency} {Index}}.
\newblock  (\bibinfo{date}{Oct.} \bibinfo{year}{2023}).
\newblock
\urldef\tempurl%
\url{http://arxiv.org/abs/2310.12941}
\showURL{%
\tempurl}
\newblock
\shownote{arXiv:2310.12941 [cs]}.


\bibitem[Bran et~al\mbox{.}(2023)]%
        {bran2023chemcrow}
\bibfield{author}{\bibinfo{person}{Andres~M Bran}, \bibinfo{person}{Sam Cox}, \bibinfo{person}{Andrew~D White}, {and} \bibinfo{person}{Philippe Schwaller}.} \bibinfo{year}{2023}\natexlab{}.
\newblock \showarticletitle{ChemCrow: Augmenting large-language models with chemistry tools}.
\newblock \bibinfo{journal}{\emph{arXiv preprint arXiv:2304.05376}} (\bibinfo{year}{2023}).
\newblock


\bibitem[Bricken et~al\mbox{.}(2023)]%
        {bricken2023monosemanticity}
\bibfield{author}{\bibinfo{person}{Trenton Bricken}, \bibinfo{person}{Adly Templeton}, \bibinfo{person}{Joshua Batson}, \bibinfo{person}{Brian Chen}, \bibinfo{person}{Adam Jermyn}, \bibinfo{person}{Tom Conerly}, \bibinfo{person}{Nick Turner}, \bibinfo{person}{Cem Anil}, \bibinfo{person}{Carson Denison}, \bibinfo{person}{Amanda Askell}, \bibinfo{person}{Robert Lasenby}, \bibinfo{person}{Yifan Wu}, \bibinfo{person}{Shauna Kravec}, \bibinfo{person}{Nicholas Schiefer}, \bibinfo{person}{Tim Maxwell}, \bibinfo{person}{Nicholas Joseph}, \bibinfo{person}{Zac Hatfield-Dodds}, \bibinfo{person}{Alex Tamkin}, \bibinfo{person}{Karina Nguyen}, \bibinfo{person}{Brayden McLean}, \bibinfo{person}{Josiah~E Burke}, \bibinfo{person}{Tristan Hume}, \bibinfo{person}{Shan Carter}, \bibinfo{person}{Tom Henighan}, {and} \bibinfo{person}{Christopher Olah}.} \bibinfo{year}{2023}\natexlab{}.
\newblock \showarticletitle{Towards Monosemanticity: Decomposing Language Models With Dictionary Learning}.
\newblock \bibinfo{journal}{\emph{Transformer Circuits Thread}} (\bibinfo{year}{2023}).
\newblock
\newblock
\shownote{https://transformer-circuits.pub/2023/monosemantic-features/index.html}.


\bibitem[Brown et~al\mbox{.}(2021)]%
        {brown2021algorithm}
\bibfield{author}{\bibinfo{person}{Shea Brown}, \bibinfo{person}{Jovana Davidovic}, {and} \bibinfo{person}{Ali Hasan}.} \bibinfo{year}{2021}\natexlab{}.
\newblock \showarticletitle{The algorithm audit: Scoring the algorithms that score us}.
\newblock \bibinfo{journal}{\emph{Big Data \& Society}} \bibinfo{volume}{8}, \bibinfo{number}{1} (\bibinfo{year}{2021}), \bibinfo{pages}{2053951720983865}.
\newblock


\bibitem[Brundage et~al\mbox{.}(2020)]%
        {brundage_toward_2020}
\bibfield{author}{\bibinfo{person}{Miles Brundage}, \bibinfo{person}{Shahar Avin}, \bibinfo{person}{Jasmine Wang}, \bibinfo{person}{Haydn Belfield}, \bibinfo{person}{Gretchen Krueger}, \bibinfo{person}{Gillian Hadfield}, \bibinfo{person}{Heidy Khlaaf}, \bibinfo{person}{Jingying Yang}, \bibinfo{person}{Helen Toner}, \bibinfo{person}{Ruth Fong}, \bibinfo{person}{Tegan Maharaj}, \bibinfo{person}{Pang~Wei Koh}, \bibinfo{person}{Sara Hooker}, \bibinfo{person}{Jade Leung}, \bibinfo{person}{Andrew Trask}, \bibinfo{person}{Emma Bluemke}, \bibinfo{person}{Jonathan Lebensold}, \bibinfo{person}{Cullen O'Keefe}, \bibinfo{person}{Mark Koren}, \bibinfo{person}{Théo Ryffel}, \bibinfo{person}{J.~B. Rubinovitz}, \bibinfo{person}{Tamay Besiroglu}, \bibinfo{person}{Federica Carugati}, \bibinfo{person}{Jack Clark}, \bibinfo{person}{Peter Eckersley}, \bibinfo{person}{Sarah de Haas}, \bibinfo{person}{Maritza Johnson}, \bibinfo{person}{Ben Laurie}, \bibinfo{person}{Alex Ingerman}, \bibinfo{person}{Igor Krawczuk},
  \bibinfo{person}{Amanda Askell}, \bibinfo{person}{Rosario Cammarota}, \bibinfo{person}{Andrew Lohn}, \bibinfo{person}{David Krueger}, \bibinfo{person}{Charlotte Stix}, \bibinfo{person}{Peter Henderson}, \bibinfo{person}{Logan Graham}, \bibinfo{person}{Carina Prunkl}, \bibinfo{person}{Bianca Martin}, \bibinfo{person}{Elizabeth Seger}, \bibinfo{person}{Noa Zilberman}, \bibinfo{person}{Seán~Ó hÉigeartaigh}, \bibinfo{person}{Frens Kroeger}, \bibinfo{person}{Girish Sastry}, \bibinfo{person}{Rebecca Kagan}, \bibinfo{person}{Adrian Weller}, \bibinfo{person}{Brian Tse}, \bibinfo{person}{Elizabeth Barnes}, \bibinfo{person}{Allan Dafoe}, \bibinfo{person}{Paul Scharre}, \bibinfo{person}{Ariel Herbert-Voss}, \bibinfo{person}{Martijn Rasser}, \bibinfo{person}{Shagun Sodhani}, \bibinfo{person}{Carrick Flynn}, \bibinfo{person}{Thomas~Krendl Gilbert}, \bibinfo{person}{Lisa Dyer}, \bibinfo{person}{Saif Khan}, \bibinfo{person}{Yoshua Bengio}, {and} \bibinfo{person}{Markus Anderljung}.} \bibinfo{year}{2020}\natexlab{}.
\newblock \showarticletitle{Toward {Trustworthy} {AI} {Development}: {Mechanisms} for {Supporting} {Verifiable} {Claims}}.
\newblock  (\bibinfo{date}{April} \bibinfo{year}{2020}).
\newblock
\urldef\tempurl%
\url{https://doi.org/10.48550/arXiv.2004.07213}
\showDOI{\tempurl}
\newblock
\shownote{arXiv:2004.07213 [cs]}.


\bibitem[Bucknall and Trager(2023)]%
        {bucknall2023structured}
\bibfield{author}{\bibinfo{person}{Benjamin~S Bucknall} {and} \bibinfo{person}{Robert~F Trager}.} \bibinfo{year}{2023}\natexlab{}.
\newblock \showarticletitle{Structured {Access} for {Third}-{Party} {Research} on {Frontier} {AI} {Models}: {Investigating} {Researchers}' {Model} {Access} {Requirements}}.
\newblock  (\bibinfo{date}{Oct.} \bibinfo{year}{2023}).
\newblock
\urldef\tempurl%
\url{https://www.oxfordmartin.ox.ac.uk/publications/structured-access-for-third-party-research-on-frontier-ai-models-investigating-researchers-model-access-requirements/}
\showURL{%
\tempurl}


\bibitem[Buiten et~al\mbox{.}(2021)]%
        {buiten_eu_2021}
\bibfield{author}{\bibinfo{person}{Miriam Buiten}, \bibinfo{person}{Alexandre de Streel}, {and} \bibinfo{person}{Martin Peitz}.} \bibinfo{year}{2021}\natexlab{}.
\newblock \showarticletitle{{EU} {Liability} {Rules} for the {Age} of {Artificial} {Intelligence}}.
\newblock  (\bibinfo{date}{April} \bibinfo{year}{2021}).
\newblock
\urldef\tempurl%
\url{https://doi.org/10.2139/ssrn.3817520}
\showDOI{\tempurl}


\bibitem[Buiten et~al\mbox{.}(2023)]%
        {buiten_vision_2023}
\bibfield{author}{\bibinfo{person}{Miriam~C. Buiten}, \bibinfo{person}{Louise~A. Dennis}, {and} \bibinfo{person}{Maike Schwammberger}.} \bibinfo{year}{2023}\natexlab{}.
\newblock \showarticletitle{A {Vision} on {What} {Explanations} of {Autonomous} {Systems} are of {Interest} to {Lawyers}}. In \bibinfo{booktitle}{\emph{2023 {IEEE} 31st {International} {Requirements} {Engineering} {Conference} {Workshops} ({REW})}}. \bibinfo{publisher}{IEEE}, \bibinfo{address}{Hannover, Germany}, \bibinfo{pages}{332--336}.
\newblock
\showISBNx{9798350326918}
\urldef\tempurl%
\url{https://doi.org/10.1109/REW57809.2023.00062}
\showDOI{\tempurl}


\bibitem[Buolamwini and Gebru(2018)]%
        {buolamwini2018gender}
\bibfield{author}{\bibinfo{person}{Joy Buolamwini} {and} \bibinfo{person}{Timnit Gebru}.} \bibinfo{year}{2018}\natexlab{}.
\newblock \showarticletitle{Gender shades: Intersectional accuracy disparities in commercial gender classification}. In \bibinfo{booktitle}{\emph{Conference on fairness, accountability and transparency}}. PMLR, \bibinfo{pages}{77--91}.
\newblock


\bibitem[Burns et~al\mbox{.}(2022)]%
        {burns2022discovering}
\bibfield{author}{\bibinfo{person}{Collin Burns}, \bibinfo{person}{Haotian Ye}, \bibinfo{person}{Dan Klein}, {and} \bibinfo{person}{Jacob Steinhardt}.} \bibinfo{year}{2022}\natexlab{}.
\newblock \showarticletitle{Discovering latent knowledge in language models without supervision}.
\newblock \bibinfo{journal}{\emph{arXiv preprint arXiv:2212.03827}} (\bibinfo{year}{2022}).
\newblock


\bibitem[Cammarata et~al\mbox{.}(2020)]%
        {cammarata2020thread}
\bibfield{author}{\bibinfo{person}{Nick Cammarata}, \bibinfo{person}{Shan Carter}, \bibinfo{person}{Gabriel Goh}, \bibinfo{person}{Chris Olah}, \bibinfo{person}{Michael Petrov}, \bibinfo{person}{Ludwig Schubert}, \bibinfo{person}{Chelsea Voss}, \bibinfo{person}{Ben Egan}, {and} \bibinfo{person}{Swee~Kiat Lim}.} \bibinfo{year}{2020}\natexlab{}.
\newblock \showarticletitle{Thread: Circuits}.
\newblock \bibinfo{journal}{\emph{Distill}} (\bibinfo{year}{2020}).
\newblock
\urldef\tempurl%
\url{https://doi.org/10.23915/distill.00024}
\showDOI{\tempurl}
\newblock
\shownote{https://distill.pub/2020/circuits}.


\bibitem[Carlini et~al\mbox{.}(2022)]%
        {carlini2022quantifying}
\bibfield{author}{\bibinfo{person}{Nicholas Carlini}, \bibinfo{person}{Daphne Ippolito}, \bibinfo{person}{Matthew Jagielski}, \bibinfo{person}{Katherine Lee}, \bibinfo{person}{Florian Tramer}, {and} \bibinfo{person}{Chiyuan Zhang}.} \bibinfo{year}{2022}\natexlab{}.
\newblock \showarticletitle{Quantifying memorization across neural language models}.
\newblock \bibinfo{journal}{\emph{arXiv preprint arXiv:2202.07646}} (\bibinfo{year}{2022}).
\newblock


\bibitem[Carlini et~al\mbox{.}(2023a)]%
        {carlini2023poisoning}
\bibfield{author}{\bibinfo{person}{Nicholas Carlini}, \bibinfo{person}{Matthew Jagielski}, \bibinfo{person}{Christopher~A Choquette-Choo}, \bibinfo{person}{Daniel Paleka}, \bibinfo{person}{Will Pearce}, \bibinfo{person}{Hyrum Anderson}, \bibinfo{person}{Andreas Terzis}, \bibinfo{person}{Kurt Thomas}, {and} \bibinfo{person}{Florian Tram{\`e}r}.} \bibinfo{year}{2023}\natexlab{a}.
\newblock \showarticletitle{Poisoning web-scale training datasets is practical}.
\newblock \bibinfo{journal}{\emph{arXiv preprint arXiv:2302.10149}} (\bibinfo{year}{2023}).
\newblock


\bibitem[Carlini et~al\mbox{.}(2023b)]%
        {carlini2023aligned}
\bibfield{author}{\bibinfo{person}{Nicholas Carlini}, \bibinfo{person}{Milad Nasr}, \bibinfo{person}{Christopher~A Choquette-Choo}, \bibinfo{person}{Matthew Jagielski}, \bibinfo{person}{Irena Gao}, \bibinfo{person}{Anas Awadalla}, \bibinfo{person}{Pang~Wei Koh}, \bibinfo{person}{Daphne Ippolito}, \bibinfo{person}{Katherine Lee}, \bibinfo{person}{Florian Tramer}, {et~al\mbox{.}}} \bibinfo{year}{2023}\natexlab{b}.
\newblock \showarticletitle{Are aligned neural networks adversarially aligned?}
\newblock \bibinfo{journal}{\emph{arXiv preprint arXiv:2306.15447}} (\bibinfo{year}{2023}).
\newblock


\bibitem[Carlini et~al\mbox{.}(2021)]%
        {carlini2021extracting}
\bibfield{author}{\bibinfo{person}{Nicholas Carlini}, \bibinfo{person}{Florian Tramer}, \bibinfo{person}{Eric Wallace}, \bibinfo{person}{Matthew Jagielski}, \bibinfo{person}{Ariel Herbert-Voss}, \bibinfo{person}{Katherine Lee}, \bibinfo{person}{Adam Roberts}, \bibinfo{person}{Tom Brown}, \bibinfo{person}{Dawn Song}, \bibinfo{person}{Ulfar Erlingsson}, {et~al\mbox{.}}} \bibinfo{year}{2021}\natexlab{}.
\newblock \showarticletitle{Extracting training data from large language models}. In \bibinfo{booktitle}{\emph{30th USENIX Security Symposium (USENIX Security 21)}}. \bibinfo{pages}{2633--2650}.
\newblock


\bibitem[Carter et~al\mbox{.}(2019)]%
        {carter2019exploring}
\bibfield{author}{\bibinfo{person}{Shan Carter}, \bibinfo{person}{Zan Armstrong}, \bibinfo{person}{Ludwig Schubert}, \bibinfo{person}{Ian Johnson}, {and} \bibinfo{person}{Chris Olah}.} \bibinfo{year}{2019}\natexlab{}.
\newblock \showarticletitle{Exploring neural networks with activation atlases}.
\newblock \bibinfo{journal}{\emph{Distill.}} (\bibinfo{year}{2019}).
\newblock


\bibitem[Carvalho et~al\mbox{.}(2019)]%
        {carvalho2019machine}
\bibfield{author}{\bibinfo{person}{Diogo~V Carvalho}, \bibinfo{person}{Eduardo~M Pereira}, {and} \bibinfo{person}{Jaime~S Cardoso}.} \bibinfo{year}{2019}\natexlab{}.
\newblock \showarticletitle{Machine learning interpretability: A survey on methods and metrics}.
\newblock \bibinfo{journal}{\emph{Electronics}} \bibinfo{volume}{8}, \bibinfo{number}{8} (\bibinfo{year}{2019}), \bibinfo{pages}{832}.
\newblock


\bibitem[Casper et~al\mbox{.}(2023a)]%
        {casper_open_2023}
\bibfield{author}{\bibinfo{person}{Stephen Casper}, \bibinfo{person}{Xander Davies}, \bibinfo{person}{Claudia Shi}, \bibinfo{person}{Thomas~Krendl Gilbert}, \bibinfo{person}{Jérémy Scheurer}, \bibinfo{person}{Javier Rando}, \bibinfo{person}{Rachel Freedman}, \bibinfo{person}{Tomasz Korbak}, \bibinfo{person}{David Lindner}, \bibinfo{person}{Pedro Freire}, \bibinfo{person}{Tony Wang}, \bibinfo{person}{Samuel Marks}, \bibinfo{person}{Charbel-Raphaël Segerie}, \bibinfo{person}{Micah Carroll}, \bibinfo{person}{Andi Peng}, \bibinfo{person}{Phillip Christoffersen}, \bibinfo{person}{Mehul Damani}, \bibinfo{person}{Stewart Slocum}, \bibinfo{person}{Usman Anwar}, \bibinfo{person}{Anand Siththaranjan}, \bibinfo{person}{Max Nadeau}, \bibinfo{person}{Eric~J. Michaud}, \bibinfo{person}{Jacob Pfau}, \bibinfo{person}{Dmitrii Krasheninnikov}, \bibinfo{person}{Xin Chen}, \bibinfo{person}{Lauro Langosco}, \bibinfo{person}{Peter Hase}, \bibinfo{person}{Erdem Bıyık}, \bibinfo{person}{Anca Dragan}, \bibinfo{person}{David
  Krueger}, \bibinfo{person}{Dorsa Sadigh}, {and} \bibinfo{person}{Dylan Hadfield-Menell}.} \bibinfo{year}{2023}\natexlab{a}.
\newblock \showarticletitle{Open {Problems} and {Fundamental} {Limitations} of {Reinforcement} {Learning} from {Human} {Feedback}}.
\newblock  (\bibinfo{date}{Sept.} \bibinfo{year}{2023}).
\newblock
\urldef\tempurl%
\url{https://doi.org/10.48550/arXiv.2307.15217}
\showDOI{\tempurl}
\newblock
\shownote{arXiv:2307.15217 [cs]}.


\bibitem[Casper et~al\mbox{.}(2023b)]%
        {casper2023measuring}
\bibfield{author}{\bibinfo{person}{Stephen Casper}, \bibinfo{person}{Zifan Guo}, \bibinfo{person}{Shreya Mogulothu}, \bibinfo{person}{Zachary Marinov}, \bibinfo{person}{Chinmay Deshpande}, \bibinfo{person}{Rui-Jie Yew}, \bibinfo{person}{Zheng Dai}, {and} \bibinfo{person}{Dylan Hadfield-Menell}.} \bibinfo{year}{2023}\natexlab{b}.
\newblock \showarticletitle{Measuring the Success of Diffusion Models at Imitating Human Artists}.
\newblock \bibinfo{journal}{\emph{arXiv preprint arXiv:2307.04028}} (\bibinfo{year}{2023}).
\newblock


\bibitem[Casper et~al\mbox{.}(2022a)]%
        {casper2022diagnostics}
\bibfield{author}{\bibinfo{person}{Stephen Casper}, \bibinfo{person}{Kaivalya Hariharan}, {and} \bibinfo{person}{Dylan Hadfield-Menell}.} \bibinfo{year}{2022}\natexlab{a}.
\newblock \showarticletitle{Diagnostics for deep neural networks with automated copy/paste attacks}. In \bibinfo{booktitle}{\emph{NeurIPS ML Safety Workshop}}.
\newblock


\bibitem[Casper et~al\mbox{.}(2023c)]%
        {casper2023red}
\bibfield{author}{\bibinfo{person}{Stephen Casper}, \bibinfo{person}{Taylor Killian}, \bibinfo{person}{Gabriel Kreiman}, {and} \bibinfo{person}{Dylan Hadfield-Menell}.} \bibinfo{year}{2023}\natexlab{c}.
\newblock \showarticletitle{Red {Teaming} with {Mind} {Reading}: {White}-{Box} {Adversarial} {Policies} {Against} {RL} {Agents}}.
\newblock  (\bibinfo{date}{Oct.} \bibinfo{year}{2023}).
\newblock
\urldef\tempurl%
\url{http://arxiv.org/abs/2209.02167}
\showURL{%
\tempurl}
\newblock
\shownote{arXiv:2209.02167 [cs]}.


\bibitem[Casper et~al\mbox{.}(2023d)]%
        {casper2023red1}
\bibfield{author}{\bibinfo{person}{Stephen Casper}, \bibinfo{person}{Yuxiao Li}, \bibinfo{person}{Jiawei Li}, \bibinfo{person}{Tong Bu}, \bibinfo{person}{Kevin Zhang}, \bibinfo{person}{Kaivalya Hariharan}, {and} \bibinfo{person}{Dylan Hadfield-Menell}.} \bibinfo{year}{2023}\natexlab{d}.
\newblock \showarticletitle{Red {Teaming} {Deep} {Neural} {Networks} with {Feature} {Synthesis} {Tools}}.
\newblock  (\bibinfo{date}{Sept.} \bibinfo{year}{2023}).
\newblock
\urldef\tempurl%
\url{http://arxiv.org/abs/2302.10894}
\showURL{%
\tempurl}
\newblock
\shownote{arXiv:2302.10894 [cs]}.


\bibitem[Casper et~al\mbox{.}(2023e)]%
        {casper2023explore}
\bibfield{author}{\bibinfo{person}{Stephen Casper}, \bibinfo{person}{Jason Lin}, \bibinfo{person}{Joe Kwon}, \bibinfo{person}{Gatlen Culp}, {and} \bibinfo{person}{Dylan Hadfield-Menell}.} \bibinfo{year}{2023}\natexlab{e}.
\newblock \showarticletitle{Explore, Establish, Exploit: Red Teaming Language Models from Scratch}.
\newblock \bibinfo{journal}{\emph{arXiv preprint arXiv:2306.09442}} (\bibinfo{year}{2023}).
\newblock


\bibitem[Casper et~al\mbox{.}(2022b)]%
        {casper2022robust}
\bibfield{author}{\bibinfo{person}{Stephen Casper}, \bibinfo{person}{Max Nadeau}, \bibinfo{person}{Dylan Hadfield-Menell}, {and} \bibinfo{person}{Gabriel Kreiman}.} \bibinfo{year}{2022}\natexlab{b}.
\newblock \showarticletitle{Robust feature-level adversaries are interpretability tools}.
\newblock \bibinfo{journal}{\emph{Advances in Neural Information Processing Systems}}  \bibinfo{volume}{35} (\bibinfo{year}{2022}), \bibinfo{pages}{33093--33106}.
\newblock


\bibitem[Casper et~al\mbox{.}(2024)]%
        {casper2024defending}
\bibfield{author}{\bibinfo{person}{Stephen Casper}, \bibinfo{person}{Lennart Schulze}, \bibinfo{person}{Oam Patel}, {and} \bibinfo{person}{Dylan Hadfield-Menell}.} \bibinfo{year}{2024}\natexlab{}.
\newblock \bibinfo{title}{Defending Against Unforeseen Failure Modes with Latent Adversarial Training}.
\newblock
\newblock
\showeprint[arxiv]{2403.05030}~[cs.CR]


\bibitem[Chan et~al\mbox{.}(2023)]%
        {chan2023harms}
\bibfield{author}{\bibinfo{person}{Alan Chan}, \bibinfo{person}{Rebecca Salganik}, \bibinfo{person}{Alva Markelius}, \bibinfo{person}{Chris Pang}, \bibinfo{person}{Nitarshan Rajkumar}, \bibinfo{person}{Dmitrii Krasheninnikov}, \bibinfo{person}{Lauro Langosco}, \bibinfo{person}{Zhonghao He}, \bibinfo{person}{Yawen Duan}, \bibinfo{person}{Micah Carroll}, \bibinfo{person}{Michelle Lin}, \bibinfo{person}{Alex Mayhew}, \bibinfo{person}{Katherine Collins}, \bibinfo{person}{Maryam Molamohammadi}, \bibinfo{person}{John Burden}, \bibinfo{person}{Wanru Zhao}, \bibinfo{person}{Shalaleh Rismani}, \bibinfo{person}{Konstantinos Voudouris}, \bibinfo{person}{Umang Bhatt}, \bibinfo{person}{Adrian Weller}, \bibinfo{person}{David Krueger}, {and} \bibinfo{person}{Tegan Maharaj}.} \bibinfo{year}{2023}\natexlab{}.
\newblock \showarticletitle{Harms from {Increasingly} {Agentic} {Algorithmic} {Systems}}. In \bibinfo{booktitle}{\emph{2023 {ACM} {Conference} on {Fairness}, {Accountability}, and {Transparency}}}. \bibinfo{pages}{651--666}.
\newblock
\urldef\tempurl%
\url{https://doi.org/10.1145/3593013.3594033}
\showDOI{\tempurl}
\newblock
\shownote{arXiv:2302.10329 [cs]}.


\bibitem[Chao et~al\mbox{.}(2023)]%
        {chao2023jailbreaking}
\bibfield{author}{\bibinfo{person}{Patrick Chao}, \bibinfo{person}{Alexander Robey}, \bibinfo{person}{Edgar Dobriban}, \bibinfo{person}{Hamed Hassani}, \bibinfo{person}{George~J Pappas}, {and} \bibinfo{person}{Eric Wong}.} \bibinfo{year}{2023}\natexlab{}.
\newblock \showarticletitle{Jailbreaking Black Box Large Language Models in Twenty Queries}.
\newblock \bibinfo{journal}{\emph{arXiv preprint arXiv:2310.08419}} (\bibinfo{year}{2023}).
\newblock


\bibitem[Charan et~al\mbox{.}(2023)]%
        {charan2023text}
\bibfield{author}{\bibinfo{person}{PV Charan}, \bibinfo{person}{Hrushikesh Chunduri}, \bibinfo{person}{P~Mohan Anand}, {and} \bibinfo{person}{Sandeep~K Shukla}.} \bibinfo{year}{2023}\natexlab{}.
\newblock \showarticletitle{From Text to MITRE Techniques: Exploring the Malicious Use of Large Language Models for Generating Cyber Attack Payloads}.
\newblock \bibinfo{journal}{\emph{arXiv preprint arXiv:2305.15336}} (\bibinfo{year}{2023}).
\newblock


\bibitem[Chen et~al\mbox{.}(2023)]%
        {chen_bias_2023}
\bibfield{author}{\bibinfo{person}{Jiawei Chen}, \bibinfo{person}{Hande Dong}, \bibinfo{person}{Xiang Wang}, \bibinfo{person}{Fuli Feng}, \bibinfo{person}{Meng Wang}, {and} \bibinfo{person}{Xiangnan He}.} \bibinfo{year}{2023}\natexlab{}.
\newblock \showarticletitle{Bias and {Debias} in {Recommender} {System}: {A} {Survey} and {Future} {Directions}}.
\newblock \bibinfo{journal}{\emph{ACM Transactions on Information Systems}} \bibinfo{volume}{41}, \bibinfo{number}{3} (\bibinfo{date}{Feb.} \bibinfo{year}{2023}), \bibinfo{pages}{67:1--67:39}.
\newblock
\showISSN{1046-8188}
\urldef\tempurl%
\url{https://doi.org/10.1145/3564284}
\showDOI{\tempurl}


\bibitem[Chen et~al\mbox{.}(2017)]%
        {chen2017targeted}
\bibfield{author}{\bibinfo{person}{Xinyun Chen}, \bibinfo{person}{Chang Liu}, \bibinfo{person}{Bo Li}, \bibinfo{person}{Kimberly Lu}, {and} \bibinfo{person}{Dawn Song}.} \bibinfo{year}{2017}\natexlab{}.
\newblock \showarticletitle{Targeted backdoor attacks on deep learning systems using data poisoning}.
\newblock \bibinfo{journal}{\emph{arXiv preprint arXiv:1712.05526}} (\bibinfo{year}{2017}).
\newblock


\bibitem[Chen et~al\mbox{.}(2022)]%
        {chen2022fairness}
\bibfield{author}{\bibinfo{person}{Zhenpeng Chen}, \bibinfo{person}{Jie~M Zhang}, \bibinfo{person}{Max Hort}, \bibinfo{person}{Federica Sarro}, {and} \bibinfo{person}{Mark Harman}.} \bibinfo{year}{2022}\natexlab{}.
\newblock \showarticletitle{Fairness testing: A comprehensive survey and analysis of trends}.
\newblock \bibinfo{journal}{\emph{arXiv preprint arXiv:2207.10223}} (\bibinfo{year}{2022}).
\newblock


\bibitem[{China Academy of Information and Communications Technology} and {JD Explore Academy}(2021)]%
        {caict2021}
\bibfield{author}{\bibinfo{person}{{China Academy of Information and Communications Technology}} {and} \bibinfo{person}{{JD Explore Academy}}.} \bibinfo{year}{2021}\natexlab{}.
\newblock \bibinfo{title}{White {Paper} on {Trustworthy} {Artificial} {Intelligence}}.
\newblock
\newblock
\urldef\tempurl%
\url{https://cset.georgetown.edu/publication/white-paper-on-trustworthy-artificial-intelligence/}
\showURL{%
\tempurl}


\bibitem[{Chinese National Information Security Standardization Technical Committee}(2023)]%
        {tc260_2023}
\bibfield{author}{\bibinfo{person}{{Chinese National Information Security Standardization Technical Committee}}.} \bibinfo{year}{2023}\natexlab{}.
\newblock \bibinfo{title}{Translation: {Basic} {Safety} {Requirements} for {Generative} {Artificial} {Intelligence} {Services} ({Draft} for {Feedback})}.
\newblock
\newblock
\urldef\tempurl%
\url{https://cset.georgetown.edu/publication/china-safety-requirements-for-generative-ai/?utm_source=substack&utm_medium=email}
\showURL{%
\tempurl}


\bibitem[Chou et~al\mbox{.}(2021)]%
        {chou_counterfactuals_2021}
\bibfield{author}{\bibinfo{person}{Yu-Liang Chou}, \bibinfo{person}{Catarina Moreira}, \bibinfo{person}{Peter Bruza}, \bibinfo{person}{Chun Ouyang}, {and} \bibinfo{person}{Joaquim Jorge}.} \bibinfo{year}{2021}\natexlab{}.
\newblock \showarticletitle{Counterfactuals and {Causability} in {Explainable} {Artificial} {Intelligence}: {Theory}, {Algorithms}, and {Applications}}.
\newblock  (\bibinfo{date}{June} \bibinfo{year}{2021}).
\newblock
\urldef\tempurl%
\url{https://doi.org/10.48550/arXiv.2103.04244}
\showDOI{\tempurl}
\newblock
\shownote{arXiv:2103.04244 [cs]}.


\bibitem[Christiano(2019)]%
        {christiano2019worst}
\bibfield{author}{\bibinfo{person}{Paul Christiano}.} \bibinfo{year}{2019}\natexlab{}.
\newblock \bibinfo{title}{Worst-case guarantees}.
\newblock
\newblock
\urldef\tempurl%
\url{https://ai-alignment.com/training-robust-corrigibility-ce0e0a3b9b4d}
\showURL{%
\tempurl}


\bibitem[Coe and Atay(2021)]%
        {coe2021evaluating}
\bibfield{author}{\bibinfo{person}{James Coe} {and} \bibinfo{person}{Mustafa Atay}.} \bibinfo{year}{2021}\natexlab{}.
\newblock \showarticletitle{Evaluating impact of race in facial recognition across machine learning and deep learning algorithms}.
\newblock \bibinfo{journal}{\emph{Computers}} \bibinfo{volume}{10}, \bibinfo{number}{9} (\bibinfo{year}{2021}), \bibinfo{pages}{113}.
\newblock


\bibitem[Company(2023)]%
        {nytimes_v_openai_2023}
\bibfield{author}{\bibinfo{person}{The New York~Times Company}.} \bibinfo{year}{2023}\natexlab{}.
\newblock \bibinfo{title}{The New York Times Company v. OpenAI}.
\newblock
\newblock
\urldef\tempurl%
\url{https://nytco-assets.nytimes.com/2023/12/NYT_Complaint_Dec2023.pdf}
\showURL{%
\tempurl}
\newblock
\shownote{Case {e 1:23-cv-11195}}.


\bibitem[Conneau et~al\mbox{.}(2018)]%
        {conneau2018you}
\bibfield{author}{\bibinfo{person}{Alexis Conneau}, \bibinfo{person}{German Kruszewski}, \bibinfo{person}{Guillaume Lample}, \bibinfo{person}{Lo{\"\i}c Barrault}, {and} \bibinfo{person}{Marco Baroni}.} \bibinfo{year}{2018}\natexlab{}.
\newblock \showarticletitle{What you can cram into a single vector: Probing sentence embeddings for linguistic properties}.
\newblock \bibinfo{journal}{\emph{arXiv preprint arXiv:1805.01070}} (\bibinfo{year}{2018}).
\newblock


\bibitem[Costanza-Chock et~al\mbox{.}(2022)]%
        {costanza-chock2022who}
\bibfield{author}{\bibinfo{person}{Sasha Costanza-Chock}, \bibinfo{person}{Inioluwa~Deborah Raji}, {and} \bibinfo{person}{Joy Buolamwini}.} \bibinfo{year}{2022}\natexlab{}.
\newblock \showarticletitle{Who {Audits} the {Auditors}? {Recommendations} from a field scan of the algorithmic auditing ecosystem}. In \bibinfo{booktitle}{\emph{Proceedings of the 2022 {ACM} {Conference} on {Fairness}, {Accountability}, and {Transparency}}} \emph{(\bibinfo{series}{{FAccT} '22})}. \bibinfo{publisher}{Association for Computing Machinery}, \bibinfo{address}{New York, NY, USA}, \bibinfo{pages}{1571--1583}.
\newblock
\showISBNx{978-1-4503-9352-2}
\urldef\tempurl%
\url{https://doi.org/10.1145/3531146.3533213}
\showDOI{\tempurl}


\bibitem[Cumbo et~al\mbox{.}(2021)]%
        {nyclaw2021}
\bibfield{author}{\bibinfo{person}{Laurie Cumbo}, \bibinfo{person}{Alicka Ampry-Samuel}, \bibinfo{person}{Helen Rosenthal}, \bibinfo{person}{Robert Cornegy}, \bibinfo{person}{Ben Kallos}, \bibinfo{person}{Adrienne Adams}, \bibinfo{person}{Farah Louis}, \bibinfo{person}{Margaret Chin}, \bibinfo{person}{Fernando Cabrera}, \bibinfo{person}{Deborah Rose}, \bibinfo{person}{Vanessa Gibson}, \bibinfo{person}{Justin Brannan}, \bibinfo{person}{Carlina Rivera}, \bibinfo{person}{Mark Levine}, \bibinfo{person}{Diana Ayala}, \bibinfo{person}{I.~Daneek Miller}, \bibinfo{person}{Stephen Levin}, {and} \bibinfo{person}{Inez Barron}.} \bibinfo{year}{2021}\natexlab{}.
\newblock \bibinfo{title}{Local {Law} 144 of 2021}.
\newblock
\newblock
\urldef\tempurl%
\url{https://legistar.council.nyc.gov/LegislationDetail.aspx?ID=4344524&GUID=B051915D-A9AC-451E-81F8-6596032FA3F9&Options=ID%7cText%7c&Search=}
\showURL{%
\tempurl}


\bibitem[Cunningham et~al\mbox{.}(2023)]%
        {cunningham2023sparse}
\bibfield{author}{\bibinfo{person}{Hoagy Cunningham}, \bibinfo{person}{Aidan Ewart}, \bibinfo{person}{Logan Riggs}, \bibinfo{person}{Robert Huben}, {and} \bibinfo{person}{Lee Sharkey}.} \bibinfo{year}{2023}\natexlab{}.
\newblock \showarticletitle{Sparse Autoencoders Find Highly Interpretable Features in Language Models}.
\newblock  (\bibinfo{year}{2023}).
\newblock
\showeprint[arxiv]{2309.08600}~[cs.LG]


\bibitem[Das and Rad(2020)]%
        {das2020opportunities}
\bibfield{author}{\bibinfo{person}{Arun Das} {and} \bibinfo{person}{Paul Rad}.} \bibinfo{year}{2020}\natexlab{}.
\newblock \showarticletitle{Opportunities and challenges in explainable artificial intelligence (xai): A survey}.
\newblock \bibinfo{journal}{\emph{arXiv preprint arXiv:2006.11371}} (\bibinfo{year}{2020}).
\newblock


\bibitem[Davidson et~al\mbox{.}(2023)]%
        {davidson_ai_2023}
\bibfield{author}{\bibinfo{person}{Tom Davidson}, \bibinfo{person}{Jean-Stanislas Denain}, \bibinfo{person}{Pablo Villalobos}, {and} \bibinfo{person}{Guillem Bas}.} \bibinfo{year}{2023}\natexlab{}.
\newblock \showarticletitle{{AI} capabilities can be significantly improved without expensive retraining}.
\newblock  (\bibinfo{date}{Dec.} \bibinfo{year}{2023}).
\newblock
\urldef\tempurl%
\url{https://arxiv.org/abs/2312.07413v1}
\showURL{%
\tempurl}


\bibitem[Deng et~al\mbox{.}(2023b)]%
        {deng2023investigating}
\bibfield{author}{\bibinfo{person}{Chunyuan Deng}, \bibinfo{person}{Yilun Zhao}, \bibinfo{person}{Xiangru Tang}, \bibinfo{person}{Mark Gerstein}, {and} \bibinfo{person}{Arman Cohan}.} \bibinfo{year}{2023}\natexlab{b}.
\newblock \showarticletitle{Investigating Data Contamination in Modern Benchmarks for Large Language Models}.
\newblock \bibinfo{journal}{\emph{arXiv preprint arXiv:2311.09783}} (\bibinfo{year}{2023}).
\newblock


\bibitem[Deng et~al\mbox{.}(2023a)]%
        {deng2023jailbreaker}
\bibfield{author}{\bibinfo{person}{Gelei Deng}, \bibinfo{person}{Yi Liu}, \bibinfo{person}{Yuekang Li}, \bibinfo{person}{Kailong Wang}, \bibinfo{person}{Ying Zhang}, \bibinfo{person}{Zefeng Li}, \bibinfo{person}{Haoyu Wang}, \bibinfo{person}{Tianwei Zhang}, {and} \bibinfo{person}{Yang Liu}.} \bibinfo{year}{2023}\natexlab{a}.
\newblock \showarticletitle{Jailbreaker: Automated Jailbreak Across Multiple Large Language Model Chatbots}.
\newblock \bibinfo{journal}{\emph{arXiv preprint arXiv:2307.08715}} (\bibinfo{year}{2023}).
\newblock


\bibitem[Deng et~al\mbox{.}(2022)]%
        {deng2022rlprompt}
\bibfield{author}{\bibinfo{person}{Mingkai Deng}, \bibinfo{person}{Jianyu Wang}, \bibinfo{person}{Cheng-Ping Hsieh}, \bibinfo{person}{Yihan Wang}, \bibinfo{person}{Han Guo}, \bibinfo{person}{Tianmin Shu}, \bibinfo{person}{Meng Song}, \bibinfo{person}{Eric~P Xing}, {and} \bibinfo{person}{Zhiting Hu}.} \bibinfo{year}{2022}\natexlab{}.
\newblock \showarticletitle{Rlprompt: Optimizing discrete text prompts with reinforcement learning}.
\newblock \bibinfo{journal}{\emph{arXiv preprint arXiv:2205.12548}} (\bibinfo{year}{2022}).
\newblock


\bibitem[Dhamala et~al\mbox{.}(2021)]%
        {dhamala2021bold}
\bibfield{author}{\bibinfo{person}{Jwala Dhamala}, \bibinfo{person}{Tony Sun}, \bibinfo{person}{Varun Kumar}, \bibinfo{person}{Satyapriya Krishna}, \bibinfo{person}{Yada Pruksachatkun}, \bibinfo{person}{Kai-Wei Chang}, {and} \bibinfo{person}{Rahul Gupta}.} \bibinfo{year}{2021}\natexlab{}.
\newblock \showarticletitle{Bold: Dataset and metrics for measuring biases in open-ended language generation}. In \bibinfo{booktitle}{\emph{Proceedings of the 2021 ACM conference on fairness, accountability, and transparency}}. \bibinfo{pages}{862--872}.
\newblock


\bibitem[Dobbe et~al\mbox{.}(2019)]%
        {dobbe2019hard}
\bibfield{author}{\bibinfo{person}{Roel Dobbe}, \bibinfo{person}{Thomas~Krendl Gilbert}, {and} \bibinfo{person}{Yonatan Mintz}.} \bibinfo{year}{2019}\natexlab{}.
\newblock \showarticletitle{Hard choices in artificial intelligence: Addressing normative uncertainty through sociotechnical commitments}.
\newblock \bibinfo{journal}{\emph{arXiv preprint arXiv:1911.09005}} (\bibinfo{year}{2019}).
\newblock


\bibitem[Dong et~al\mbox{.}(2017)]%
        {dong2017towards}
\bibfield{author}{\bibinfo{person}{Yinpeng Dong}, \bibinfo{person}{Hang Su}, \bibinfo{person}{Jun Zhu}, {and} \bibinfo{person}{Fan Bao}.} \bibinfo{year}{2017}\natexlab{}.
\newblock \showarticletitle{Towards interpretable deep neural networks by leveraging adversarial examples}.
\newblock \bibinfo{journal}{\emph{arXiv preprint arXiv:1708.05493}} (\bibinfo{year}{2017}).
\newblock


\bibitem[Du et~al\mbox{.}(2023)]%
        {du2023shortcut}
\bibfield{author}{\bibinfo{person}{Mengnan Du}, \bibinfo{person}{Fengxiang He}, \bibinfo{person}{Na Zou}, \bibinfo{person}{Dacheng Tao}, {and} \bibinfo{person}{Xia Hu}.} \bibinfo{year}{2023}\natexlab{}.
\newblock \showarticletitle{Shortcut learning of large language models in natural language understanding}.
\newblock \bibinfo{journal}{\emph{Communications of the ACM (CACM)}} (\bibinfo{year}{2023}).
\newblock


\bibitem[Dwivedi et~al\mbox{.}(2023)]%
        {dwivedi2023explainable}
\bibfield{author}{\bibinfo{person}{Rudresh Dwivedi}, \bibinfo{person}{Devam Dave}, \bibinfo{person}{Het Naik}, \bibinfo{person}{Smiti Singhal}, \bibinfo{person}{Rana Omer}, \bibinfo{person}{Pankesh Patel}, \bibinfo{person}{Bin Qian}, \bibinfo{person}{Zhenyu Wen}, \bibinfo{person}{Tejal Shah}, \bibinfo{person}{Graham Morgan}, {et~al\mbox{.}}} \bibinfo{year}{2023}\natexlab{}.
\newblock \showarticletitle{Explainable AI (XAI): Core ideas, techniques, and solutions}.
\newblock \bibinfo{journal}{\emph{Comput. Surveys}} \bibinfo{volume}{55}, \bibinfo{number}{9} (\bibinfo{year}{2023}), \bibinfo{pages}{1--33}.
\newblock


\bibitem[Ebrahimi et~al\mbox{.}(2017)]%
        {ebrahimi2017hotflip}
\bibfield{author}{\bibinfo{person}{Javid Ebrahimi}, \bibinfo{person}{Anyi Rao}, \bibinfo{person}{Daniel Lowd}, {and} \bibinfo{person}{Dejing Dou}.} \bibinfo{year}{2017}\natexlab{}.
\newblock \showarticletitle{Hotflip: White-box adversarial examples for text classification}.
\newblock \bibinfo{journal}{\emph{arXiv preprint arXiv:1712.06751}} (\bibinfo{year}{2017}).
\newblock


\bibitem[Eckhouse et~al\mbox{.}(2019)]%
        {eckhouse2019layers}
\bibfield{author}{\bibinfo{person}{Laurel Eckhouse}, \bibinfo{person}{Kristian Lum}, \bibinfo{person}{Cynthia Conti-Cook}, {and} \bibinfo{person}{Julie Ciccolini}.} \bibinfo{year}{2019}\natexlab{}.
\newblock \showarticletitle{Layers of bias: A unified approach for understanding problems with risk assessment}.
\newblock \bibinfo{journal}{\emph{Criminal Justice and Behavior}} \bibinfo{volume}{46}, \bibinfo{number}{2} (\bibinfo{year}{2019}), \bibinfo{pages}{185--209}.
\newblock


\bibitem[Edelman(1992)]%
        {edelman1992legal}
\bibfield{author}{\bibinfo{person}{Lauren~B. Edelman}.} \bibinfo{year}{1992}\natexlab{}.
\newblock \showarticletitle{Legal {Ambiguity} and {Symbolic} {Structures}: {Organizational} {Mediation} of {Civil} {Rights} {Law}}.
\newblock \bibinfo{journal}{\emph{Amer. J. Sociology}} \bibinfo{volume}{97}, \bibinfo{number}{6} (\bibinfo{date}{May} \bibinfo{year}{1992}), \bibinfo{pages}{1531--1576}.
\newblock
\showISSN{0002-9602}
\urldef\tempurl%
\url{https://doi.org/10.1086/229939}
\showDOI{\tempurl}
\newblock
\shownote{Publisher: The University of Chicago Press}.


\bibitem[Edelman(2016)]%
        {edelman2016working}
\bibfield{author}{\bibinfo{person}{Lauren~B. Edelman}.} \bibinfo{year}{2016}\natexlab{}.
\newblock \bibinfo{booktitle}{\emph{Working {Law}: {Courts}, {Corporations}, and {Symbolic} {Civil} {Rights}}}.
\newblock \bibinfo{publisher}{University of Chicago Press}, \bibinfo{address}{Chicago, IL}.
\newblock
\showISBNx{978-0-226-40076-1}
\urldef\tempurl%
\url{https://press.uchicago.edu/ucp/books/book/chicago/W/bo24550454.html}
\showURL{%
\tempurl}


\bibitem[Elazar et~al\mbox{.}(2021)]%
        {elazar2021amnesic}
\bibfield{author}{\bibinfo{person}{Yanai Elazar}, \bibinfo{person}{Shauli Ravfogel}, \bibinfo{person}{Alon Jacovi}, {and} \bibinfo{person}{Yoav Goldberg}.} \bibinfo{year}{2021}\natexlab{}.
\newblock \showarticletitle{Amnesic probing: Behavioral explanation with amnesic counterfactuals}.
\newblock \bibinfo{journal}{\emph{Transactions of the Association for Computational Linguistics}}  \bibinfo{volume}{9} (\bibinfo{year}{2021}), \bibinfo{pages}{160--175}.
\newblock


\bibitem[{European Commission}(2021)]%
        {european2021laying}
\bibfield{author}{\bibinfo{person}{{European Commission}}.} \bibinfo{year}{2021}\natexlab{}.
\newblock \showarticletitle{Laying down harmonised rules on artificial intelligence (Artificial Intelligence Act) and amending certain union legislative acts}.
\newblock \bibinfo{journal}{\emph{Eur Comm}}  \bibinfo{volume}{106} (\bibinfo{year}{2021}), \bibinfo{pages}{1--108}.
\newblock


\bibitem[{European Union}(2016)]%
        {gdpr2016}
\bibfield{author}{\bibinfo{person}{{European Union}}.} \bibinfo{year}{2016}\natexlab{}.
\newblock \bibinfo{title}{General {Data} {Protection} {Regulation}}.
\newblock
\newblock
\urldef\tempurl%
\url{https://gdpr-info.eu/}
\showURL{%
\tempurl}


\bibitem[{European Union}(2021)]%
        {aiact2021}
\bibfield{author}{\bibinfo{person}{{European Union}}.} \bibinfo{year}{2021}\natexlab{}.
\newblock \bibinfo{title}{Artificial {Intelligence} {Act}}.
\newblock
\newblock
\urldef\tempurl%
\url{https://eur-lex.europa.eu/legal-content/EN/TXT/?uri=CELEX%3A52021PC0206}
\showURL{%
\tempurl}


\bibitem[{European Union}(2022)]%
        {dma2022}
\bibfield{author}{\bibinfo{person}{{European Union}}.} \bibinfo{year}{2022}\natexlab{}.
\newblock \bibinfo{title}{Digital Markets Act}.
\newblock
\newblock
\urldef\tempurl%
\url{https://eur-lex.europa.eu/legal-content/EN/TXT/?uri=CELEX%3A32022R1925}
\showURL{%
\tempurl}


\bibitem[EY(2019)]%
        {eycode}
\bibfield{author}{\bibinfo{person}{EY}.} \bibinfo{year}{2019}\natexlab{}.
\newblock \bibinfo{title}{EY Global Code of Conduct}.
\newblock \bibinfo{howpublished}{Online}.
\newblock
\newblock
\shownote{Retrieved from: \url{https://assets.ey.com/content/dam/ey-sites/ey-com/en_gl/generic/EY_Code_of_Conduct.pdf}}.


\bibitem[Farrell and Rabin(1996)]%
        {farrell1996cheap}
\bibfield{author}{\bibinfo{person}{Joseph Farrell} {and} \bibinfo{person}{Matthew Rabin}.} \bibinfo{year}{1996}\natexlab{}.
\newblock \showarticletitle{Cheap {Talk}}.
\newblock \bibinfo{journal}{\emph{Journal of Economic Perspectives}} \bibinfo{volume}{10}, \bibinfo{number}{3} (\bibinfo{date}{Sept.} \bibinfo{year}{1996}), \bibinfo{pages}{103--118}.
\newblock
\showISSN{0895-3309}
\urldef\tempurl%
\url{https://doi.org/10.1257/jep.10.3.103}
\showDOI{\tempurl}


\bibitem[Feffer et~al\mbox{.}(2024)]%
        {feffer_red-teaming_2024}
\bibfield{author}{\bibinfo{person}{Michael Feffer}, \bibinfo{person}{Anusha Sinha}, \bibinfo{person}{Zachary~C. Lipton}, {and} \bibinfo{person}{Hoda Heidari}.} \bibinfo{year}{2024}\natexlab{}.
\newblock \bibinfo{title}{Red-{Teaming} for {Generative} {AI}: {Silver} {Bullet} or {Security} {Theater}?}
\newblock
\newblock
\urldef\tempurl%
\url{http://arxiv.org/abs/2401.15897}
\showURL{%
\tempurl}
\newblock
\shownote{arXiv:2401.15897 [cs]}.


\bibitem[Fiotto-Kaufmann et~al\mbox{.}(2023)]%
        {nnsight}
\bibfield{author}{\bibinfo{person}{Jaden Fiotto-Kaufmann}, \bibinfo{person}{Arnab Sen-Sharma}, \bibinfo{person}{Caden Juang}, \bibinfo{person}{David Bau}, \bibinfo{person}{Eric Todd}, \bibinfo{person}{Francesca Lucchetti}, {and} \bibinfo{person}{Will Brockman}.} \bibinfo{year}{2023}\natexlab{}.
\newblock \bibinfo{title}{nnsight}.
\newblock
\newblock
\urldef\tempurl%
\url{https://nnsight.net/}
\showURL{%
\tempurl}


\bibitem[Fuerman(2009)]%
        {fuerman2009bernard}
\bibfield{author}{\bibinfo{person}{Ross~D Fuerman}.} \bibinfo{year}{2009}\natexlab{}.
\newblock \showarticletitle{Bernard Madoff and the solo auditor red flag}.
\newblock \bibinfo{journal}{\emph{Journal of Forensic \& Investigative Accounting}} \bibinfo{volume}{1}, \bibinfo{number}{1} (\bibinfo{year}{2009}), \bibinfo{pages}{1--38}.
\newblock


\bibitem[{G7}(2023)]%
        {hiroshima2023}
\bibfield{author}{\bibinfo{person}{{G7}}.} \bibinfo{year}{2023}\natexlab{}.
\newblock \bibinfo{title}{Hiroshima {Process} {International} {Code} of {Conduct} for {Organizations} {Developing} {Advanced} {AI} {Systems}}.
\newblock
\newblock
\urldef\tempurl%
\url{https://digital-strategy.ec.europa.eu/en/library/hiroshima-process-international-code-conduct-advanced-ai-systems}
\showURL{%
\tempurl}


\bibitem[Gandelsman et~al\mbox{.}(2023)]%
        {gandelsman2023interpreting}
\bibfield{author}{\bibinfo{person}{Yossi Gandelsman}, \bibinfo{person}{Alexei~A Efros}, {and} \bibinfo{person}{Jacob Steinhardt}.} \bibinfo{year}{2023}\natexlab{}.
\newblock \showarticletitle{Interpreting CLIP's Image Representation via Text-Based Decomposition}.
\newblock \bibinfo{journal}{\emph{arXiv preprint arXiv:2310.05916}} (\bibinfo{year}{2023}).
\newblock


\bibitem[Ganguli et~al\mbox{.}(2022)]%
        {ganguli2022red}
\bibfield{author}{\bibinfo{person}{Deep Ganguli}, \bibinfo{person}{Liane Lovitt}, \bibinfo{person}{Jackson Kernion}, \bibinfo{person}{Amanda Askell}, \bibinfo{person}{Yuntao Bai}, \bibinfo{person}{Saurav Kadavath}, \bibinfo{person}{Ben Mann}, \bibinfo{person}{Ethan Perez}, \bibinfo{person}{Nicholas Schiefer}, \bibinfo{person}{Kamal Ndousse}, {et~al\mbox{.}}} \bibinfo{year}{2022}\natexlab{}.
\newblock \showarticletitle{Red teaming language models to reduce harms: Methods, scaling behaviors, and lessons learned}.
\newblock \bibinfo{journal}{\emph{arXiv preprint arXiv:2209.07858}} (\bibinfo{year}{2022}).
\newblock


\bibitem[Geirhos et~al\mbox{.}(2020a)]%
        {geirhos2020shortcut}
\bibfield{author}{\bibinfo{person}{Robert Geirhos}, \bibinfo{person}{J{\"o}rn-Henrik Jacobsen}, \bibinfo{person}{Claudio Michaelis}, \bibinfo{person}{Richard Zemel}, \bibinfo{person}{Wieland Brendel}, \bibinfo{person}{Matthias Bethge}, {and} \bibinfo{person}{Felix~A Wichmann}.} \bibinfo{year}{2020}\natexlab{a}.
\newblock \showarticletitle{Shortcut learning in deep neural networks}.
\newblock \bibinfo{journal}{\emph{Nature Machine Intelligence}} \bibinfo{volume}{2}, \bibinfo{number}{11} (\bibinfo{year}{2020}), \bibinfo{pages}{665--673}.
\newblock


\bibitem[Geirhos et~al\mbox{.}(2020b)]%
        {Geirhos2020ShortcutLI}
\bibfield{author}{\bibinfo{person}{Robert Geirhos}, \bibinfo{person}{J{\"o}rn-Henrik Jacobsen}, \bibinfo{person}{Claudio Michaelis}, \bibinfo{person}{Richard~S. Zemel}, \bibinfo{person}{Wieland Brendel}, \bibinfo{person}{Matthias Bethge}, {and} \bibinfo{person}{Felix Wichmann}.} \bibinfo{year}{2020}\natexlab{b}.
\newblock \showarticletitle{Shortcut learning in deep neural networks}.
\newblock \bibinfo{journal}{\emph{Nature Machine Intelligence}}  \bibinfo{volume}{2} (\bibinfo{year}{2020}), \bibinfo{pages}{665 -- 673}.
\newblock
\urldef\tempurl%
\url{https://api.semanticscholar.org/CorpusID:215786368}
\showURL{%
\tempurl}


\bibitem[Geva et~al\mbox{.}(2023)]%
        {geva2023dissecting}
\bibfield{author}{\bibinfo{person}{Mor Geva}, \bibinfo{person}{Jasmijn Bastings}, \bibinfo{person}{Katja Filippova}, {and} \bibinfo{person}{Amir Globerson}.} \bibinfo{year}{2023}\natexlab{}.
\newblock \showarticletitle{Dissecting recall of factual associations in auto-regressive language models}.
\newblock \bibinfo{journal}{\emph{arXiv preprint arXiv:2304.14767}} (\bibinfo{year}{2023}).
\newblock


\bibitem[Geva et~al\mbox{.}(2020)]%
        {geva2020transformer}
\bibfield{author}{\bibinfo{person}{Mor Geva}, \bibinfo{person}{Roei Schuster}, \bibinfo{person}{Jonathan Berant}, {and} \bibinfo{person}{Omer Levy}.} \bibinfo{year}{2020}\natexlab{}.
\newblock \showarticletitle{Transformer feed-forward layers are key-value memories}.
\newblock \bibinfo{journal}{\emph{arXiv preprint arXiv:2012.14913}} (\bibinfo{year}{2020}).
\newblock


\bibitem[Ghorbani and Zou(2020)]%
        {ghorbani2020neuron}
\bibfield{author}{\bibinfo{person}{Amirata Ghorbani} {and} \bibinfo{person}{James Zou}.} \bibinfo{year}{2020}\natexlab{}.
\newblock \showarticletitle{Neuron Shapley: Discovering the Responsible Neurons}.
\newblock  (\bibinfo{year}{2020}).
\newblock
\showeprint[arxiv]{2002.09815}~[stat.ML]


\bibitem[Gichoya et~al\mbox{.}(2022)]%
        {gichoya2022ai}
\bibfield{author}{\bibinfo{person}{Judy~Wawira Gichoya}, \bibinfo{person}{Imon Banerjee}, \bibinfo{person}{Ananth~Reddy Bhimireddy}, \bibinfo{person}{John~L Burns}, \bibinfo{person}{Leo~Anthony Celi}, \bibinfo{person}{Li-Ching Chen}, \bibinfo{person}{Ramon Correa}, \bibinfo{person}{Natalie Dullerud}, \bibinfo{person}{Marzyeh Ghassemi}, \bibinfo{person}{Shih-Cheng Huang}, {et~al\mbox{.}}} \bibinfo{year}{2022}\natexlab{}.
\newblock \showarticletitle{AI recognition of patient race in medical imaging: a modelling study}.
\newblock \bibinfo{journal}{\emph{The Lancet Digital Health}} \bibinfo{volume}{4}, \bibinfo{number}{6} (\bibinfo{year}{2022}), \bibinfo{pages}{e406--e414}.
\newblock


\bibitem[Gilpin et~al\mbox{.}(2019)]%
        {gilpin2019explaining}
\bibfield{author}{\bibinfo{person}{Leilani~H. Gilpin}, \bibinfo{person}{David Bau}, \bibinfo{person}{Ben~Z. Yuan}, \bibinfo{person}{Ayesha Bajwa}, \bibinfo{person}{Michael Specter}, {and} \bibinfo{person}{Lalana Kagal}.} \bibinfo{year}{2019}\natexlab{}.
\newblock \showarticletitle{Explaining {Explanations}: {An} {Overview} of {Interpretability} of {Machine} {Learning}}.
\newblock  (\bibinfo{date}{Feb.} \bibinfo{year}{2019}).
\newblock
\urldef\tempurl%
\url{http://arxiv.org/abs/1806.00069}
\showURL{%
\tempurl}
\newblock
\shownote{arXiv:1806.00069 [cs, stat]}.


\bibitem[Goh and Li(2013)]%
        {goh_disciplining_2013}
\bibfield{author}{\bibinfo{person}{Beng~Wee Goh} {and} \bibinfo{person}{Dan Li}.} \bibinfo{year}{2013}\natexlab{}.
\newblock \showarticletitle{The {Disciplining} {Effect} of the {Internal} {Control} {Provisions} of the {Sarbanes}–{Oxley} {Act} on the {Governance} {Structures} of {Firms}}.
\newblock \bibinfo{journal}{\emph{The International Journal of Accounting}} \bibinfo{volume}{48}, \bibinfo{number}{2} (\bibinfo{date}{June} \bibinfo{year}{2013}), \bibinfo{pages}{248--278}.
\newblock
\showISSN{0020-7063}
\urldef\tempurl%
\url{https://doi.org/10.1016/j.intacc.2013.04.004}
\showDOI{\tempurl}


\bibitem[Golchin and Surdeanu(2023)]%
        {golchin2023time}
\bibfield{author}{\bibinfo{person}{Shahriar Golchin} {and} \bibinfo{person}{Mihai Surdeanu}.} \bibinfo{year}{2023}\natexlab{}.
\newblock \showarticletitle{Time travel in llms: Tracing data contamination in large language models}.
\newblock \bibinfo{journal}{\emph{arXiv preprint arXiv:2308.08493}} (\bibinfo{year}{2023}).
\newblock


\bibitem[Goldman and Barlev(1974)]%
        {goldman1974auditor-firm}
\bibfield{author}{\bibinfo{person}{Arieh Goldman} {and} \bibinfo{person}{Benzion Barlev}.} \bibinfo{year}{1974}\natexlab{}.
\newblock \showarticletitle{The {Auditor}-{Firm} {Conflict} of {Interests}: {Its} {Implications} for {Independence}}.
\newblock \bibinfo{journal}{\emph{The Accounting Review}} \bibinfo{volume}{49}, \bibinfo{number}{4} (\bibinfo{year}{1974}), \bibinfo{pages}{707--718}.
\newblock
\showISSN{0001-4826}
\urldef\tempurl%
\url{https://www.jstor.org/stable/245049}
\showURL{%
\tempurl}
\newblock
\shownote{Publisher: American Accounting Association}.


\bibitem[Goodfellow et~al\mbox{.}(2014)]%
        {goodfellow2014explaining}
\bibfield{author}{\bibinfo{person}{Ian~J Goodfellow}, \bibinfo{person}{Jonathon Shlens}, {and} \bibinfo{person}{Christian Szegedy}.} \bibinfo{year}{2014}\natexlab{}.
\newblock \showarticletitle{Explaining and harnessing adversarial examples}.
\newblock \bibinfo{journal}{\emph{arXiv preprint arXiv:1412.6572}} (\bibinfo{year}{2014}).
\newblock


\bibitem[{Google}(2021)]%
        {google_consultation_2021}
\bibfield{author}{\bibinfo{person}{{Google}}.} \bibinfo{year}{2021}\natexlab{}.
\newblock \bibinfo{title}{Consultation on the {EU} {AI} {Act} {Proposal}}.
\newblock
\newblock
\urldef\tempurl%
\url{https://ec.europa.eu/info/law/better-regulation/have-your-say/initiatives/12527-Artificial-intelligence-ethical-and-legal-requirements/F2662492_en}
\showURL{%
\tempurl}


\bibitem[Goyal et~al\mbox{.}(2022)]%
        {goyal2022fairness}
\bibfield{author}{\bibinfo{person}{Priya Goyal}, \bibinfo{person}{Adriana~Romero Soriano}, \bibinfo{person}{Caner Hazirbas}, \bibinfo{person}{Levent Sagun}, {and} \bibinfo{person}{Nicolas Usunier}.} \bibinfo{year}{2022}\natexlab{}.
\newblock \showarticletitle{Fairness indicators for systematic assessments of visual feature extractors}. In \bibinfo{booktitle}{\emph{Proceedings of the 2022 ACM Conference on Fairness, Accountability, and Transparency}}. \bibinfo{pages}{70--88}.
\newblock


\bibitem[Grathwohl et~al\mbox{.}(2017)]%
        {grathwohl2017backpropagation}
\bibfield{author}{\bibinfo{person}{Will Grathwohl}, \bibinfo{person}{Dami Choi}, \bibinfo{person}{Yuhuai Wu}, \bibinfo{person}{Geoffrey Roeder}, {and} \bibinfo{person}{David Duvenaud}.} \bibinfo{year}{2017}\natexlab{}.
\newblock \showarticletitle{Backpropagation through the void: Optimizing control variates for black-box gradient estimation}.
\newblock \bibinfo{journal}{\emph{arXiv preprint arXiv:1711.00123}} (\bibinfo{year}{2017}).
\newblock


\bibitem[Gryz and Rojszczak(2021)]%
        {gryz2021black}
\bibfield{author}{\bibinfo{person}{Jarek Gryz} {and} \bibinfo{person}{Marcin Rojszczak}.} \bibinfo{year}{2021}\natexlab{}.
\newblock \showarticletitle{Black box algorithms and the rights of individuals: No easy solution to the" explainability" problem}.
\newblock \bibinfo{journal}{\emph{Internet Policy Review}} \bibinfo{volume}{10}, \bibinfo{number}{2} (\bibinfo{year}{2021}), \bibinfo{pages}{1--24}.
\newblock


\bibitem[Guo et~al\mbox{.}(2021)]%
        {guo2021gradient}
\bibfield{author}{\bibinfo{person}{Chuan Guo}, \bibinfo{person}{Alexandre Sablayrolles}, \bibinfo{person}{Herv{\'e} J{\'e}gou}, {and} \bibinfo{person}{Douwe Kiela}.} \bibinfo{year}{2021}\natexlab{}.
\newblock \showarticletitle{Gradient-based adversarial attacks against text transformers}.
\newblock \bibinfo{journal}{\emph{arXiv preprint arXiv:2104.13733}} (\bibinfo{year}{2021}).
\newblock


\bibitem[Gurnee and Tegmark(2023)]%
        {gurnee2023language}
\bibfield{author}{\bibinfo{person}{Wes Gurnee} {and} \bibinfo{person}{Max Tegmark}.} \bibinfo{year}{2023}\natexlab{}.
\newblock \showarticletitle{Language Models Represent Space and Time}.
\newblock  (\bibinfo{year}{2023}).
\newblock
\showeprint[arxiv]{2310.02207}~[cs.LG]


\bibitem[Hacker(2023)]%
        {hacker_european_2023}
\bibfield{author}{\bibinfo{person}{Philipp Hacker}.} \bibinfo{year}{2023}\natexlab{}.
\newblock \showarticletitle{The {European} {AI} liability directives – {Critique} of a half-hearted approach and lessons for the future}.
\newblock \bibinfo{journal}{\emph{Computer Law \& Security Review}}  \bibinfo{volume}{51} (\bibinfo{date}{Nov.} \bibinfo{year}{2023}), \bibinfo{pages}{105871}.
\newblock
\showISSN{0267-3649}
\urldef\tempurl%
\url{https://doi.org/10.1016/j.clsr.2023.105871}
\showDOI{\tempurl}


\bibitem[Hacker et~al\mbox{.}(2023)]%
        {hacker_regulating_2023-1}
\bibfield{author}{\bibinfo{person}{Philipp Hacker}, \bibinfo{person}{Johann Cordes}, {and} \bibinfo{person}{Janina Rochon}.} \bibinfo{year}{2023}\natexlab{}.
\newblock \showarticletitle{Regulating {Gatekeeper} {AI} and {Data}: {Transparency}, {Access}, and {Fairness} under the {DMA}, the {GDPR}, and beyond}.
\newblock  (\bibinfo{date}{Aug.} \bibinfo{year}{2023}).
\newblock
\urldef\tempurl%
\url{http://arxiv.org/abs/2212.04997}
\showURL{%
\tempurl}
\newblock
\shownote{arXiv:2212.04997 [cs]}.


\bibitem[Hacker and Passoth(2022)]%
        {hacker_varieties_2022}
\bibfield{author}{\bibinfo{person}{Philipp Hacker} {and} \bibinfo{person}{Jan-Hendrik Passoth}.} \bibinfo{year}{2022}\natexlab{}.
\newblock \showarticletitle{Varieties of {AI} {Explanations} {Under} the {Law}. {From} the {GDPR} to the {AIA}, and {Beyond}}.
\newblock In \bibinfo{booktitle}{\emph{{xxAI} - {Beyond} {Explainable} {AI}: {International} {Workshop}, {Held} in {Conjunction} with {ICML} 2020, {July} 18, 2020, {Vienna}, {Austria}, {Revised} and {Extended} {Papers}}}, \bibfield{editor}{\bibinfo{person}{Andreas Holzinger}, \bibinfo{person}{Randy Goebel}, \bibinfo{person}{Ruth Fong}, \bibinfo{person}{Taesup Moon}, \bibinfo{person}{Klaus-Robert Müller}, {and} \bibinfo{person}{Wojciech Samek}} (Eds.). \bibinfo{publisher}{Springer International Publishing}, \bibinfo{address}{Cham}, \bibinfo{pages}{343--373}.
\newblock
\showISBNx{978-3-031-04083-2}
\urldef\tempurl%
\url{https://doi.org/10.1007/978-3-031-04083-2_17}
\showDOI{\tempurl}


\bibitem[Hamon et~al\mbox{.}(2022)]%
        {hamon_bridging_2022}
\bibfield{author}{\bibinfo{person}{Ronan Hamon}, \bibinfo{person}{Henrik Junklewitz}, \bibinfo{person}{Ignacio Sanchez}, \bibinfo{person}{Gianclaudio Malgieri}, {and} \bibinfo{person}{Paul De~Hert}.} \bibinfo{year}{2022}\natexlab{}.
\newblock \showarticletitle{Bridging the {Gap} {Between} {AI} and {Explainability} in the {GDPR}: {Towards} {Trustworthiness}-by-{Design} in {Automated} {Decision}-{Making}}.
\newblock \bibinfo{journal}{\emph{IEEE Computational Intelligence Magazine}} \bibinfo{volume}{17}, \bibinfo{number}{1} (\bibinfo{date}{Feb.} \bibinfo{year}{2022}), \bibinfo{pages}{72--85}.
\newblock
\showISSN{1556-6048}
\urldef\tempurl%
\url{https://doi.org/10.1109/MCI.2021.3129960}
\showDOI{\tempurl}
\newblock
\shownote{Conference Name: IEEE Computational Intelligence Magazine}.


\bibitem[Hazell(2023)]%
        {hazell2023large}
\bibfield{author}{\bibinfo{person}{Julian Hazell}.} \bibinfo{year}{2023}\natexlab{}.
\newblock \showarticletitle{Large language models can be used to effectively scale spear phishing campaigns}.
\newblock \bibinfo{journal}{\emph{arXiv preprint arXiv:2305.06972}} (\bibinfo{year}{2023}).
\newblock


\bibitem[He et~al\mbox{.}(2020)]%
        {he2020deberta}
\bibfield{author}{\bibinfo{person}{Pengcheng He}, \bibinfo{person}{Xiaodong Liu}, \bibinfo{person}{Jianfeng Gao}, {and} \bibinfo{person}{Weizhu Chen}.} \bibinfo{year}{2020}\natexlab{}.
\newblock \showarticletitle{Deberta: Decoding-enhanced bert with disentangled attention}.
\newblock \bibinfo{journal}{\emph{arXiv preprint arXiv:2006.03654}} (\bibinfo{year}{2020}).
\newblock


\bibitem[Henderson et~al\mbox{.}(2023)]%
        {henderson2023foundation}
\bibfield{author}{\bibinfo{person}{Peter Henderson}, \bibinfo{person}{Xuechen Li}, \bibinfo{person}{Dan Jurafsky}, \bibinfo{person}{Tatsunori Hashimoto}, \bibinfo{person}{Mark~A Lemley}, {and} \bibinfo{person}{Percy Liang}.} \bibinfo{year}{2023}\natexlab{}.
\newblock \showarticletitle{Foundation models and fair use}.
\newblock \bibinfo{journal}{\emph{arXiv preprint arXiv:2303.15715}} (\bibinfo{year}{2023}).
\newblock


\bibitem[Hendrycks et~al\mbox{.}(2020)]%
        {hendrycks2020measuring}
\bibfield{author}{\bibinfo{person}{Dan Hendrycks}, \bibinfo{person}{Collin Burns}, \bibinfo{person}{Steven Basart}, \bibinfo{person}{Andy Zou}, \bibinfo{person}{Mantas Mazeika}, \bibinfo{person}{Dawn Song}, {and} \bibinfo{person}{Jacob Steinhardt}.} \bibinfo{year}{2020}\natexlab{}.
\newblock \showarticletitle{Measuring massive multitask language understanding}.
\newblock \bibinfo{journal}{\emph{arXiv preprint arXiv:2009.03300}} (\bibinfo{year}{2020}).
\newblock


\bibitem[Hendrycks et~al\mbox{.}(2021)]%
        {hendrycks2021natural}
\bibfield{author}{\bibinfo{person}{Dan Hendrycks}, \bibinfo{person}{Kevin Zhao}, \bibinfo{person}{Steven Basart}, \bibinfo{person}{Jacob Steinhardt}, {and} \bibinfo{person}{Dawn Song}.} \bibinfo{year}{2021}\natexlab{}.
\newblock \showarticletitle{Natural Adversarial Examples}.
\newblock  (\bibinfo{year}{2021}).
\newblock
\showeprint[arxiv]{1907.07174}~[cs.LG]


\bibitem[Hernandez et~al\mbox{.}(2021)]%
        {hernandez2021natural}
\bibfield{author}{\bibinfo{person}{Evan Hernandez}, \bibinfo{person}{Sarah Schwettmann}, \bibinfo{person}{David Bau}, \bibinfo{person}{Teona Bagashvili}, \bibinfo{person}{Antonio Torralba}, {and} \bibinfo{person}{Jacob Andreas}.} \bibinfo{year}{2021}\natexlab{}.
\newblock \showarticletitle{Natural language descriptions of deep visual features}. In \bibinfo{booktitle}{\emph{International Conference on Learning Representations}}.
\newblock


\bibitem[Hess(2019)]%
        {hess2019transparency}
\bibfield{author}{\bibinfo{person}{David Hess}.} \bibinfo{year}{2019}\natexlab{}.
\newblock \showarticletitle{The {Transparency} {Trap}: {Non}-{Financial} {Disclosure} and the {Responsibility} of {Business} to {Respect} {Human} {Rights}}.
\newblock \bibinfo{journal}{\emph{American Business Law Journal}} \bibinfo{volume}{56}, \bibinfo{number}{1} (\bibinfo{year}{2019}), \bibinfo{pages}{5--53}.
\newblock
\showISSN{1744-1714}
\urldef\tempurl%
\url{https://doi.org/10.1111/ablj.12134}
\showDOI{\tempurl}
\newblock
\shownote{\_eprint: https://onlinelibrary.wiley.com/doi/pdf/10.1111/ablj.12134}.


\bibitem[Huang et~al\mbox{.}(2023)]%
        {huang2023survey}
\bibfield{author}{\bibinfo{person}{Lei Huang}, \bibinfo{person}{Weijiang Yu}, \bibinfo{person}{Weitao Ma}, \bibinfo{person}{Weihong Zhong}, \bibinfo{person}{Zhangyin Feng}, \bibinfo{person}{Haotian Wang}, \bibinfo{person}{Qianglong Chen}, \bibinfo{person}{Weihua Peng}, \bibinfo{person}{Xiaocheng Feng}, \bibinfo{person}{Bing Qin}, {and} \bibinfo{person}{Ting Liu}.} \bibinfo{year}{2023}\natexlab{}.
\newblock \showarticletitle{A Survey on Hallucination in Large Language Models: Principles, Taxonomy, Challenges, and Open Questions}.
\newblock  (\bibinfo{year}{2023}).
\newblock
\showeprint[arxiv]{2311.05232}~[cs.CL]


\bibitem[Hubinger(2020)]%
        {hubinger2020overview}
\bibfield{author}{\bibinfo{person}{Evan Hubinger}.} \bibinfo{year}{2020}\natexlab{}.
\newblock \showarticletitle{An overview of 11 proposals for building safe advanced ai}.
\newblock \bibinfo{journal}{\emph{arXiv preprint arXiv:2012.07532}} (\bibinfo{year}{2020}).
\newblock


\bibitem[Hubinger et~al\mbox{.}(2024)]%
        {hubinger2024sleeper}
\bibfield{author}{\bibinfo{person}{Evan Hubinger}, \bibinfo{person}{Carson Denison}, \bibinfo{person}{Jesse Mu}, \bibinfo{person}{Mike Lambert}, \bibinfo{person}{Meg Tong}, \bibinfo{person}{Monte MacDiarmid}, \bibinfo{person}{Tamera Lanham}, \bibinfo{person}{Daniel~M Ziegler}, \bibinfo{person}{Tim Maxwell}, \bibinfo{person}{Newton Cheng}, {et~al\mbox{.}}} \bibinfo{year}{2024}\natexlab{}.
\newblock \showarticletitle{Sleeper Agents: Training Deceptive LLMs that Persist Through Safety Training}.
\newblock \bibinfo{journal}{\emph{arXiv preprint arXiv:2401.05566}} (\bibinfo{year}{2024}).
\newblock


\bibitem[Ilyas et~al\mbox{.}(2018)]%
        {ilyas2018black}
\bibfield{author}{\bibinfo{person}{Andrew Ilyas}, \bibinfo{person}{Logan Engstrom}, \bibinfo{person}{Anish Athalye}, {and} \bibinfo{person}{Jessy Lin}.} \bibinfo{year}{2018}\natexlab{}.
\newblock \showarticletitle{Black-box adversarial attacks with limited queries and information}. In \bibinfo{booktitle}{\emph{International conference on machine learning}}. PMLR, \bibinfo{pages}{2137--2146}.
\newblock


\bibitem[{International Atomic Energy Agency}(2016)]%
        {IAEASafeguardsInspector2016}
\bibfield{author}{\bibinfo{person}{{International Atomic Energy Agency}}.} \bibinfo{year}{2016}\natexlab{}.
\newblock \bibinfo{title}{A Day in the Life of a Safeguards Inspector}.
\newblock
\newblock
\urldef\tempurl%
\url{https://www.iaea.org/newscenter/news/a-day-in-the-life-of-a-safeguards-inspector}
\showURL{%
\tempurl}
\newblock
\shownote{Accessed: 2024-04-15}.


\bibitem[{International Atomic Energy Agency}(2023)]%
        {IAEA2023}
\bibfield{author}{\bibinfo{person}{{International Atomic Energy Agency}}.} \bibinfo{year}{2023}\natexlab{}.
\newblock \bibinfo{title}{IAEA Safeguards Overview: Comprehensive Safeguards Agreements and Additional Protocols}.
\newblock
\newblock
\urldef\tempurl%
\url{https://www.iaea.org/publications/factsheets/iaea-safeguards-overview}
\showURL{%
\tempurl}


\bibitem[Jacovi et~al\mbox{.}(2023)]%
        {jacovi2023stop}
\bibfield{author}{\bibinfo{person}{Alon Jacovi}, \bibinfo{person}{Avi Caciularu}, \bibinfo{person}{Omer Goldman}, {and} \bibinfo{person}{Yoav Goldberg}.} \bibinfo{year}{2023}\natexlab{}.
\newblock \showarticletitle{Stop uploading test data in plain text: Practical strategies for mitigating data contamination by evaluation benchmarks}.
\newblock \bibinfo{journal}{\emph{arXiv preprint arXiv:2305.10160}} (\bibinfo{year}{2023}).
\newblock


\bibitem[Jain et~al\mbox{.}(2023)]%
        {jain2023mechanistically}
\bibfield{author}{\bibinfo{person}{Samyak Jain}, \bibinfo{person}{Robert Kirk}, \bibinfo{person}{Ekdeep~Singh Lubana}, \bibinfo{person}{Robert~P Dick}, \bibinfo{person}{Hidenori Tanaka}, \bibinfo{person}{Edward Grefenstette}, \bibinfo{person}{Tim Rockt{\"a}schel}, {and} \bibinfo{person}{David~Scott Krueger}.} \bibinfo{year}{2023}\natexlab{}.
\newblock \showarticletitle{Mechanistically analyzing the effects of fine-tuning on procedurally defined tasks}.
\newblock \bibinfo{journal}{\emph{arXiv preprint arXiv:2311.12786}} (\bibinfo{year}{2023}).
\newblock


\bibitem[Ji et~al\mbox{.}(2023)]%
        {ji2023survey}
\bibfield{author}{\bibinfo{person}{Ziwei Ji}, \bibinfo{person}{Nayeon Lee}, \bibinfo{person}{Rita Frieske}, \bibinfo{person}{Tiezheng Yu}, \bibinfo{person}{Dan Su}, \bibinfo{person}{Yan Xu}, \bibinfo{person}{Etsuko Ishii}, \bibinfo{person}{Ye~Jin Bang}, \bibinfo{person}{Andrea Madotto}, {and} \bibinfo{person}{Pascale Fung}.} \bibinfo{year}{2023}\natexlab{}.
\newblock \showarticletitle{Survey of hallucination in natural language generation}.
\newblock \bibinfo{journal}{\emph{Comput. Surveys}} \bibinfo{volume}{55}, \bibinfo{number}{12} (\bibinfo{year}{2023}), \bibinfo{pages}{1--38}.
\newblock


\bibitem[Jiang et~al\mbox{.}(2019)]%
        {jiang2019smart}
\bibfield{author}{\bibinfo{person}{Haoming Jiang}, \bibinfo{person}{Pengcheng He}, \bibinfo{person}{Weizhu Chen}, \bibinfo{person}{Xiaodong Liu}, \bibinfo{person}{Jianfeng Gao}, {and} \bibinfo{person}{Tuo Zhao}.} \bibinfo{year}{2019}\natexlab{}.
\newblock \showarticletitle{Smart: Robust and efficient fine-tuning for pre-trained natural language models through principled regularized optimization}.
\newblock \bibinfo{journal}{\emph{arXiv preprint arXiv:1911.03437}} (\bibinfo{year}{2019}).
\newblock


\bibitem[Johnston and Fusi(2023)]%
        {johnston2023abstract}
\bibfield{author}{\bibinfo{person}{W~Jeffrey Johnston} {and} \bibinfo{person}{Stefano Fusi}.} \bibinfo{year}{2023}\natexlab{}.
\newblock \showarticletitle{Abstract representations emerge naturally in neural networks trained to perform multiple tasks}.
\newblock \bibinfo{journal}{\emph{Nature Communications}} \bibinfo{volume}{14}, \bibinfo{number}{1} (\bibinfo{year}{2023}), \bibinfo{pages}{1040}.
\newblock


\bibitem[Jones et~al\mbox{.}(2023)]%
        {jones2023automatically}
\bibfield{author}{\bibinfo{person}{Erik Jones}, \bibinfo{person}{Anca Dragan}, \bibinfo{person}{Aditi Raghunathan}, {and} \bibinfo{person}{Jacob Steinhardt}.} \bibinfo{year}{2023}\natexlab{}.
\newblock \showarticletitle{Automatically Auditing Large Language Models via Discrete Optimization}.
\newblock \bibinfo{journal}{\emph{arXiv preprint arXiv:2303.04381}} (\bibinfo{year}{2023}).
\newblock


\bibitem[Jose et~al\mbox{.}(2021)]%
        {jose2021fairness}
\bibfield{author}{\bibinfo{person}{Joemon~M Jose} {et~al\mbox{.}}} \bibinfo{year}{2021}\natexlab{}.
\newblock \showarticletitle{On fairness and interpretability}.
\newblock \bibinfo{journal}{\emph{arXiv preprint arXiv:2106.13271}} (\bibinfo{year}{2021}).
\newblock


\bibitem[Kapoor and Narayanan(2023)]%
        {kapoor2023leakage}
\bibfield{author}{\bibinfo{person}{Sayash Kapoor} {and} \bibinfo{person}{Arvind Narayanan}.} \bibinfo{year}{2023}\natexlab{}.
\newblock \showarticletitle{Leakage and the reproducibility crisis in machine-learning-based science}.
\newblock \bibinfo{journal}{\emph{Patterns}} \bibinfo{volume}{4}, \bibinfo{number}{9} (\bibinfo{date}{Sept.} \bibinfo{year}{2023}), \bibinfo{pages}{100804}.
\newblock
\showISSN{26663899}
\urldef\tempurl%
\url{https://doi.org/10.1016/j.patter.2023.100804}
\showDOI{\tempurl}


\bibitem[Karamolegkou et~al\mbox{.}(2023)]%
        {karamolegkou2023copyright}
\bibfield{author}{\bibinfo{person}{Antonia Karamolegkou}, \bibinfo{person}{Jiaang Li}, \bibinfo{person}{Li Zhou}, {and} \bibinfo{person}{Anders S{\o}gaard}.} \bibinfo{year}{2023}\natexlab{}.
\newblock \showarticletitle{Copyright Violations and Large Language Models}.
\newblock \bibinfo{journal}{\emph{arXiv preprint arXiv:2310.13771}} (\bibinfo{year}{2023}).
\newblock


\bibitem[Karanjai(2022)]%
        {karanjai2022targeted}
\bibfield{author}{\bibinfo{person}{Rabimba Karanjai}.} \bibinfo{year}{2022}\natexlab{}.
\newblock \showarticletitle{Targeted phishing campaigns using large scale language models}.
\newblock \bibinfo{journal}{\emph{arXiv preprint arXiv:2301.00665}} (\bibinfo{year}{2022}).
\newblock


\bibitem[Kaufmann et~al\mbox{.}(2023)]%
        {kaufmann2023testing}
\bibfield{author}{\bibinfo{person}{Max Kaufmann}, \bibinfo{person}{Daniel Kang}, \bibinfo{person}{Yi Sun}, \bibinfo{person}{Steven Basart}, \bibinfo{person}{Xuwang Yin}, \bibinfo{person}{Mantas Mazeika}, \bibinfo{person}{Akul Arora}, \bibinfo{person}{Adam Dziedzic}, \bibinfo{person}{Franziska Boenisch}, \bibinfo{person}{Tom Brown}, \bibinfo{person}{Jacob Steinhardt}, {and} \bibinfo{person}{Dan Hendrycks}.} \bibinfo{year}{2023}\natexlab{}.
\newblock \showarticletitle{Testing Robustness Against Unforeseen Adversaries}.
\newblock  (\bibinfo{year}{2023}).
\newblock
\showeprint[arxiv]{1908.08016}~[cs.LG]


\bibitem[Kazim et~al\mbox{.}(2021)]%
        {kazim_systematizing_2021}
\bibfield{author}{\bibinfo{person}{Emre Kazim}, \bibinfo{person}{Adriano~Soares Koshiyama}, \bibinfo{person}{Airlie Hilliard}, {and} \bibinfo{person}{Roseline Polle}.} \bibinfo{year}{2021}\natexlab{}.
\newblock \showarticletitle{Systematizing {Audit} in {Algorithmic} {Recruitment}}.
\newblock \bibinfo{journal}{\emph{Journal of Intelligence}} \bibinfo{volume}{9}, \bibinfo{number}{3} (\bibinfo{date}{Sept.} \bibinfo{year}{2021}), \bibinfo{pages}{46}.
\newblock
\showISSN{2079-3200}
\urldef\tempurl%
\url{https://doi.org/10.3390/jintelligence9030046}
\showDOI{\tempurl}
\newblock
\shownote{Number: 3 Publisher: Multidisciplinary Digital Publishing Institute}.


\bibitem[Khalil et~al\mbox{.}(2020)]%
        {khalil_investigating_2020}
\bibfield{author}{\bibinfo{person}{Ashraf Khalil}, \bibinfo{person}{Soha~Glal Ahmed}, \bibinfo{person}{Asad~Masood Khattak}, {and} \bibinfo{person}{Nabeel Al-Qirim}.} \bibinfo{year}{2020}\natexlab{}.
\newblock \showarticletitle{Investigating {Bias} in {Facial} {Analysis} {Systems}: {A} {Systematic} {Review}}.
\newblock \bibinfo{journal}{\emph{IEEE Access}}  \bibinfo{volume}{8} (\bibinfo{year}{2020}), \bibinfo{pages}{130751--130761}.
\newblock
\showISSN{2169-3536}
\urldef\tempurl%
\url{https://doi.org/10.1109/ACCESS.2020.3006051}
\showDOI{\tempurl}
\newblock
\shownote{Conference Name: IEEE Access}.


\bibitem[Khan and Khan(2012)]%
        {khan2012comparative}
\bibfield{author}{\bibinfo{person}{Mohd~Ehmer Khan} {and} \bibinfo{person}{Farmeena Khan}.} \bibinfo{year}{2012}\natexlab{}.
\newblock \showarticletitle{A comparative study of white box, black box and grey box testing techniques}.
\newblock \bibinfo{journal}{\emph{International Journal of Advanced Computer Science and Applications}} \bibinfo{volume}{3}, \bibinfo{number}{6} (\bibinfo{year}{2012}).
\newblock


\bibitem[Khlaaf(2023)]%
        {khlaaf_how_2023}
\bibfield{author}{\bibinfo{person}{Heidy Khlaaf}.} \bibinfo{year}{2023}\natexlab{}.
\newblock \showarticletitle{How {AI} {Can} {Be} {Regulated} {Like} {Nuclear} {Energy}}.
\newblock \bibinfo{journal}{\emph{TIME}} (\bibinfo{date}{Oct.} \bibinfo{year}{2023}).
\newblock
\urldef\tempurl%
\url{https://time.com/6327635/ai-needs-to-be-regulated-like-nuclear-weapons/}
\showURL{%
\tempurl}


\bibitem[Kim et~al\mbox{.}(2018)]%
        {kim_interpretability_2018}
\bibfield{author}{\bibinfo{person}{Been Kim}, \bibinfo{person}{Martin Wattenberg}, \bibinfo{person}{Justin Gilmer}, \bibinfo{person}{Carrie Cai}, \bibinfo{person}{James Wexler}, \bibinfo{person}{Fernanda Viegas}, {and} \bibinfo{person}{Rory Sayres}.} \bibinfo{year}{2018}\natexlab{}.
\newblock \showarticletitle{Interpretability {Beyond} {Feature} {Attribution}: {Quantitative} {Testing} with {Concept} {Activation} {Vectors} ({TCAV})}. In \bibinfo{booktitle}{\emph{Proceedings of the 35th {International} {Conference} on {Machine} {Learning}}}. \bibinfo{publisher}{PMLR}, \bibinfo{pages}{2668--2677}.
\newblock
\urldef\tempurl%
\url{https://proceedings.mlr.press/v80/kim18d.html}
\showURL{%
\tempurl}
\newblock
\shownote{ISSN: 2640-3498}.


\bibitem[Kinniment et~al\mbox{.}(2023)]%
        {kinniment2023evaluating}
\bibfield{author}{\bibinfo{person}{Megan Kinniment}, \bibinfo{person}{Lucas~Jun Koba~Sato}, \bibinfo{person}{Haoxing Du}, \bibinfo{person}{Brian Goodrich}, \bibinfo{person}{Max Hasin}, \bibinfo{person}{Lawrence Chan}, \bibinfo{person}{Luke~Harold Miles}, \bibinfo{person}{Tao~R Lin}, \bibinfo{person}{Hjalmar Wijk}, \bibinfo{person}{Joel Burget}, \bibinfo{person}{Aaron Ho}, \bibinfo{person}{Elizabeth Barnes}, {and} \bibinfo{person}{Paul Christiano}.} \bibinfo{year}{2023}\natexlab{}.
\newblock \showarticletitle{Evaluating Language-Model Agents on Realistic Autonomous Tasks}.
\newblock \bibinfo{howpublished}{\url{https://evals.alignment.org/language-model-pilot-report}}.
\newblock  (\bibinfo{date}{July} \bibinfo{year}{2023}).
\newblock


\bibitem[Kitada and Iyatomi(2023)]%
        {kitada2023making}
\bibfield{author}{\bibinfo{person}{Shunsuke Kitada} {and} \bibinfo{person}{Hitoshi Iyatomi}.} \bibinfo{year}{2023}\natexlab{}.
\newblock \showarticletitle{Making attention mechanisms more robust and interpretable with virtual adversarial training}.
\newblock \bibinfo{journal}{\emph{Applied Intelligence}} \bibinfo{volume}{53}, \bibinfo{number}{12} (\bibinfo{year}{2023}), \bibinfo{pages}{15802--15817}.
\newblock


\bibitem[Koessler and Schuett(2023)]%
        {koessler_risk_2023}
\bibfield{author}{\bibinfo{person}{Leonie Koessler} {and} \bibinfo{person}{Jonas Schuett}.} \bibinfo{year}{2023}\natexlab{}.
\newblock \showarticletitle{Risk assessment at {AGI} companies: {A} review of popular risk assessment techniques from other safety-critical industries}.
\newblock  (\bibinfo{date}{July} \bibinfo{year}{2023}).
\newblock
\urldef\tempurl%
\url{https://arxiv.org/abs/2307.08823v1}
\showURL{%
\tempurl}


\bibitem[Kolt(2023)]%
        {kolt2023algorithmic}
\bibfield{author}{\bibinfo{person}{Noam Kolt}.} \bibinfo{year}{2023}\natexlab{}.
\newblock \showarticletitle{Algorithmic black swans}.
\newblock \bibinfo{journal}{\emph{Washington University Law Review}}  \bibinfo{volume}{101} (\bibinfo{year}{2023}).
\newblock


\bibitem[Koshiyama et~al\mbox{.}(2022)]%
        {koshiyama2022algorithm}
\bibfield{author}{\bibinfo{person}{Adriano Koshiyama}, \bibinfo{person}{Emre Kazim}, {and} \bibinfo{person}{Philip Treleaven}.} \bibinfo{year}{2022}\natexlab{}.
\newblock \showarticletitle{Algorithm auditing: Managing the legal, ethical, and technological risks of artificial intelligence, machine learning, and associated algorithms}.
\newblock \bibinfo{journal}{\emph{Computer}} \bibinfo{volume}{55}, \bibinfo{number}{4} (\bibinfo{year}{2022}), \bibinfo{pages}{40--50}.
\newblock


\bibitem[Krawiec(2003)]%
        {krawiec2003cosmetic}
\bibfield{author}{\bibinfo{person}{Kimberly~D. Krawiec}.} \bibinfo{year}{2003}\natexlab{}.
\newblock \showarticletitle{Cosmetic {Compliance} and the {Failure} of {Negotiated} {Governance}}.
\newblock \bibinfo{journal}{\emph{SSRN Electronic Journal}} (\bibinfo{year}{2003}).
\newblock
\showISSN{1556-5068}
\urldef\tempurl%
\url{https://doi.org/10.2139/ssrn.448221}
\showDOI{\tempurl}


\bibitem[Krishna et~al\mbox{.}(2024)]%
        {DBLP:journals/corr/abs-2402-06625}
\bibfield{author}{\bibinfo{person}{Satyapriya Krishna}, \bibinfo{person}{Chirag Agarwal}, {and} \bibinfo{person}{Himabindu Lakkaraju}.} \bibinfo{year}{2024}\natexlab{}.
\newblock \showarticletitle{Understanding the Effects of Iterative Prompting on Truthfulness}.
\newblock \bibinfo{journal}{\emph{CoRR}}  \bibinfo{volume}{abs/2402.06625} (\bibinfo{year}{2024}).
\newblock
\urldef\tempurl%
\url{https://doi.org/10.48550/ARXIV.2402.06625}
\showDOI{\tempurl}
\showeprint[arXiv]{2402.06625}


\bibitem[Krishna et~al\mbox{.}(2022)]%
        {krishna2022measuring}
\bibfield{author}{\bibinfo{person}{Satyapriya Krishna}, \bibinfo{person}{Rahul Gupta}, \bibinfo{person}{Apurv Verma}, \bibinfo{person}{Jwala Dhamala}, \bibinfo{person}{Yada Pruksachatkun}, {and} \bibinfo{person}{Kai-Wei Chang}.} \bibinfo{year}{2022}\natexlab{}.
\newblock \showarticletitle{Measuring Fairness of Text Classifiers via Prediction Sensitivity}. In \bibinfo{booktitle}{\emph{Proceedings of the 60th Annual Meeting of the Association for Computational Linguistics (Volume 1: Long Papers)}}. \bibinfo{pages}{5830--5842}.
\newblock


\bibitem[Krishnan(2020)]%
        {krishnan2020against}
\bibfield{author}{\bibinfo{person}{Maya Krishnan}.} \bibinfo{year}{2020}\natexlab{}.
\newblock \showarticletitle{Against interpretability: a critical examination of the interpretability problem in machine learning}.
\newblock \bibinfo{journal}{\emph{Philosophy \& Technology}} \bibinfo{volume}{33}, \bibinfo{number}{3} (\bibinfo{year}{2020}), \bibinfo{pages}{487--502}.
\newblock


\bibitem[Kuang and Bharti({[n.\,d.]})]%
        {kuangscale}
\bibfield{author}{\bibinfo{person}{Yilun Kuang} {and} \bibinfo{person}{Yash Bharti}.} \bibinfo{year}{[n.\,d.]}\natexlab{}.
\newblock \showarticletitle{Scale-invariant-Fine-Tuning (SiFT) for Improved Generalization in Classification}.
\newblock  (\bibinfo{year}{[n.\,d.]}).
\newblock


\bibitem[Kumar et~al\mbox{.}(2022)]%
        {kumar2022gradient}
\bibfield{author}{\bibinfo{person}{Sachin Kumar}, \bibinfo{person}{Biswajit Paria}, {and} \bibinfo{person}{Yulia Tsvetkov}.} \bibinfo{year}{2022}\natexlab{}.
\newblock \showarticletitle{Gradient-based constrained sampling from language models}. In \bibinfo{booktitle}{\emph{Proceedings of the 2022 Conference on Empirical Methods in Natural Language Processing}}. \bibinfo{pages}{2251--2277}.
\newblock


\bibitem[Kumari et~al\mbox{.}(2019)]%
        {kumari2019harnessing}
\bibfield{author}{\bibinfo{person}{Nupur Kumari}, \bibinfo{person}{Mayank Singh}, \bibinfo{person}{Abhishek Sinha}, \bibinfo{person}{Harshitha Machiraju}, \bibinfo{person}{Balaji Krishnamurthy}, {and} \bibinfo{person}{Vineeth~N Balasubramanian}.} \bibinfo{year}{2019}\natexlab{}.
\newblock \showarticletitle{Harnessing the vulnerability of latent layers in adversarially trained models}. In \bibinfo{booktitle}{\emph{Proceedings of the 28th International Joint Conference on Artificial Intelligence}}. \bibinfo{pages}{2779--2785}.
\newblock


\bibitem[Lambert et~al\mbox{.}(2023)]%
        {lambert2023entangled}
\bibfield{author}{\bibinfo{person}{Nathan Lambert}, \bibinfo{person}{Thomas~Krendl Gilbert}, {and} \bibinfo{person}{Tom Zick}.} \bibinfo{year}{2023}\natexlab{}.
\newblock \showarticletitle{Entangled Preferences: The History and Risks of Reinforcement Learning and Human Feedback}.
\newblock  (\bibinfo{year}{2023}).
\newblock
\showeprint[arxiv]{2310.13595}~[cs.CY]


\bibitem[Landers and Behrend(2023)]%
        {landers2023auditing}
\bibfield{author}{\bibinfo{person}{Richard~N Landers} {and} \bibinfo{person}{Tara~S Behrend}.} \bibinfo{year}{2023}\natexlab{}.
\newblock \showarticletitle{Auditing the AI auditors: A framework for evaluating fairness and bias in high stakes AI predictive models.}
\newblock \bibinfo{journal}{\emph{American Psychologist}} \bibinfo{volume}{78}, \bibinfo{number}{1} (\bibinfo{year}{2023}), \bibinfo{pages}{36}.
\newblock


\bibitem[Lanz(2023)]%
        {lanz2023stable}
\bibfield{author}{\bibinfo{person}{Jose~Antonio Lanz}.} \bibinfo{year}{2023}\natexlab{}.
\newblock \bibinfo{title}{Stable Diffusion XL v0.9 Leaks Early, Generating Raves From Users}.
\newblock
\newblock
\urldef\tempurl%
\url{https://decrypt.co/147612/stable-diffusion-xl-v0-9-leaks-early-generating-raves-from-users}
\showURL{%
\tempurl}


\bibitem[Lapid et~al\mbox{.}(2023)]%
        {lapid2023open}
\bibfield{author}{\bibinfo{person}{Raz Lapid}, \bibinfo{person}{Ron Langberg}, {and} \bibinfo{person}{Moshe Sipper}.} \bibinfo{year}{2023}\natexlab{}.
\newblock \showarticletitle{Open Sesame! Universal Black Box Jailbreaking of Large Language Models}.
\newblock \bibinfo{journal}{\emph{arXiv preprint arXiv:2309.01446}} (\bibinfo{year}{2023}).
\newblock


\bibitem[Lazar and Nelson(2023)]%
        {lazar2023ai}
\bibfield{author}{\bibinfo{person}{Seth Lazar} {and} \bibinfo{person}{Alondra Nelson}.} \bibinfo{year}{2023}\natexlab{}.
\newblock \bibinfo{title}{AI safety on whose terms?}
\newblock , \bibinfo{numpages}{138--138}~pages.
\newblock


\bibitem[Lee et~al\mbox{.}(2024)]%
        {lee2024mechanistic}
\bibfield{author}{\bibinfo{person}{Andrew Lee}, \bibinfo{person}{Xiaoyan Bai}, \bibinfo{person}{Itamar Pres}, \bibinfo{person}{Martin Wattenberg}, \bibinfo{person}{Jonathan~K Kummerfeld}, {and} \bibinfo{person}{Rada Mihalcea}.} \bibinfo{year}{2024}\natexlab{}.
\newblock \showarticletitle{A Mechanistic Understanding of Alignment Algorithms: A Case Study on DPO and Toxicity}.
\newblock \bibinfo{journal}{\emph{arXiv preprint arXiv:2401.01967}} (\bibinfo{year}{2024}).
\newblock


\bibitem[Lee et~al\mbox{.}(2023)]%
        {sharkey2023auditing}
\bibfield{author}{\bibinfo{person}{Sharkey Lee}, \bibinfo{person}{Ghuidhir Clíodhna~Ní}, \bibinfo{person}{Dan Braun}, \bibinfo{person}{Scheurer Jérémy}, \bibinfo{person}{Mikita Balesni}, \bibinfo{person}{Bushnaq Lucius}, \bibinfo{person}{Stix Charlotte}, {and} \bibinfo{person}{Marius Hobbhahn}.} \bibinfo{year}{2023}\natexlab{}.
\newblock \showarticletitle{A causal framework for AI Regulation and Auditing}.
\newblock  (\bibinfo{year}{2023}).
\newblock


\bibitem[Lemley et~al\mbox{.}(2023)]%
        {henderson2023wheres}
\bibfield{author}{\bibinfo{person}{Mark~A. Lemley}, \bibinfo{person}{Peter Henderson}, {and} \bibinfo{person}{Tatsunori Hashimoto}.} \bibinfo{year}{2023}\natexlab{}.
\newblock \showarticletitle{Where's the {Liability} in {Harmful} {AI} {Speech}?}
\newblock \bibinfo{journal}{\emph{SSRN Electronic Journal}} (\bibinfo{year}{2023}).
\newblock
\showISSN{1556-5068}
\urldef\tempurl%
\url{https://doi.org/10.2139/ssrn.4531029}
\showDOI{\tempurl}


\bibitem[Lennox(2000)]%
        {lennox2000companies}
\bibfield{author}{\bibinfo{person}{Clive Lennox}.} \bibinfo{year}{2000}\natexlab{}.
\newblock \showarticletitle{Do companies successfully engage in opinion-shopping? {Evidence} from the {UK}}.
\newblock \bibinfo{journal}{\emph{Journal of Accounting and Economics}} \bibinfo{volume}{29}, \bibinfo{number}{3} (\bibinfo{date}{June} \bibinfo{year}{2000}), \bibinfo{pages}{321--337}.
\newblock
\showISSN{0165-4101}
\urldef\tempurl%
\url{https://doi.org/10.1016/S0165-4101(00)00025-2}
\showDOI{\tempurl}


\bibitem[Lermen et~al\mbox{.}(2023)]%
        {lermen2023lora}
\bibfield{author}{\bibinfo{person}{Simon Lermen}, \bibinfo{person}{Charlie Rogers-Smith}, {and} \bibinfo{person}{Jeffrey Ladish}.} \bibinfo{year}{2023}\natexlab{}.
\newblock \showarticletitle{LoRA Fine-tuning Efficiently Undoes Safety Training in Llama 2-Chat 70B}.
\newblock  (\bibinfo{year}{2023}).
\newblock
\showeprint[arxiv]{2310.20624}~[cs.LG]


\bibitem[Li et~al\mbox{.}(2023)]%
        {li2023multi}
\bibfield{author}{\bibinfo{person}{Haoran Li}, \bibinfo{person}{Dadi Guo}, \bibinfo{person}{Wei Fan}, \bibinfo{person}{Mingshi Xu}, {and} \bibinfo{person}{Yangqiu Song}.} \bibinfo{year}{2023}\natexlab{}.
\newblock \showarticletitle{Multi-step Jailbreaking Privacy Attacks on ChatGPT}.
\newblock \bibinfo{journal}{\emph{arXiv preprint arXiv:2304.05197}} (\bibinfo{year}{2023}).
\newblock


\bibitem[Li et~al\mbox{.}(2018)]%
        {li2018textbugger}
\bibfield{author}{\bibinfo{person}{Jinfeng Li}, \bibinfo{person}{Shouling Ji}, \bibinfo{person}{Tianyu Du}, \bibinfo{person}{Bo Li}, {and} \bibinfo{person}{Ting Wang}.} \bibinfo{year}{2018}\natexlab{}.
\newblock \showarticletitle{Textbugger: Generating adversarial text against real-world applications}.
\newblock \bibinfo{journal}{\emph{arXiv preprint arXiv:1812.05271}} (\bibinfo{year}{2018}).
\newblock


\bibitem[Li and Qiu(2021)]%
        {li2021token}
\bibfield{author}{\bibinfo{person}{Linyang Li} {and} \bibinfo{person}{Xipeng Qiu}.} \bibinfo{year}{2021}\natexlab{}.
\newblock \showarticletitle{Token-aware virtual adversarial training in natural language understanding}. In \bibinfo{booktitle}{\emph{Proceedings of the AAAI Conference on Artificial Intelligence}}, Vol.~\bibinfo{volume}{35}. \bibinfo{pages}{8410--8418}.
\newblock


\bibitem[Liang et~al\mbox{.}(2022)]%
        {liang2022holistic}
\bibfield{author}{\bibinfo{person}{Percy Liang}, \bibinfo{person}{Rishi Bommasani}, \bibinfo{person}{Tony Lee}, \bibinfo{person}{Dimitris Tsipras}, \bibinfo{person}{Dilara Soylu}, \bibinfo{person}{Michihiro Yasunaga}, \bibinfo{person}{Yian Zhang}, \bibinfo{person}{Deepak Narayanan}, \bibinfo{person}{Yuhuai Wu}, \bibinfo{person}{Ananya Kumar}, {et~al\mbox{.}}} \bibinfo{year}{2022}\natexlab{}.
\newblock \showarticletitle{Holistic evaluation of language models}.
\newblock \bibinfo{journal}{\emph{arXiv preprint arXiv:2211.09110}} (\bibinfo{year}{2022}).
\newblock


\bibitem[Linardatos et~al\mbox{.}(2020)]%
        {linardatos2020explainable}
\bibfield{author}{\bibinfo{person}{Pantelis Linardatos}, \bibinfo{person}{Vasilis Papastefanopoulos}, {and} \bibinfo{person}{Sotiris Kotsiantis}.} \bibinfo{year}{2020}\natexlab{}.
\newblock \showarticletitle{Explainable ai: A review of machine learning interpretability methods}.
\newblock \bibinfo{journal}{\emph{Entropy}} \bibinfo{volume}{23}, \bibinfo{number}{1} (\bibinfo{year}{2020}), \bibinfo{pages}{18}.
\newblock


\bibitem[Linthicum et~al\mbox{.}(2010)]%
        {linthicum2010social}
\bibfield{author}{\bibinfo{person}{Cheryl Linthicum}, \bibinfo{person}{Austin~L Reitenga}, {and} \bibinfo{person}{Juan~Manuel Sanchez}.} \bibinfo{year}{2010}\natexlab{}.
\newblock \showarticletitle{Social responsibility and corporate reputation: The case of the Arthur Andersen Enron audit failure}.
\newblock \bibinfo{journal}{\emph{Journal of Accounting and Public Policy}} \bibinfo{volume}{29}, \bibinfo{number}{2} (\bibinfo{year}{2010}), \bibinfo{pages}{160--176}.
\newblock


\bibitem[Liu et~al\mbox{.}(2022b)]%
        {Liu2022CharacterlevelWA}
\bibfield{author}{\bibinfo{person}{Aiwei Liu}, \bibinfo{person}{Honghai Yu}, \bibinfo{person}{Xuming Hu}, \bibinfo{person}{Shuang Li}, \bibinfo{person}{Li Lin}, \bibinfo{person}{Fukun Ma}, \bibinfo{person}{Yawen Yang}, {and} \bibinfo{person}{Lijie Wen}.} \bibinfo{year}{2022}\natexlab{b}.
\newblock \showarticletitle{Character-level White-Box Adversarial Attacks against Transformers via Attachable Subwords Substitution}.
\newblock \bibinfo{journal}{\emph{ArXiv}}  \bibinfo{volume}{abs/2210.17004} (\bibinfo{year}{2022}).
\newblock
\urldef\tempurl%
\url{https://api.semanticscholar.org/CorpusID:253236900}
\showURL{%
\tempurl}


\bibitem[Liu et~al\mbox{.}(2020)]%
        {liu2020adversarial}
\bibfield{author}{\bibinfo{person}{Xiaodong Liu}, \bibinfo{person}{Hao Cheng}, \bibinfo{person}{Pengcheng He}, \bibinfo{person}{Weizhu Chen}, \bibinfo{person}{Yu Wang}, \bibinfo{person}{Hoifung Poon}, {and} \bibinfo{person}{Jianfeng Gao}.} \bibinfo{year}{2020}\natexlab{}.
\newblock \showarticletitle{Adversarial training for large neural language models}.
\newblock \bibinfo{journal}{\emph{arXiv preprint arXiv:2004.08994}} (\bibinfo{year}{2020}).
\newblock


\bibitem[Liu et~al\mbox{.}(2022a)]%
        {liu_medical_2022}
\bibfield{author}{\bibinfo{person}{Xiaoxuan Liu}, \bibinfo{person}{Ben Glocker}, \bibinfo{person}{Melissa~M. McCradden}, \bibinfo{person}{Marzyeh Ghassemi}, \bibinfo{person}{Alastair~K. Denniston}, {and} \bibinfo{person}{Lauren Oakden-Rayner}.} \bibinfo{year}{2022}\natexlab{a}.
\newblock \showarticletitle{The medical algorithmic audit}.
\newblock \bibinfo{journal}{\emph{The Lancet Digital Health}} \bibinfo{volume}{4}, \bibinfo{number}{5} (\bibinfo{date}{May} \bibinfo{year}{2022}), \bibinfo{pages}{e384--e397}.
\newblock
\showISSN{2589-7500}
\urldef\tempurl%
\url{https://doi.org/10.1016/S2589-7500(22)00003-6}
\showDOI{\tempurl}
\newblock
\shownote{Publisher: Elsevier}.


\bibitem[Liu et~al\mbox{.}(2023b)]%
        {liu2023latent}
\bibfield{author}{\bibinfo{person}{Xingbin Liu}, \bibinfo{person}{Huafeng Kuang}, \bibinfo{person}{Hong Liu}, \bibinfo{person}{Xianming Lin}, \bibinfo{person}{Yongjian Wu}, {and} \bibinfo{person}{Rongrong Ji}.} \bibinfo{year}{2023}\natexlab{b}.
\newblock \showarticletitle{Latent Feature Relation Consistency for Adversarial Robustness}.
\newblock \bibinfo{journal}{\emph{arXiv preprint arXiv:2303.16697}} (\bibinfo{year}{2023}).
\newblock


\bibitem[Liu et~al\mbox{.}(2016)]%
        {liu2016delving}
\bibfield{author}{\bibinfo{person}{Yanpei Liu}, \bibinfo{person}{Xinyun Chen}, \bibinfo{person}{Chang Liu}, {and} \bibinfo{person}{Dawn Song}.} \bibinfo{year}{2016}\natexlab{}.
\newblock \showarticletitle{Delving into transferable adversarial examples and black-box attacks}.
\newblock \bibinfo{journal}{\emph{arXiv preprint arXiv:1611.02770}} (\bibinfo{year}{2016}).
\newblock


\bibitem[Liu et~al\mbox{.}(2023a)]%
        {liu2023jailbreaking}
\bibfield{author}{\bibinfo{person}{Yi Liu}, \bibinfo{person}{Gelei Deng}, \bibinfo{person}{Zhengzi Xu}, \bibinfo{person}{Yuekang Li}, \bibinfo{person}{Yaowen Zheng}, \bibinfo{person}{Ying Zhang}, \bibinfo{person}{Lida Zhao}, \bibinfo{person}{Tianwei Zhang}, {and} \bibinfo{person}{Yang Liu}.} \bibinfo{year}{2023}\natexlab{a}.
\newblock \showarticletitle{Jailbreaking chatgpt via prompt engineering: An empirical study}.
\newblock \bibinfo{journal}{\emph{arXiv preprint arXiv:2305.13860}} (\bibinfo{year}{2023}).
\newblock


\bibitem[Lucaj et~al\mbox{.}(2023)]%
        {Lucaj2023AIRI}
\bibfield{author}{\bibinfo{person}{Laura Lucaj}, \bibinfo{person}{Patrick van~der Smagt}, {and} \bibinfo{person}{Djalel Benbouzid}.} \bibinfo{year}{2023}\natexlab{}.
\newblock \showarticletitle{AI Regulation Is (not) All You Need}.
\newblock \bibinfo{journal}{\emph{Proceedings of the 2023 ACM Conference on Fairness, Accountability, and Transparency}} (\bibinfo{year}{2023}).
\newblock
\urldef\tempurl%
\url{https://api.semanticscholar.org/CorpusID:259139804}
\showURL{%
\tempurl}


\bibitem[Luccioni and Viviano(2021)]%
        {luccioni2021s}
\bibfield{author}{\bibinfo{person}{Alexandra~Sasha Luccioni} {and} \bibinfo{person}{Joseph~D Viviano}.} \bibinfo{year}{2021}\natexlab{}.
\newblock \showarticletitle{What's in the Box? A Preliminary Analysis of Undesirable Content in the Common Crawl Corpus}.
\newblock \bibinfo{journal}{\emph{arXiv preprint arXiv:2105.02732}} (\bibinfo{year}{2021}).
\newblock


\bibitem[Mahajan et~al\mbox{.}(2020)]%
        {mahajan_algorithmic_2020}
\bibfield{author}{\bibinfo{person}{Vidur Mahajan}, \bibinfo{person}{Vasantha~Kumar Venugopal}, \bibinfo{person}{Murali Murugavel}, {and} \bibinfo{person}{Harsh Mahajan}.} \bibinfo{year}{2020}\natexlab{}.
\newblock \showarticletitle{The {Algorithmic} {Audit}: {Working} with {Vendors} to {Validate} {Radiology}-{AI} {Algorithms}—{How} {We} {Do} {It}}.
\newblock \bibinfo{journal}{\emph{Academic Radiology}} \bibinfo{volume}{27}, \bibinfo{number}{1} (\bibinfo{date}{Jan.} \bibinfo{year}{2020}), \bibinfo{pages}{132--135}.
\newblock
\showISSN{1076-6332}
\urldef\tempurl%
\url{https://doi.org/10.1016/j.acra.2019.09.009}
\showDOI{\tempurl}


\bibitem[Marks et~al\mbox{.}(2024)]%
        {marks2024sparse}
\bibfield{author}{\bibinfo{person}{Samuel Marks}, \bibinfo{person}{Can Rager}, \bibinfo{person}{Eric~J. Michaud}, \bibinfo{person}{Yonatan Belinkov}, \bibinfo{person}{David Bau}, {and} \bibinfo{person}{Aaron Mueller}.} \bibinfo{year}{2024}\natexlab{}.
\newblock \bibinfo{title}{Sparse Feature Circuits: Discovering and Editing Interpretable Causal Graphs in Language Models}.
\newblock
\newblock
\showeprint[arxiv]{2403.19647}~[cs.LG]


\bibitem[Marks and Tegmark(2023)]%
        {marks2023geometry}
\bibfield{author}{\bibinfo{person}{Samuel Marks} {and} \bibinfo{person}{Max Tegmark}.} \bibinfo{year}{2023}\natexlab{}.
\newblock \showarticletitle{The Geometry of Truth: Emergent Linear Structure in Large Language Model Representations of True/False Datasets}.
\newblock  (\bibinfo{year}{2023}).
\newblock
\showeprint[arxiv]{2310.06824}~[cs.AI]


\bibitem[Marquis et~al\mbox{.}(2016)]%
        {marquis2016scrutiny}
\bibfield{author}{\bibinfo{person}{Christopher Marquis}, \bibinfo{person}{Michael~W. Toffel}, {and} \bibinfo{person}{Yanhua Zhou}.} \bibinfo{year}{2016}\natexlab{}.
\newblock \showarticletitle{Scrutiny, {Norms}, and {Selective} {Disclosure}: {A} {Global} {Study} of {Greenwashing}}.
\newblock \bibinfo{journal}{\emph{Organization Science}} \bibinfo{volume}{27}, \bibinfo{number}{2} (\bibinfo{date}{March} \bibinfo{year}{2016}), \bibinfo{pages}{483--504}.
\newblock
\showISSN{1047-7039}
\urldef\tempurl%
\url{https://doi.org/10.1287/orsc.2015.1039}
\showDOI{\tempurl}
\newblock
\shownote{Publisher: INFORMS}.


\bibitem[Martic et~al\mbox{.}(2018)]%
        {martic2018scaling}
\bibfield{author}{\bibinfo{person}{Miljan Martic}, \bibinfo{person}{Jan Leike}, \bibinfo{person}{Andrew Trask}, \bibinfo{person}{Matteo Hessel}, \bibinfo{person}{Shane Legg}, {and} \bibinfo{person}{Pushmeet Kohli}.} \bibinfo{year}{2018}\natexlab{}.
\newblock \showarticletitle{Scaling shared model governance via model splitting}.
\newblock \bibinfo{journal}{\emph{arXiv preprint arXiv:1812.05979}} (\bibinfo{year}{2018}).
\newblock


\bibitem[Mehrabi et~al\mbox{.}(2021)]%
        {mehrabi2021survey}
\bibfield{author}{\bibinfo{person}{Ninareh Mehrabi}, \bibinfo{person}{Fred Morstatter}, \bibinfo{person}{Nripsuta Saxena}, \bibinfo{person}{Kristina Lerman}, {and} \bibinfo{person}{Aram Galstyan}.} \bibinfo{year}{2021}\natexlab{}.
\newblock \showarticletitle{A survey on bias and fairness in machine learning}.
\newblock \bibinfo{journal}{\emph{ACM computing surveys (CSUR)}} \bibinfo{volume}{54}, \bibinfo{number}{6} (\bibinfo{year}{2021}), \bibinfo{pages}{1--35}.
\newblock


\bibitem[Meng et~al\mbox{.}(2022)]%
        {meng2022locating}
\bibfield{author}{\bibinfo{person}{Kevin Meng}, \bibinfo{person}{David Bau}, \bibinfo{person}{Alex Andonian}, {and} \bibinfo{person}{Yonatan Belinkov}.} \bibinfo{year}{2022}\natexlab{}.
\newblock \showarticletitle{Locating and editing factual associations in GPT}.
\newblock \bibinfo{journal}{\emph{Advances in Neural Information Processing Systems}}  \bibinfo{volume}{35} (\bibinfo{year}{2022}), \bibinfo{pages}{17359--17372}.
\newblock


\bibitem[Metcalf et~al\mbox{.}(2022)]%
        {metcalf2022relationship}
\bibfield{author}{\bibinfo{person}{Jacob Metcalf}, \bibinfo{person}{Emanuel Moss}, \bibinfo{person}{Ranjit Singh}, \bibinfo{person}{Emnet Tafese}, {and} \bibinfo{person}{Elizabeth~Anne Watkins}.} \bibinfo{year}{2022}\natexlab{}.
\newblock \showarticletitle{A relationship and not a thing: A relational approach to algorithmic accountability and assessment documentation}.
\newblock \bibinfo{journal}{\emph{arXiv preprint arXiv:2203.01455}} (\bibinfo{year}{2022}).
\newblock


\bibitem[Metcalf et~al\mbox{.}(2021)]%
        {metcalf2021algorithmic}
\bibfield{author}{\bibinfo{person}{Jacob Metcalf}, \bibinfo{person}{Emanuel Moss}, \bibinfo{person}{Elizabeth~Anne Watkins}, \bibinfo{person}{Ranjit Singh}, {and} \bibinfo{person}{Madeleine~Clare Elish}.} \bibinfo{year}{2021}\natexlab{}.
\newblock \showarticletitle{Algorithmic impact assessments and accountability: The co-construction of impacts}. In \bibinfo{booktitle}{\emph{Proceedings of the 2021 ACM conference on fairness, accountability, and transparency}}. \bibinfo{pages}{735--746}.
\newblock


\bibitem[METR(2023)]%
        {metr}
\bibfield{author}{\bibinfo{person}{METR}.} \bibinfo{year}{2023}\natexlab{}.
\newblock \bibinfo{title}{METR}.
\newblock
\newblock
\urldef\tempurl%
\url{https://evals.alignment.org/}
\showURL{%
\tempurl}


\bibitem[Miller(2019)]%
        {miller2019explanation}
\bibfield{author}{\bibinfo{person}{Tim Miller}.} \bibinfo{year}{2019}\natexlab{}.
\newblock \showarticletitle{Explanation in artificial intelligence: Insights from the social sciences}.
\newblock \bibinfo{journal}{\emph{Artificial intelligence}}  \bibinfo{volume}{267} (\bibinfo{year}{2019}), \bibinfo{pages}{1--38}.
\newblock


\bibitem[Miotti and Wasil(2023)]%
        {miotti2023taking}
\bibfield{author}{\bibinfo{person}{Andrea Miotti} {and} \bibinfo{person}{Akash Wasil}.} \bibinfo{year}{2023}\natexlab{}.
\newblock \showarticletitle{Taking control: Policies to address extinction risks from advanced AI}.
\newblock \bibinfo{journal}{\emph{arXiv preprint arXiv:2310.20563}} (\bibinfo{year}{2023}).
\newblock


\bibitem[Mitchell et~al\mbox{.}(2019)]%
        {mitchell2019model}
\bibfield{author}{\bibinfo{person}{Margaret Mitchell}, \bibinfo{person}{Simone Wu}, \bibinfo{person}{Andrew Zaldivar}, \bibinfo{person}{Parker Barnes}, \bibinfo{person}{Lucy Vasserman}, \bibinfo{person}{Ben Hutchinson}, \bibinfo{person}{Elena Spitzer}, \bibinfo{person}{Inioluwa~Deborah Raji}, {and} \bibinfo{person}{Timnit Gebru}.} \bibinfo{year}{2019}\natexlab{}.
\newblock \showarticletitle{Model cards for model reporting}. In \bibinfo{booktitle}{\emph{Proceedings of the conference on fairness, accountability, and transparency}}. \bibinfo{pages}{220--229}.
\newblock


\bibitem[M{\"o}kander(2023)]%
        {mokander2023auditing}
\bibfield{author}{\bibinfo{person}{Jakob M{\"o}kander}.} \bibinfo{year}{2023}\natexlab{}.
\newblock \showarticletitle{Auditing of AI: Legal, Ethical and Technical Approaches}.
\newblock \bibinfo{journal}{\emph{Digital Society}} \bibinfo{volume}{2}, \bibinfo{number}{3} (\bibinfo{year}{2023}), \bibinfo{pages}{49}.
\newblock


\bibitem[Moore et~al\mbox{.}(2006)]%
        {moore2006conflicts}
\bibfield{author}{\bibinfo{person}{Don~A. Moore}, \bibinfo{person}{Philip~E. Tetlock}, \bibinfo{person}{Lloyd Tanlu}, {and} \bibinfo{person}{Max~H. Bazerman}.} \bibinfo{year}{2006}\natexlab{}.
\newblock \showarticletitle{Conflicts of {Interest} and the {Case} of {Auditor} {Independence}: {Moral} {Seduction} and {Strategic} {Issue} {Cycling}}.
\newblock \bibinfo{journal}{\emph{The Academy of Management Review}} \bibinfo{volume}{31}, \bibinfo{number}{1} (\bibinfo{year}{2006}), \bibinfo{pages}{10--29}.
\newblock
\showISSN{0363-7425}
\urldef\tempurl%
\url{https://www.jstor.org/stable/20159182}
\showURL{%
\tempurl}
\newblock
\shownote{Publisher: Academy of Management}.


\bibitem[Mouton et~al\mbox{.}(2023)]%
        {mouton2023operational}
\bibfield{author}{\bibinfo{person}{Christopher~A Mouton}, \bibinfo{person}{Caleb Lucas}, {and} \bibinfo{person}{Ella Guest}.} \bibinfo{year}{2023}\natexlab{}.
\newblock \showarticletitle{The Operational Risks of AI in Large-Scale Biological Attacks: A Red-Team Approach}.
\newblock  (\bibinfo{year}{2023}).
\newblock


\bibitem[Mu and Andreas(2020)]%
        {mu2020compositional}
\bibfield{author}{\bibinfo{person}{Jesse Mu} {and} \bibinfo{person}{Jacob Andreas}.} \bibinfo{year}{2020}\natexlab{}.
\newblock \showarticletitle{Compositional explanations of neurons}.
\newblock \bibinfo{journal}{\emph{Advances in Neural Information Processing Systems}}  \bibinfo{volume}{33} (\bibinfo{year}{2020}), \bibinfo{pages}{17153--17163}.
\newblock


\bibitem[Mökander and Floridi(2021)]%
        {mokander_ethics-based_2021}
\bibfield{author}{\bibinfo{person}{Jakob Mökander} {and} \bibinfo{person}{Luciano Floridi}.} \bibinfo{year}{2021}\natexlab{}.
\newblock \showarticletitle{Ethics-{Based} {Auditing} to {Develop} {Trustworthy} {AI}}.
\newblock \bibinfo{journal}{\emph{Minds and Machines}} \bibinfo{volume}{31}, \bibinfo{number}{2} (\bibinfo{date}{June} \bibinfo{year}{2021}), \bibinfo{pages}{323--327}.
\newblock
\showISSN{1572-8641}
\urldef\tempurl%
\url{https://doi.org/10.1007/s11023-021-09557-8}
\showDOI{\tempurl}


\bibitem[Mökander et~al\mbox{.}(2023a)]%
        {M_kander_2023}
\bibfield{author}{\bibinfo{person}{Jakob Mökander}, \bibinfo{person}{Jonas Schuett}, \bibinfo{person}{Hannah~Rose Kirk}, {and} \bibinfo{person}{Luciano Floridi}.} \bibinfo{year}{2023}\natexlab{a}.
\newblock \showarticletitle{Auditing large language models: a three-layered approach}.
\newblock \bibinfo{journal}{\emph{{AI} and Ethics}} (\bibinfo{date}{may} \bibinfo{year}{2023}).
\newblock
\urldef\tempurl%
\url{https://doi.org/10.1007/s43681-023-00289-2}
\showDOI{\tempurl}


\bibitem[Mökander et~al\mbox{.}(2023b)]%
        {mokander_auditing_2023}
\bibfield{author}{\bibinfo{person}{Jakob Mökander}, \bibinfo{person}{Jonas Schuett}, \bibinfo{person}{Hannah~Rose Kirk}, {and} \bibinfo{person}{Luciano Floridi}.} \bibinfo{year}{2023}\natexlab{b}.
\newblock \showarticletitle{Auditing large language models: a three-layered approach}.
\newblock \bibinfo{journal}{\emph{AI and Ethics}} (\bibinfo{date}{May} \bibinfo{year}{2023}).
\newblock
\showISSN{2730-5953, 2730-5961}
\urldef\tempurl%
\url{https://doi.org/10.1007/s43681-023-00289-2}
\showDOI{\tempurl}
\newblock
\shownote{arXiv:2302.08500 [cs]}.


\bibitem[Naihin et~al\mbox{.}(2023)]%
        {naihin_testing_2023}
\bibfield{author}{\bibinfo{person}{Silen Naihin}, \bibinfo{person}{David Atkinson}, \bibinfo{person}{Marc Green}, \bibinfo{person}{Merwane Hamadi}, \bibinfo{person}{Craig Swift}, \bibinfo{person}{Douglas Schonholtz}, \bibinfo{person}{Adam~Tauman Kalai}, {and} \bibinfo{person}{David Bau}.} \bibinfo{year}{2023}\natexlab{}.
\newblock \showarticletitle{Testing {Language} {Model} {Agents} {Safely} in the {Wild}}.
\newblock  (\bibinfo{date}{Dec.} \bibinfo{year}{2023}).
\newblock
\urldef\tempurl%
\url{https://doi.org/10.48550/arXiv.2311.10538}
\showDOI{\tempurl}
\newblock
\shownote{arXiv:2311.10538 [cs]}.


\bibitem[Nanda et~al\mbox{.}(2023)]%
        {nanda2023progress}
\bibfield{author}{\bibinfo{person}{Neel Nanda}, \bibinfo{person}{Lawrence Chan}, \bibinfo{person}{Tom Lieberum}, \bibinfo{person}{Jess Smith}, {and} \bibinfo{person}{Jacob Steinhardt}.} \bibinfo{year}{2023}\natexlab{}.
\newblock \showarticletitle{Progress measures for grokking via mechanistic interpretability}.
\newblock  (\bibinfo{year}{2023}).
\newblock
\showeprint[arxiv]{2301.05217}~[cs.LG]


\bibitem[Narayanan and Kapoor(2023)]%
        {narayanan2023evaluating}
\bibfield{author}{\bibinfo{person}{Arvind Narayanan} {and} \bibinfo{person}{Sayash Kapoor}.} \bibinfo{year}{2023}\natexlab{}.
\newblock \bibinfo{title}{Evaluating {LLMs} is a minefield}.
\newblock
\newblock
\urldef\tempurl%
\url{https://www.cs.princeton.edu/~arvindn/talks/evaluating_llms_minefield/#/8}
\showURL{%
\tempurl}


\bibitem[Nasr et~al\mbox{.}(2023)]%
        {nasr_scalable_2023}
\bibfield{author}{\bibinfo{person}{Milad Nasr}, \bibinfo{person}{Nicholas Carlini}, \bibinfo{person}{Jonathan Hayase}, \bibinfo{person}{Matthew Jagielski}, \bibinfo{person}{A.~Feder Cooper}, \bibinfo{person}{Daphne Ippolito}, \bibinfo{person}{Christopher~A. Choquette-Choo}, \bibinfo{person}{Eric Wallace}, \bibinfo{person}{Florian Tramèr}, {and} \bibinfo{person}{Katherine Lee}.} \bibinfo{year}{2023}\natexlab{}.
\newblock \showarticletitle{Scalable {Extraction} of {Training} {Data} from ({Production}) {Language} {Models}}.
\newblock  (\bibinfo{date}{Nov.} \bibinfo{year}{2023}).
\newblock
\urldef\tempurl%
\url{https://doi.org/10.48550/arXiv.2311.17035}
\showDOI{\tempurl}
\newblock
\shownote{arXiv:2311.17035 [cs]}.


\bibitem[{National Institute for Standards and Technology}(2023)]%
        {nist2023request}
\bibfield{author}{\bibinfo{person}{{National Institute for Standards and Technology}}.} \bibinfo{year}{2023}\natexlab{}.
\newblock \bibinfo{title}{Request for Information (RFI) Related to NIST's Assignments Under Sections 4.1, 4.5 and 11 of the Executive Order Concerning Artificial Intelligence (Sections 4.1, 4.5, and 11)}.
\newblock
\newblock
\urldef\tempurl%
\url{https://www.federalregister.gov/documents/2023/12/21/2023-28232/request-for-information-rfi-related-to-nists-assignments-under-sections-41-45-and-11-of-the}
\showURL{%
\tempurl}


\bibitem[{National New Generation Artificial Intelligence Governance Expert Committee}(2019)]%
        {nngai2023}
\bibfield{author}{\bibinfo{person}{{National New Generation Artificial Intelligence Governance Expert Committee}}.} \bibinfo{year}{2019}\natexlab{}.
\newblock \bibinfo{title}{Translation: {Chinese} {Expert} {Group} {Offers} '{Governance} {Principles}' for '{Responsible} {AI}'}.
\newblock
\newblock
\urldef\tempurl%
\url{https://digichina.stanford.edu/work/translation-chinese-expert-group-offers-governance-principles-for-responsible-ai/}
\showURL{%
\tempurl}


\bibitem[{National New Generation Artificial Intelligence Governance Specialist Committee}(2021)]%
        {nngai2021}
\bibfield{author}{\bibinfo{person}{{National New Generation Artificial Intelligence Governance Specialist Committee}}.} \bibinfo{year}{2021}\natexlab{}.
\newblock \bibinfo{title}{"{Ethical} {Norms} for {New} {Generation} {Artificial} {Intelligence}" {Released}}.
\newblock
\newblock
\urldef\tempurl%
\url{https://cset.georgetown.edu/publication/ethical-norms-for-new-generation-artificial-intelligence-released/}
\showURL{%
\tempurl}


\bibitem[Nevo et~al\mbox{.}(2023)]%
        {nevo_securing_2023}
\bibfield{author}{\bibinfo{person}{Sella Nevo}, \bibinfo{person}{Dan Lahav}, \bibinfo{person}{Ajay Karpur}, \bibinfo{person}{Jeff Alstott}, {and} \bibinfo{person}{Jason Matheny}.} \bibinfo{year}{2023}\natexlab{}.
\newblock \bibinfo{booktitle}{\emph{Securing {Artificial} {Intelligence} {Model} {Weights}: {Interim} {Report}}}.
\newblock \bibinfo{type}{{T}echnical {R}eport}. \bibinfo{institution}{RAND Corporation}.
\newblock
\urldef\tempurl%
\url{https://www.rand.org/pubs/working_papers/WRA2849-1.html}
\showURL{%
\tempurl}


\bibitem[Ng et~al\mbox{.}(2023)]%
        {chinaaiml2023}
\bibfield{author}{\bibinfo{person}{Kwan~Yee Ng}, \bibinfo{person}{Jason Zhou}, \bibinfo{person}{Ben Murphy}, \bibinfo{person}{Rogier Creemers}, {and} \bibinfo{person}{Hunter Dorwart}.} \bibinfo{year}{2023}\natexlab{}.
\newblock \bibinfo{title}{Translation: {Artificial} {Intelligence} {Law}, {Model} {Law} v. 1.0 ({Expert} {Suggestion} {Draft}) – {Aug}. 2023}.  (\bibinfo{date}{Aug.} \bibinfo{year}{2023}).
\newblock
\urldef\tempurl%
\url{https://digichina.stanford.edu/work/translation-artificial-intelligence-law-model-law-v-1-0-expert-suggestion-draft-aug-2023/}
\showURL{%
\tempurl}


\bibitem[Ngo et~al\mbox{.}(2022)]%
        {ngo2022alignment}
\bibfield{author}{\bibinfo{person}{Richard Ngo}, \bibinfo{person}{Lawrence Chan}, {and} \bibinfo{person}{S{\"o}ren Mindermann}.} \bibinfo{year}{2022}\natexlab{}.
\newblock \showarticletitle{The alignment problem from a deep learning perspective}.
\newblock \bibinfo{journal}{\emph{arXiv preprint arXiv:2209.00626}} (\bibinfo{year}{2022}).
\newblock


\bibitem[Nielson(2018)]%
        {nielson2018sticky}
\bibfield{author}{\bibinfo{person}{Aaron~L Nielson}.} \bibinfo{year}{2018}\natexlab{}.
\newblock \showarticletitle{Sticky Regulations}.
\newblock \bibinfo{journal}{\emph{U. Chi. L. Rev.}}  \bibinfo{volume}{85} (\bibinfo{year}{2018}), \bibinfo{pages}{85}.
\newblock


\bibitem[{OECD}(2019)]%
        {oecd2019}
\bibfield{author}{\bibinfo{person}{{OECD}}.} \bibinfo{year}{2019}\natexlab{}.
\newblock \bibinfo{title}{Recommendation of the {Council} on {Artificial} {Intelligence}}.
\newblock
\newblock
\urldef\tempurl%
\url{https://legalinstruments.oecd.org/en/instruments/OECD-LEGAL-0449}
\showURL{%
\tempurl}


\bibitem[of~Federal~Regulations(2023)]%
        {sec-reg-m}
\bibfield{author}{\bibinfo{person}{Electronic~Code of Federal~Regulations}.} \bibinfo{year}{2023}\natexlab{}.
\newblock \bibinfo{title}{Regulation M}.
\newblock \bibinfo{howpublished}{Code of Federal Regulations}.
\newblock
\urldef\tempurl%
\url{https://www.ecfr.gov/current/title-17/chapter-II/part-242/subject-group-ECFR3dd95cf4d3f6730}
\showURL{%
\tempurl}
\newblock
\shownote{17 CFR Part 242}.


\bibitem[{Office of Science and Technology Policy}(2022)]%
        {usaibor2020}
\bibfield{author}{\bibinfo{person}{{Office of Science and Technology Policy}}.} \bibinfo{year}{2022}\natexlab{}.
\newblock \bibinfo{title}{Notice and {Explanation}}.
\newblock
\newblock
\urldef\tempurl%
\url{https://www.whitehouse.gov/ostp/ai-bill-of-rights/notice-and-explanation/}
\showURL{%
\tempurl}


\bibitem[{Office of the President of the United States}(2023)]%
        {biden2023executive}
\bibfield{author}{\bibinfo{person}{{Office of the President of the United States}}.} \bibinfo{year}{2023}\natexlab{}.
\newblock \bibinfo{title}{Executive {Order} on the {Safe}, {Secure}, and {Trustworthy} {Development} and {Use} of {Artificial} {Intelligence}}.
\newblock
\newblock
\urldef\tempurl%
\url{https://www.whitehouse.gov/briefing-room/presidential-actions/2023/10/30/executive-order-on-the-safe-secure-and-trustworthy-development-and-use-of-artificial-intelligence/}
\showURL{%
\tempurl}


\bibitem[Ojewale et~al\mbox{.}(2024)]%
        {ojewale2024towards}
\bibfield{author}{\bibinfo{person}{Victor Ojewale}, \bibinfo{person}{Ryan Steed}, \bibinfo{person}{Briana Vecchione}, \bibinfo{person}{Abeba Birhane}, {and} \bibinfo{person}{Inioluwa~Deborah Raji}.} \bibinfo{year}{2024}\natexlab{}.
\newblock \showarticletitle{Towards AI Accountability Infrastructure: Gaps and Opportunities in AI Audit Tooling}.
\newblock \bibinfo{journal}{\emph{arXiv preprint arXiv:2402.17861}} (\bibinfo{year}{2024}).
\newblock


\bibitem[Oliver(1991)]%
        {oliver1991strategic}
\bibfield{author}{\bibinfo{person}{Christine Oliver}.} \bibinfo{year}{1991}\natexlab{}.
\newblock \showarticletitle{Strategic responses to institutional processes}.
\newblock \bibinfo{journal}{\emph{Academy of Management Review}} \bibinfo{volume}{16}, \bibinfo{number}{1} (\bibinfo{date}{Jan.} \bibinfo{year}{1991}), \bibinfo{pages}{145--179}.
\newblock
\showISSN{0363-7425}
\urldef\tempurl%
\url{https://doi.org/10.5465/amr.1991.4279002}
\showDOI{\tempurl}
\newblock
\shownote{Publisher: Academy of Management}.


\bibitem[Oneal(2023)]%
        {oneal2023}
\bibfield{author}{\bibinfo{person}{A.J. Oneal}.} \bibinfo{year}{2023}\natexlab{}.
\newblock \bibinfo{title}{Chat GPT "DAN" (and other "Jailbreaks")}.
\newblock \bibinfo{howpublished}{\url{https://gist.github.com/coolaj86/6f4f7b30129b0251f61fa7baaa881516}}.
\newblock


\bibitem[OpenAI(2023a)]%
        {openai2023gpt}
\bibfield{author}{\bibinfo{person}{OpenAI}.} \bibinfo{year}{2023}\natexlab{a}.
\newblock \bibinfo{title}{GPT-3.5 Turbo fine-tuning and API updates}.
\newblock
\newblock
\urldef\tempurl%
\url{https://openai.com/blog/gpt-3-5-turbo-fine-tuning-and-api-updates}
\showURL{%
\tempurl}


\bibitem[OpenAI(2023b)]%
        {openai2023gpt4}
\bibfield{author}{\bibinfo{person}{OpenAI}.} \bibinfo{year}{2023}\natexlab{b}.
\newblock \showarticletitle{GPT-4 Technical Report}.
\newblock  (\bibinfo{year}{2023}).
\newblock
\showeprint[arxiv]{2303.08774}~[cs.CL]


\bibitem[OpenAI(2023c)]%
        {oai_prep_challenge}
\bibfield{author}{\bibinfo{person}{OpenAI}.} \bibinfo{year}{2023}\natexlab{c}.
\newblock \bibinfo{title}{OpenAI Preparedness Challenge}.
\newblock
\newblock
\urldef\tempurl%
\url{https://openai.com/form/preparedness-challenge}
\showURL{%
\tempurl}


\bibitem[OpenAI(2023d)]%
        {oai_red_teaming_network}
\bibfield{author}{\bibinfo{person}{OpenAI}.} \bibinfo{year}{2023}\natexlab{d}.
\newblock \bibinfo{title}{OpenAI Red Teaming Network}.
\newblock
\newblock
\urldef\tempurl%
\url{https://openai.com/blog/red-teaming-network}
\showURL{%
\tempurl}


\bibitem[Openmined(2023)]%
        {openmined_how_2023}
\bibfield{author}{\bibinfo{person}{Openmined}.} \bibinfo{year}{2023}\natexlab{}.
\newblock \showarticletitle{How to audit an {AI} model owned by someone else (part 1)}.
\newblock \bibinfo{journal}{\emph{OpenMined Blog}} (\bibinfo{date}{June} \bibinfo{year}{2023}).
\newblock
\urldef\tempurl%
\url{https://blog.openmined.org/ai-audit-part-1/}
\showURL{%
\tempurl}


\bibitem[Osada et~al\mbox{.}(2022)]%
        {osada2022latent}
\bibfield{author}{\bibinfo{person}{Genki Osada}, \bibinfo{person}{Budrul Ahsan}, \bibinfo{person}{Revoti~Prasad Bora}, {and} \bibinfo{person}{Takashi Nishide}.} \bibinfo{year}{2022}\natexlab{}.
\newblock \showarticletitle{Latent Space Virtual Adversarial Training for Supervised and Semi-Supervised Learning}.
\newblock \bibinfo{journal}{\emph{IEICE TRANSACTIONS on Information and Systems}} \bibinfo{volume}{105}, \bibinfo{number}{3} (\bibinfo{year}{2022}), \bibinfo{pages}{667--678}.
\newblock


\bibitem[Pan et~al\mbox{.}(2022)]%
        {pan2022improved}
\bibfield{author}{\bibinfo{person}{Lin Pan}, \bibinfo{person}{Chung-Wei Hang}, \bibinfo{person}{Avirup Sil}, {and} \bibinfo{person}{Saloni Potdar}.} \bibinfo{year}{2022}\natexlab{}.
\newblock \showarticletitle{Improved text classification via contrastive adversarial training}. In \bibinfo{booktitle}{\emph{Proceedings of the AAAI Conference on Artificial Intelligence}}, Vol.~\bibinfo{volume}{36}. \bibinfo{pages}{11130--11138}.
\newblock


\bibitem[Papernot et~al\mbox{.}(2016)]%
        {papernot2016technical}
\bibfield{author}{\bibinfo{person}{Nicolas Papernot}, \bibinfo{person}{Fartash Faghri}, \bibinfo{person}{Nicholas Carlini}, \bibinfo{person}{Ian Goodfellow}, \bibinfo{person}{Reuben Feinman}, \bibinfo{person}{Alexey Kurakin}, \bibinfo{person}{Cihang Xie}, \bibinfo{person}{Yash Sharma}, \bibinfo{person}{Tom Brown}, \bibinfo{person}{Aurko Roy}, {et~al\mbox{.}}} \bibinfo{year}{2016}\natexlab{}.
\newblock \showarticletitle{Technical report on the cleverhans v2. 1.0 adversarial examples library}.
\newblock \bibinfo{journal}{\emph{arXiv preprint arXiv:1610.00768}} (\bibinfo{year}{2016}).
\newblock


\bibitem[Park and Lee(2021)]%
        {park2021reliably}
\bibfield{author}{\bibinfo{person}{Geon~Yeong Park} {and} \bibinfo{person}{Sang~Wan Lee}.} \bibinfo{year}{2021}\natexlab{}.
\newblock \showarticletitle{Reliably fast adversarial training via latent adversarial perturbation}. In \bibinfo{booktitle}{\emph{Proceedings of the IEEE/CVF International Conference on Computer Vision}}. \bibinfo{pages}{7758--7767}.
\newblock


\bibitem[Park et~al\mbox{.}(2023b)]%
        {park_generative_2023}
\bibfield{author}{\bibinfo{person}{Joon~Sung Park}, \bibinfo{person}{Joseph~C. O'Brien}, \bibinfo{person}{Carrie~J. Cai}, \bibinfo{person}{Meredith~Ringel Morris}, \bibinfo{person}{Percy Liang}, {and} \bibinfo{person}{Michael~S. Bernstein}.} \bibinfo{year}{2023}\natexlab{b}.
\newblock \showarticletitle{Generative {Agents}: {Interactive} {Simulacra} of {Human} {Behavior}}.
\newblock  (\bibinfo{date}{Aug.} \bibinfo{year}{2023}).
\newblock
\urldef\tempurl%
\url{https://doi.org/10.48550/arXiv.2304.03442}
\showDOI{\tempurl}
\newblock
\shownote{arXiv:2304.03442 [cs]}.


\bibitem[Park et~al\mbox{.}(2023a)]%
        {park2023ai}
\bibfield{author}{\bibinfo{person}{Peter~S. Park}, \bibinfo{person}{Simon Goldstein}, \bibinfo{person}{Aidan O'Gara}, \bibinfo{person}{Michael Chen}, {and} \bibinfo{person}{Dan Hendrycks}.} \bibinfo{year}{2023}\natexlab{a}.
\newblock \showarticletitle{AI Deception: A Survey of Examples, Risks, and Potential Solutions}.
\newblock  (\bibinfo{year}{2023}).
\newblock
\showeprint[arxiv]{2308.14752}~[cs.CY]


\bibitem[PCAOB(2002)]%
        {sarbanesoxley2002}
\bibfield{author}{\bibinfo{person}{PCAOB}.} \bibinfo{year}{2002}\natexlab{}.
\newblock \bibinfo{title}{Sarbanes-Oxley Act of 2002}.
\newblock
\newblock
\urldef\tempurl%
\url{https://pcaobus.org/About/History/Documents/PDFs/Sarbanes_Oxley_Act_of_2002.pdf}
\showURL{%
\tempurl}
\newblock
\shownote{Public Law 107-204, 116 Stat. 745}.


\bibitem[Perez et~al\mbox{.}(2022a)]%
        {perez2022red}
\bibfield{author}{\bibinfo{person}{Ethan Perez}, \bibinfo{person}{Saffron Huang}, \bibinfo{person}{Francis Song}, \bibinfo{person}{Trevor Cai}, \bibinfo{person}{Roman Ring}, \bibinfo{person}{John Aslanides}, \bibinfo{person}{Amelia Glaese}, \bibinfo{person}{Nat McAleese}, {and} \bibinfo{person}{Geoffrey Irving}.} \bibinfo{year}{2022}\natexlab{a}.
\newblock \showarticletitle{Red teaming language models with language models}.
\newblock \bibinfo{journal}{\emph{arXiv preprint arXiv:2202.03286}} (\bibinfo{year}{2022}).
\newblock


\bibitem[Perez et~al\mbox{.}(2022b)]%
        {perez2022discovering}
\bibfield{author}{\bibinfo{person}{Ethan Perez}, \bibinfo{person}{Sam Ringer}, \bibinfo{person}{Kamil{\.e} Luko{\v{s}}i{\=u}t{\.e}}, \bibinfo{person}{Karina Nguyen}, \bibinfo{person}{Edwin Chen}, \bibinfo{person}{Scott Heiner}, \bibinfo{person}{Craig Pettit}, \bibinfo{person}{Catherine Olsson}, \bibinfo{person}{Sandipan Kundu}, \bibinfo{person}{Saurav Kadavath}, {et~al\mbox{.}}} \bibinfo{year}{2022}\natexlab{b}.
\newblock \showarticletitle{Discovering Language Model Behaviors with Model-Written Evaluations}.
\newblock \bibinfo{journal}{\emph{arXiv preprint arXiv:2212.09251}} (\bibinfo{year}{2022}).
\newblock


\bibitem[{Personal Data Protection Commission Singapore}(2020)]%
        {sg2020}
\bibfield{author}{\bibinfo{person}{{Personal Data Protection Commission Singapore}}.} \bibinfo{year}{2020}\natexlab{}.
\newblock \bibinfo{title}{Model {Artificial} {Intelligence} {Governance} {Framework}, {Second} {Edition}}.
\newblock
\newblock
\urldef\tempurl%
\url{https://www.pdpc.gov.sg/-/media/Files/PDPC/PDF-Files/Resource-for-Organisation/AI/SGModelAIGovFramework2.pdf}
\showURL{%
\tempurl}


\bibitem[Phillips et~al\mbox{.}(2021a)]%
        {nistxai2021}
\bibfield{author}{\bibinfo{person}{P~Jonathon Phillips}, \bibinfo{person}{Carina~A Hahn}, \bibinfo{person}{Peter~C Fontana}, \bibinfo{person}{Amy~N Yates}, \bibinfo{person}{Kristen Greene}, \bibinfo{person}{David~A Broniatowski}, {and} \bibinfo{person}{Mark~A Przybocki}.} \bibinfo{year}{2021}\natexlab{a}.
\newblock \bibinfo{booktitle}{\emph{Four principles of explainable artificial intelligence}}.
\newblock \bibinfo{type}{{T}echnical {R}eport} NIST IR 8312. \bibinfo{institution}{National Institute of Standards and Technology (U.S.)}, \bibinfo{address}{Gaithersburg, MD}. \bibinfo{pages}{NIST IR 8312} pages.
\newblock
\urldef\tempurl%
\url{https://doi.org/10.6028/NIST.IR.8312}
\showDOI{\tempurl}


\bibitem[Phillips et~al\mbox{.}(2021b)]%
        {phillips_four_2021}
\bibfield{author}{\bibinfo{person}{P.~Jonathon Phillips}, \bibinfo{person}{Carina~A. Hahn}, \bibinfo{person}{Peter~C. Fontana}, \bibinfo{person}{Amy~N. Yates}, \bibinfo{person}{Kristen Greene}, \bibinfo{person}{David~A. Broniatowski}, {and} \bibinfo{person}{Mark~A. Przybocki}.} \bibinfo{year}{2021}\natexlab{b}.
\newblock \bibinfo{booktitle}{\emph{Four {Principles} of {Explainable} {Artificial} {Intelligence}}}.
\newblock \bibinfo{type}{Interagency or {Internal} {Report}} 8312. \bibinfo{institution}{National Institute for Standards and Technology}.
\newblock


\bibitem[Ploug and Holm(2021)]%
        {ploug2021right}
\bibfield{author}{\bibinfo{person}{Thomas Ploug} {and} \bibinfo{person}{S{\o}ren Holm}.} \bibinfo{year}{2021}\natexlab{}.
\newblock \showarticletitle{Right to Contest AI Diagnostics: Defining Transparency and Explainability Requirements from a Patient’s Perspective}.
\newblock In \bibinfo{booktitle}{\emph{Artificial Intelligence in Medicine}}. \bibinfo{publisher}{Springer}, \bibinfo{pages}{1--12}.
\newblock


\bibitem[Pozzobon et~al\mbox{.}(2023)]%
        {pozzobon2023challenges}
\bibfield{author}{\bibinfo{person}{Luiza Pozzobon}, \bibinfo{person}{Beyza Ermis}, \bibinfo{person}{Patrick Lewis}, {and} \bibinfo{person}{Sara Hooker}.} \bibinfo{year}{2023}\natexlab{}.
\newblock \showarticletitle{On the Challenges of Using Black-Box APIs for Toxicity Evaluation in Research}.
\newblock \bibinfo{journal}{\emph{arXiv preprint arXiv:2304.12397}} (\bibinfo{year}{2023}).
\newblock


\bibitem[Prasad et~al\mbox{.}(2022)]%
        {prasad2022grips}
\bibfield{author}{\bibinfo{person}{Archiki Prasad}, \bibinfo{person}{Peter Hase}, \bibinfo{person}{Xiang Zhou}, {and} \bibinfo{person}{Mohit Bansal}.} \bibinfo{year}{2022}\natexlab{}.
\newblock \showarticletitle{Grips: Gradient-free, edit-based instruction search for prompting large language models}.
\newblock \bibinfo{journal}{\emph{arXiv preprint arXiv:2203.07281}} (\bibinfo{year}{2022}).
\newblock


\bibitem[Qi et~al\mbox{.}(2023a)]%
        {qi2023visual}
\bibfield{author}{\bibinfo{person}{Xiangyu Qi}, \bibinfo{person}{Kaixuan Huang}, \bibinfo{person}{Ashwinee Panda}, \bibinfo{person}{Mengdi Wang}, {and} \bibinfo{person}{Prateek Mittal}.} \bibinfo{year}{2023}\natexlab{a}.
\newblock \showarticletitle{Visual Adversarial Examples Jailbreak Large Language Models}.
\newblock \bibinfo{journal}{\emph{arXiv preprint arXiv:2306.13213}} (\bibinfo{year}{2023}).
\newblock


\bibitem[Qi et~al\mbox{.}(2023b)]%
        {qi2023fine}
\bibfield{author}{\bibinfo{person}{Xiangyu Qi}, \bibinfo{person}{Yi Zeng}, \bibinfo{person}{Tinghao Xie}, \bibinfo{person}{Pin-Yu Chen}, \bibinfo{person}{Ruoxi Jia}, \bibinfo{person}{Prateek Mittal}, {and} \bibinfo{person}{Peter Henderson}.} \bibinfo{year}{2023}\natexlab{b}.
\newblock \showarticletitle{Fine-tuning Aligned Language Models Compromises Safety, Even When Users Do Not Intend To!}
\newblock \bibinfo{journal}{\emph{arXiv preprint arXiv:2310.03693}} (\bibinfo{year}{2023}).
\newblock


\bibitem[Qian et~al\mbox{.}(2021)]%
        {qian2021towards}
\bibfield{author}{\bibinfo{person}{Yaguan Qian}, \bibinfo{person}{Qiqi Shao}, \bibinfo{person}{Tengteng Yao}, \bibinfo{person}{Bin Wang}, \bibinfo{person}{Shouling Ji}, \bibinfo{person}{Shaoning Zeng}, \bibinfo{person}{Zhaoquan Gu}, {and} \bibinfo{person}{Wassim Swaileh}.} \bibinfo{year}{2021}\natexlab{}.
\newblock \showarticletitle{Towards Speeding up Adversarial Training in Latent Spaces}.
\newblock \bibinfo{journal}{\emph{arXiv preprint arXiv:2102.00662}} (\bibinfo{year}{2021}).
\newblock


\bibitem[Qu et~al\mbox{.}(2023)]%
        {qu2023unsafe}
\bibfield{author}{\bibinfo{person}{Yiting Qu}, \bibinfo{person}{Xinyue Shen}, \bibinfo{person}{Xinlei He}, \bibinfo{person}{Michael Backes}, \bibinfo{person}{Savvas Zannettou}, {and} \bibinfo{person}{Yang Zhang}.} \bibinfo{year}{2023}\natexlab{}.
\newblock \showarticletitle{Unsafe diffusion: On the generation of unsafe images and hateful memes from text-to-image models}.
\newblock \bibinfo{journal}{\emph{arXiv preprint arXiv:2305.13873}} (\bibinfo{year}{2023}).
\newblock


\bibitem[Raghavan et~al\mbox{.}(2020)]%
        {raghavan_mitigating_2020}
\bibfield{author}{\bibinfo{person}{Manish Raghavan}, \bibinfo{person}{Solon Barocas}, \bibinfo{person}{Jon Kleinberg}, {and} \bibinfo{person}{Karen Levy}.} \bibinfo{year}{2020}\natexlab{}.
\newblock \showarticletitle{Mitigating bias in algorithmic hiring: evaluating claims and practices}. In \bibinfo{booktitle}{\emph{Proceedings of the 2020 {Conference} on {Fairness}, {Accountability}, and {Transparency}}} \emph{(\bibinfo{series}{{FAT}* '20})}. \bibinfo{publisher}{Association for Computing Machinery}, \bibinfo{address}{New York, NY, USA}, \bibinfo{pages}{469--481}.
\newblock
\showISBNx{978-1-4503-6936-7}
\urldef\tempurl%
\url{https://doi.org/10.1145/3351095.3372828}
\showDOI{\tempurl}


\bibitem[Raghavan and Kim(2023)]%
        {raghavan_limitations_2023}
\bibfield{author}{\bibinfo{person}{Manish Raghavan} {and} \bibinfo{person}{Pauline Kim}.} \bibinfo{year}{2023}\natexlab{}.
\newblock \showarticletitle{Limitations of the “{Four}-{Fifths} {Rule}” and {Statistical} {Parity} {Tests} for {Measuring} {Fairness}}.
\newblock
\urldef\tempurl%
\url{https://openreview.net/forum?id=M2aNjwX4Ec&referrer=%5Bthe%20profile%20of%20Manish%20Raghavan%5D(%2Fprofile%3Fid%3D~Manish_Raghavan1)}
\showURL{%
\tempurl}


\bibitem[Raji(2022)]%
        {raji2022anatomy}
\bibfield{author}{\bibinfo{person}{Inioluwa~Deborah Raji}.} \bibinfo{year}{2022}\natexlab{}.
\newblock \showarticletitle{The Anatomy of AI Audits: Form, Process, and Consequences}.
\newblock  (\bibinfo{year}{2022}).
\newblock


\bibitem[Raji and Buolamwini(2019)]%
        {raji2019actionable}
\bibfield{author}{\bibinfo{person}{Inioluwa~Deborah Raji} {and} \bibinfo{person}{Joy Buolamwini}.} \bibinfo{year}{2019}\natexlab{}.
\newblock \showarticletitle{Actionable auditing: Investigating the impact of publicly naming biased performance results of commercial ai products}. In \bibinfo{booktitle}{\emph{Proceedings of the 2019 AAAI/ACM Conference on AI, Ethics, and Society}}. \bibinfo{pages}{429--435}.
\newblock


\bibitem[Raji and Buolamwini(2022)]%
        {raji2022actionable}
\bibfield{author}{\bibinfo{person}{Inioluwa~Deborah Raji} {and} \bibinfo{person}{Joy Buolamwini}.} \bibinfo{year}{2022}\natexlab{}.
\newblock \showarticletitle{Actionable Auditing Revisited: Investigating the Impact of Publicly Naming Biased Performance Results of Commercial AI Products}.
\newblock \bibinfo{journal}{\emph{Commun. ACM}} \bibinfo{volume}{66}, \bibinfo{number}{1} (\bibinfo{year}{2022}), \bibinfo{pages}{101--108}.
\newblock


\bibitem[Raji et~al\mbox{.}(2020a)]%
        {raji_saving_2020}
\bibfield{author}{\bibinfo{person}{Inioluwa~Deborah Raji}, \bibinfo{person}{Timnit Gebru}, \bibinfo{person}{Margaret Mitchell}, \bibinfo{person}{Joy Buolamwini}, \bibinfo{person}{Joonseok Lee}, {and} \bibinfo{person}{Emily Denton}.} \bibinfo{year}{2020}\natexlab{a}.
\newblock \showarticletitle{Saving {Face}: {Investigating} the {Ethical} {Concerns} of {Facial} {Recognition} {Auditing}}. In \bibinfo{booktitle}{\emph{Proceedings of the {AAAI}/{ACM} {Conference} on {AI}, {Ethics}, and {Society}}} \emph{(\bibinfo{series}{{AIES} '20})}. \bibinfo{publisher}{Association for Computing Machinery}, \bibinfo{address}{New York, NY, USA}, \bibinfo{pages}{145--151}.
\newblock
\showISBNx{978-1-4503-7110-0}
\urldef\tempurl%
\url{https://doi.org/10.1145/3375627.3375820}
\showDOI{\tempurl}


\bibitem[Raji et~al\mbox{.}(2020b)]%
        {raji2020closing}
\bibfield{author}{\bibinfo{person}{Inioluwa~Deborah Raji}, \bibinfo{person}{Andrew Smart}, \bibinfo{person}{Rebecca~N. White}, \bibinfo{person}{Margaret Mitchell}, \bibinfo{person}{Timnit Gebru}, \bibinfo{person}{Ben Hutchinson}, \bibinfo{person}{Jamila Smith-Loud}, \bibinfo{person}{Daniel Theron}, {and} \bibinfo{person}{Parker Barnes}.} \bibinfo{year}{2020}\natexlab{b}.
\newblock \showarticletitle{Closing the {AI} accountability gap: defining an end-to-end framework for internal algorithmic auditing}. In \bibinfo{booktitle}{\emph{Proceedings of the 2020 {Conference} on {Fairness}, {Accountability}, and {Transparency}}}. \bibinfo{publisher}{ACM}, \bibinfo{address}{Barcelona Spain}, \bibinfo{pages}{33--44}.
\newblock
\showISBNx{978-1-4503-6936-7}
\urldef\tempurl%
\url{https://doi.org/10.1145/3351095.3372873}
\showDOI{\tempurl}


\bibitem[Raji et~al\mbox{.}(2022)]%
        {raji2022outsider}
\bibfield{author}{\bibinfo{person}{Inioluwa~Deborah Raji}, \bibinfo{person}{Peggy Xu}, \bibinfo{person}{Colleen Honigsberg}, {and} \bibinfo{person}{Daniel Ho}.} \bibinfo{year}{2022}\natexlab{}.
\newblock \showarticletitle{Outsider oversight: Designing a third party audit ecosystem for ai governance}. In \bibinfo{booktitle}{\emph{Proceedings of the 2022 AAAI/ACM Conference on AI, Ethics, and Society}}. \bibinfo{pages}{557--571}.
\newblock


\bibitem[Rando et~al\mbox{.}(2022)]%
        {rando2022red}
\bibfield{author}{\bibinfo{person}{Javier Rando}, \bibinfo{person}{Daniel Paleka}, \bibinfo{person}{David Lindner}, \bibinfo{person}{Lennart Heim}, {and} \bibinfo{person}{Florian Tram{\`e}r}.} \bibinfo{year}{2022}\natexlab{}.
\newblock \showarticletitle{Red-teaming the stable diffusion safety filter}.
\newblock \bibinfo{journal}{\emph{arXiv preprint arXiv:2210.04610}} (\bibinfo{year}{2022}).
\newblock


\bibitem[Rando and Tramèr(2023)]%
        {rando2023universal}
\bibfield{author}{\bibinfo{person}{Javier Rando} {and} \bibinfo{person}{Florian Tramèr}.} \bibinfo{year}{2023}\natexlab{}.
\newblock \showarticletitle{Universal Jailbreak Backdoors from Poisoned Human Feedback}.
\newblock  (\bibinfo{year}{2023}).
\newblock
\showeprint[arxiv]{2311.14455}~[cs.AI]


\bibitem[Rao et~al\mbox{.}(2023)]%
        {rao2023tricking}
\bibfield{author}{\bibinfo{person}{Abhinav Rao}, \bibinfo{person}{Sachin Vashistha}, \bibinfo{person}{Atharva Naik}, \bibinfo{person}{Somak Aditya}, {and} \bibinfo{person}{Monojit Choudhury}.} \bibinfo{year}{2023}\natexlab{}.
\newblock \showarticletitle{Tricking LLMs into Disobedience: Understanding, Analyzing, and Preventing Jailbreaks}.
\newblock  (\bibinfo{year}{2023}).
\newblock
\showeprint[arxiv]{2305.14965}~[cs.CL]


\bibitem[R{\"a}uker et~al\mbox{.}(2023)]%
        {rauker2023toward}
\bibfield{author}{\bibinfo{person}{Tilman R{\"a}uker}, \bibinfo{person}{Anson Ho}, \bibinfo{person}{Stephen Casper}, {and} \bibinfo{person}{Dylan Hadfield-Menell}.} \bibinfo{year}{2023}\natexlab{}.
\newblock \showarticletitle{Toward transparent ai: A survey on interpreting the inner structures of deep neural networks}. In \bibinfo{booktitle}{\emph{2023 IEEE Conference on Secure and Trustworthy Machine Learning (SaTML)}}. IEEE, \bibinfo{pages}{464--483}.
\newblock


\bibitem[Ravichander et~al\mbox{.}(2020)]%
        {ravichander2020probing}
\bibfield{author}{\bibinfo{person}{Abhilasha Ravichander}, \bibinfo{person}{Yonatan Belinkov}, {and} \bibinfo{person}{Eduard Hovy}.} \bibinfo{year}{2020}\natexlab{}.
\newblock \showarticletitle{Probing the probing paradigm: Does probing accuracy entail task relevance?}
\newblock \bibinfo{journal}{\emph{arXiv preprint arXiv:2005.00719}} (\bibinfo{year}{2020}).
\newblock


\bibitem[Ren et~al\mbox{.}(2019)]%
        {ren2019generating}
\bibfield{author}{\bibinfo{person}{Shuhuai Ren}, \bibinfo{person}{Yihe Deng}, \bibinfo{person}{Kun He}, {and} \bibinfo{person}{Wanxiang Che}.} \bibinfo{year}{2019}\natexlab{}.
\newblock \showarticletitle{Generating natural language adversarial examples through probability weighted word saliency}. In \bibinfo{booktitle}{\emph{Proceedings of the 57th annual meeting of the association for computational linguistics}}. \bibinfo{pages}{1085--1097}.
\newblock


\bibitem[Ribeiro et~al\mbox{.}(2016)]%
        {ribeiro2016model}
\bibfield{author}{\bibinfo{person}{Marco~Tulio Ribeiro}, \bibinfo{person}{Sameer Singh}, {and} \bibinfo{person}{Carlos Guestrin}.} \bibinfo{year}{2016}\natexlab{}.
\newblock \showarticletitle{Model-agnostic interpretability of machine learning}.
\newblock \bibinfo{journal}{\emph{arXiv preprint arXiv:1606.05386}} (\bibinfo{year}{2016}).
\newblock


\bibitem[Robertson et~al\mbox{.}(2018)]%
        {robertson_auditing_2018}
\bibfield{author}{\bibinfo{person}{Ronald~E. Robertson}, \bibinfo{person}{David Lazer}, {and} \bibinfo{person}{Christo Wilson}.} \bibinfo{year}{2018}\natexlab{}.
\newblock \showarticletitle{Auditing the {Personalization} and {Composition} of {Politically}-{Related} {Search} {Engine} {Results} {Pages}}. In \bibinfo{booktitle}{\emph{Proceedings of the 2018 {World} {Wide} {Web} {Conference}}} \emph{(\bibinfo{series}{{WWW} '18})}. \bibinfo{publisher}{International World Wide Web Conferences Steering Committee}, \bibinfo{address}{Republic and Canton of Geneva, CHE}, \bibinfo{pages}{955--965}.
\newblock
\showISBNx{978-1-4503-5639-8}
\urldef\tempurl%
\url{https://doi.org/10.1145/3178876.3186143}
\showDOI{\tempurl}


\bibitem[Rodriguez~Maffioli(2023)]%
        {rodriguez2023copyright}
\bibfield{author}{\bibinfo{person}{Daniel Rodriguez~Maffioli}.} \bibinfo{year}{2023}\natexlab{}.
\newblock \showarticletitle{Copyright in Generative AI training: Balancing Fair Use through Standardization and Transparency}.
\newblock \bibinfo{journal}{\emph{Available at SSRN 4579322}} (\bibinfo{year}{2023}).
\newblock


\bibitem[Roth(2023)]%
        {roth_new_2023}
\bibfield{author}{\bibinfo{person}{Emma Roth}.} \bibinfo{year}{2023}\natexlab{}.
\newblock \showarticletitle{The {New} {York} {Times} is suing {OpenAI} and {Microsoft} for copyright infringement}.
\newblock \bibinfo{journal}{\emph{The Verge}} (\bibinfo{date}{Dec.} \bibinfo{year}{2023}).
\newblock
\urldef\tempurl%
\url{https://www.theverge.com/2023/12/27/24016212/new-york-times-openai-microsoft-lawsuit-copyright-infringement}
\showURL{%
\tempurl}


\bibitem[Roth et~al\mbox{.}(2021)]%
        {Roth2021TokenModificationAA}
\bibfield{author}{\bibinfo{person}{Tom Roth}, \bibinfo{person}{Yansong Gao}, \bibinfo{person}{Alsharif Abuadbba}, \bibinfo{person}{Surya Nepal}, {and} \bibinfo{person}{Wei Liu}.} \bibinfo{year}{2021}\natexlab{}.
\newblock \showarticletitle{Token-Modification Adversarial Attacks for Natural Language Processing: A Survey}.
\newblock \bibinfo{journal}{\emph{ArXiv}}  \bibinfo{volume}{abs/2103.00676} (\bibinfo{year}{2021}).
\newblock
\urldef\tempurl%
\url{https://api.semanticscholar.org/CorpusID:232075640}
\showURL{%
\tempurl}


\bibitem[Rudin(2018)]%
        {rudin2018please}
\bibfield{author}{\bibinfo{person}{Cynthia Rudin}.} \bibinfo{year}{2018}\natexlab{}.
\newblock \showarticletitle{Please stop explaining black box models for high stakes decisions}.
\newblock \bibinfo{journal}{\emph{Stat}}  \bibinfo{volume}{1050} (\bibinfo{year}{2018}), \bibinfo{pages}{26}.
\newblock


\bibitem[Sae-Lim and Phoomvuthisarn(2022)]%
        {sae2022weighted}
\bibfield{author}{\bibinfo{person}{Teerapong Sae-Lim} {and} \bibinfo{person}{Suronapee Phoomvuthisarn}.} \bibinfo{year}{2022}\natexlab{}.
\newblock \showarticletitle{Weighted Token-Level Virtual Adversarial Training in Text Classification}. In \bibinfo{booktitle}{\emph{2022 3rd International Conference on Pattern Recognition and Machine Learning (PRML)}}. IEEE, \bibinfo{pages}{117--123}.
\newblock


\bibitem[Sandbrink(2023)]%
        {sandbrink2023artificial}
\bibfield{author}{\bibinfo{person}{Jonas~B Sandbrink}.} \bibinfo{year}{2023}\natexlab{}.
\newblock \showarticletitle{Artificial intelligence and biological misuse: Differentiating risks of language models and biological design tools}.
\newblock \bibinfo{journal}{\emph{arXiv preprint arXiv:2306.13952}} (\bibinfo{year}{2023}).
\newblock


\bibitem[Sankaranarayanan et~al\mbox{.}(2018)]%
        {sankaranarayanan2018regularizing}
\bibfield{author}{\bibinfo{person}{Swami Sankaranarayanan}, \bibinfo{person}{Arpit Jain}, \bibinfo{person}{Rama Chellappa}, {and} \bibinfo{person}{Ser~Nam Lim}.} \bibinfo{year}{2018}\natexlab{}.
\newblock \showarticletitle{Regularizing deep networks using efficient layerwise adversarial training}. In \bibinfo{booktitle}{\emph{Proceedings of the AAAI Conference on Artificial Intelligence}}, Vol.~\bibinfo{volume}{32}.
\newblock


\bibitem[Santurkar et~al\mbox{.}(2023)]%
        {santurkar2023whose}
\bibfield{author}{\bibinfo{person}{Shibani Santurkar}, \bibinfo{person}{Esin Durmus}, \bibinfo{person}{Faisal Ladhak}, \bibinfo{person}{Cinoo Lee}, \bibinfo{person}{Percy Liang}, {and} \bibinfo{person}{Tatsunori Hashimoto}.} \bibinfo{year}{2023}\natexlab{}.
\newblock \showarticletitle{Whose opinions do language models reflect?}
\newblock \bibinfo{journal}{\emph{arXiv preprint arXiv:2303.17548}} (\bibinfo{year}{2023}).
\newblock


\bibitem[Sarlin et~al\mbox{.}(2020)]%
        {sarlin2020superglue}
\bibfield{author}{\bibinfo{person}{Paul-Edouard Sarlin}, \bibinfo{person}{Daniel DeTone}, \bibinfo{person}{Tomasz Malisiewicz}, {and} \bibinfo{person}{Andrew Rabinovich}.} \bibinfo{year}{2020}\natexlab{}.
\newblock \showarticletitle{Superglue: Learning feature matching with graph neural networks}. In \bibinfo{booktitle}{\emph{Proceedings of the IEEE/CVF conference on computer vision and pattern recognition}}. \bibinfo{pages}{4938--4947}.
\newblock


\bibitem[Schaeffer et~al\mbox{.}(2023)]%
        {schaeffer2023emergent}
\bibfield{author}{\bibinfo{person}{Rylan Schaeffer}, \bibinfo{person}{Brando Miranda}, {and} \bibinfo{person}{Sanmi Koyejo}.} \bibinfo{year}{2023}\natexlab{}.
\newblock \showarticletitle{Are Emergent Abilities of Large Language Models a Mirage?}
\newblock  (\bibinfo{year}{2023}).
\newblock
\showeprint[arxiv]{2304.15004}~[cs.AI]


\bibitem[Scheurer et~al\mbox{.}(2023)]%
        {scheurer2023technical}
\bibfield{author}{\bibinfo{person}{J{\'e}r{\'e}my Scheurer}, \bibinfo{person}{Mikita Balesni}, {and} \bibinfo{person}{Marius Hobbhahn}.} \bibinfo{year}{2023}\natexlab{}.
\newblock \showarticletitle{Technical Report: Large Language Models can Strategically Deceive their Users when Put Under Pressure}.
\newblock \bibinfo{journal}{\emph{arXiv preprint arXiv:2311.07590}} (\bibinfo{year}{2023}).
\newblock


\bibitem[Schuett(2022)]%
        {schuett2022three}
\bibfield{author}{\bibinfo{person}{Jonas Schuett}.} \bibinfo{year}{2022}\natexlab{}.
\newblock \showarticletitle{Three lines of defense against risks from AI}.
\newblock \bibinfo{journal}{\emph{arXiv preprint arXiv:2212.08364}} (\bibinfo{year}{2022}).
\newblock


\bibitem[Schuett(2023)]%
        {schuett_agi_2023}
\bibfield{author}{\bibinfo{person}{Jonas Schuett}.} \bibinfo{year}{2023}\natexlab{}.
\newblock \showarticletitle{{AGI} labs need an internal audit function}.
\newblock  (\bibinfo{date}{May} \bibinfo{year}{2023}).
\newblock
\urldef\tempurl%
\url{https://arxiv.org/abs/2305.17038v1}
\showURL{%
\tempurl}


\bibitem[Schuett et~al\mbox{.}(2023)]%
        {schuett2023towards}
\bibfield{author}{\bibinfo{person}{Jonas Schuett}, \bibinfo{person}{Noemi Dreksler}, \bibinfo{person}{Markus Anderljung}, \bibinfo{person}{David McCaffary}, \bibinfo{person}{Lennart Heim}, \bibinfo{person}{Emma Bluemke}, {and} \bibinfo{person}{Ben Garfinkel}.} \bibinfo{year}{2023}\natexlab{}.
\newblock \showarticletitle{Towards best practices in AGI safety and governance: A survey of expert opinion}.
\newblock \bibinfo{journal}{\emph{arXiv preprint arXiv:2305.07153}} (\bibinfo{year}{2023}).
\newblock


\bibitem[Schwinn et~al\mbox{.}(2023)]%
        {schwinn2023adversarial}
\bibfield{author}{\bibinfo{person}{Leo Schwinn}, \bibinfo{person}{David Dobre}, \bibinfo{person}{Stephan Günnemann}, {and} \bibinfo{person}{Gauthier Gidel}.} \bibinfo{year}{2023}\natexlab{}.
\newblock \showarticletitle{Adversarial Attacks and Defenses in Large Language Models: Old and New Threats}.
\newblock  (\bibinfo{year}{2023}).
\newblock
\showeprint[arxiv]{2310.19737}~[cs.AI]


\bibitem[Seger et~al\mbox{.}(2023)]%
        {seger2023open}
\bibfield{author}{\bibinfo{person}{Elizabeth Seger}, \bibinfo{person}{Noemi Dreksler}, \bibinfo{person}{Richard Moulange}, \bibinfo{person}{Emily Dardaman}, \bibinfo{person}{Jonas Schuett}, \bibinfo{person}{K Wei}, \bibinfo{person}{Christoph Winter}, \bibinfo{person}{Mackenzie Arnold}, \bibinfo{person}{Se{\'a}n {\'O}~h{\'E}igeartaigh}, \bibinfo{person}{Anton Korinek}, {et~al\mbox{.}}} \bibinfo{year}{2023}\natexlab{}.
\newblock \showarticletitle{Open-Sourcing Highly Capable Foundation Models: An Evaluation of Risks, Benefits, and Alternative Methods for Pursuing Open-Source Objectives}.
\newblock  (\bibinfo{year}{2023}).
\newblock


\bibitem[Shah et~al\mbox{.}(2023)]%
        {shah2023scalable}
\bibfield{author}{\bibinfo{person}{Rusheb Shah}, \bibinfo{person}{Quentin Feuillade-Montixi}, \bibinfo{person}{Soroush Pour}, \bibinfo{person}{Arush Tagade}, \bibinfo{person}{Stephen Casper}, {and} \bibinfo{person}{Javier Rando}.} \bibinfo{year}{2023}\natexlab{}.
\newblock \showarticletitle{Scalable and Transferable Black-Box Jailbreaks for Language Models via Persona Modulation}.
\newblock  (\bibinfo{year}{2023}).
\newblock
\showeprint[arxiv]{2311.03348}~[cs.CL]


\bibitem[Shahbazi et~al\mbox{.}(2023)]%
        {shahbazi2023representation}
\bibfield{author}{\bibinfo{person}{Nima Shahbazi}, \bibinfo{person}{Yin Lin}, \bibinfo{person}{Abolfazl Asudeh}, {and} \bibinfo{person}{HV Jagadish}.} \bibinfo{year}{2023}\natexlab{}.
\newblock \showarticletitle{Representation Bias in Data: A Survey on Identification and Resolution Techniques}.
\newblock \bibinfo{journal}{\emph{Comput. Surveys}} (\bibinfo{year}{2023}).
\newblock


\bibitem[Sharkey et~al\mbox{.}(2024)]%
        {sharkey2024causal}
\bibfield{author}{\bibinfo{person}{Lee Sharkey}, \bibinfo{person}{Cl{\'\i}odhna~N{\'\i} Ghuidhir}, \bibinfo{person}{Dan Braun}, \bibinfo{person}{J{\'e}r{\'e}my Scheurer}, \bibinfo{person}{Mikita Balesni}, \bibinfo{person}{Lucius Bushnaq}, \bibinfo{person}{Charlotte Stix}, {and} \bibinfo{person}{Marius Hobbhahn}.} \bibinfo{year}{2024}\natexlab{}.
\newblock \showarticletitle{A Causal Framework for AI Regulation and Auditing}.
\newblock  (\bibinfo{year}{2024}).
\newblock


\bibitem[Sharma et~al\mbox{.}(2023)]%
        {sharma2023understanding}
\bibfield{author}{\bibinfo{person}{Mrinank Sharma}, \bibinfo{person}{Meg Tong}, \bibinfo{person}{Tomasz Korbak}, \bibinfo{person}{David Duvenaud}, \bibinfo{person}{Amanda Askell}, \bibinfo{person}{Samuel~R. Bowman}, \bibinfo{person}{Newton Cheng}, \bibinfo{person}{Esin Durmus}, \bibinfo{person}{Zac Hatfield-Dodds}, \bibinfo{person}{Scott~R. Johnston}, \bibinfo{person}{Shauna Kravec}, \bibinfo{person}{Timothy Maxwell}, \bibinfo{person}{Sam McCandlish}, \bibinfo{person}{Kamal Ndousse}, \bibinfo{person}{Oliver Rausch}, \bibinfo{person}{Nicholas Schiefer}, \bibinfo{person}{Da Yan}, \bibinfo{person}{Miranda Zhang}, {and} \bibinfo{person}{Ethan Perez}.} \bibinfo{year}{2023}\natexlab{}.
\newblock \showarticletitle{Towards Understanding Sycophancy in Language Models}.
\newblock  (\bibinfo{year}{2023}).
\newblock
\showeprint[arxiv]{2310.13548}~[cs.CL]


\bibitem[Shayegani et~al\mbox{.}(2023)]%
        {shayegani2023survey}
\bibfield{author}{\bibinfo{person}{Erfan Shayegani}, \bibinfo{person}{Md~Abdullah~Al Mamun}, \bibinfo{person}{Yu Fu}, \bibinfo{person}{Pedram Zaree}, \bibinfo{person}{Yue Dong}, {and} \bibinfo{person}{Nael Abu-Ghazaleh}.} \bibinfo{year}{2023}\natexlab{}.
\newblock \showarticletitle{Survey of Vulnerabilities in Large Language Models Revealed by Adversarial Attacks}.
\newblock \bibinfo{journal}{\emph{arXiv preprint arXiv:2310.10844}} (\bibinfo{year}{2023}).
\newblock


\bibitem[Shen et~al\mbox{.}(2023)]%
        {shen2023anything}
\bibfield{author}{\bibinfo{person}{Xinyue Shen}, \bibinfo{person}{Zeyuan Chen}, \bibinfo{person}{Michael Backes}, \bibinfo{person}{Yun Shen}, {and} \bibinfo{person}{Yang Zhang}.} \bibinfo{year}{2023}\natexlab{}.
\newblock \showarticletitle{" Do Anything Now": Characterizing and Evaluating In-The-Wild Jailbreak Prompts on Large Language Models}.
\newblock \bibinfo{journal}{\emph{arXiv preprint arXiv:2308.03825}} (\bibinfo{year}{2023}).
\newblock


\bibitem[Shevlane(2022)]%
        {shevlane2022structured}
\bibfield{author}{\bibinfo{person}{Toby Shevlane}.} \bibinfo{year}{2022}\natexlab{}.
\newblock \showarticletitle{Structured access: an emerging paradigm for safe AI deployment}.
\newblock  (\bibinfo{year}{2022}).
\newblock
\showeprint[arxiv]{2201.05159}~[cs.AI]


\bibitem[Shevlane et~al\mbox{.}(2023)]%
        {shevlane2023model}
\bibfield{author}{\bibinfo{person}{Toby Shevlane}, \bibinfo{person}{Sebastian Farquhar}, \bibinfo{person}{Ben Garfinkel}, \bibinfo{person}{Mary Phuong}, \bibinfo{person}{Jess Whittlestone}, \bibinfo{person}{Jade Leung}, \bibinfo{person}{Daniel Kokotajlo}, \bibinfo{person}{Nahema Marchal}, \bibinfo{person}{Markus Anderljung}, \bibinfo{person}{Noam Kolt}, {et~al\mbox{.}}} \bibinfo{year}{2023}\natexlab{}.
\newblock \showarticletitle{Model evaluation for extreme risks}.
\newblock \bibinfo{journal}{\emph{arXiv preprint arXiv:2305.15324}} (\bibinfo{year}{2023}).
\newblock


\bibitem[Shi et~al\mbox{.}(2023)]%
        {shi2023detecting}
\bibfield{author}{\bibinfo{person}{Weijia Shi}, \bibinfo{person}{Anirudh Ajith}, \bibinfo{person}{Mengzhou Xia}, \bibinfo{person}{Yangsibo Huang}, \bibinfo{person}{Daogao Liu}, \bibinfo{person}{Terra Blevins}, \bibinfo{person}{Danqi Chen}, {and} \bibinfo{person}{Luke Zettlemoyer}.} \bibinfo{year}{2023}\natexlab{}.
\newblock \showarticletitle{Detecting pretraining data from large language models}.
\newblock \bibinfo{journal}{\emph{arXiv preprint arXiv:2310.16789}} (\bibinfo{year}{2023}).
\newblock


\bibitem[Shi et~al\mbox{.}(2022)]%
        {shi2022toward}
\bibfield{author}{\bibinfo{person}{Weijia Shi}, \bibinfo{person}{Xiaochuang Han}, \bibinfo{person}{Hila Gonen}, \bibinfo{person}{Ari Holtzman}, \bibinfo{person}{Yulia Tsvetkov}, {and} \bibinfo{person}{Luke Zettlemoyer}.} \bibinfo{year}{2022}\natexlab{}.
\newblock \showarticletitle{Toward Human Readable Prompt Tuning: Kubrick's The Shining is a good movie, and a good prompt too?}
\newblock \bibinfo{journal}{\emph{arXiv preprint arXiv:2212.10539}} (\bibinfo{year}{2022}).
\newblock


\bibitem[Shin et~al\mbox{.}(2020)]%
        {shin2020autoprompt}
\bibfield{author}{\bibinfo{person}{Taylor Shin}, \bibinfo{person}{Yasaman Razeghi}, \bibinfo{person}{Robert~L Logan~IV}, \bibinfo{person}{Eric Wallace}, {and} \bibinfo{person}{Sameer Singh}.} \bibinfo{year}{2020}\natexlab{}.
\newblock \showarticletitle{Autoprompt: Eliciting knowledge from language models with automatically generated prompts}.
\newblock \bibinfo{journal}{\emph{arXiv preprint arXiv:2010.15980}} (\bibinfo{year}{2020}).
\newblock


\bibitem[Shur-Ofry(2023)]%
        {shur2023multiplicity}
\bibfield{author}{\bibinfo{person}{Michal Shur-Ofry}.} \bibinfo{year}{2023}\natexlab{}.
\newblock \showarticletitle{Multiplicity as an AI Governance Principle}.
\newblock \bibinfo{journal}{\emph{Available at SSRN 4444354}} (\bibinfo{year}{2023}).
\newblock


\bibitem[Singhal et~al\mbox{.}(2022)]%
        {singhal2022large}
\bibfield{author}{\bibinfo{person}{Karan Singhal}, \bibinfo{person}{Shekoofeh Azizi}, \bibinfo{person}{Tao Tu}, \bibinfo{person}{S~Sara Mahdavi}, \bibinfo{person}{Jason Wei}, \bibinfo{person}{Hyung~Won Chung}, \bibinfo{person}{Nathan Scales}, \bibinfo{person}{Ajay Tanwani}, \bibinfo{person}{Heather Cole-Lewis}, \bibinfo{person}{Stephen Pfohl}, {et~al\mbox{.}}} \bibinfo{year}{2022}\natexlab{}.
\newblock \showarticletitle{Large language models encode clinical knowledge}.
\newblock \bibinfo{journal}{\emph{arXiv preprint arXiv:2212.13138}} (\bibinfo{year}{2022}).
\newblock


\bibitem[Slack et~al\mbox{.}(2020)]%
        {slack_fooling_2020}
\bibfield{author}{\bibinfo{person}{Dylan Slack}, \bibinfo{person}{Sophie Hilgard}, \bibinfo{person}{Emily Jia}, \bibinfo{person}{Sameer Singh}, {and} \bibinfo{person}{Himabindu Lakkaraju}.} \bibinfo{year}{2020}\natexlab{}.
\newblock \showarticletitle{Fooling {LIME} and {SHAP}: {Adversarial} {Attacks} on {Post} hoc {Explanation} {Methods}}. In \bibinfo{booktitle}{\emph{Proceedings of the {AAAI}/{ACM} {Conference} on {AI}, {Ethics}, and {Society}}} \emph{(\bibinfo{series}{{AIES} '20})}. \bibinfo{publisher}{Association for Computing Machinery}, \bibinfo{address}{New York, NY, USA}, \bibinfo{pages}{180--186}.
\newblock
\showISBNx{978-1-4503-7110-0}
\urldef\tempurl%
\url{https://doi.org/10.1145/3375627.3375830}
\showDOI{\tempurl}


\bibitem[Smith et~al\mbox{.}(2023)]%
        {smith2023identifying}
\bibfield{author}{\bibinfo{person}{Victoria Smith}, \bibinfo{person}{Ali~Shahin Shamsabadi}, \bibinfo{person}{Carolyn Ashurst}, {and} \bibinfo{person}{Adrian Weller}.} \bibinfo{year}{2023}\natexlab{}.
\newblock \showarticletitle{Identifying and Mitigating Privacy Risks Stemming from Language Models: A Survey}.
\newblock \bibinfo{journal}{\emph{arXiv preprint arXiv:2310.01424}} (\bibinfo{year}{2023}).
\newblock


\bibitem[Soice et~al\mbox{.}(2023)]%
        {soice2023can}
\bibfield{author}{\bibinfo{person}{Emily~H Soice}, \bibinfo{person}{Rafael Rocha}, \bibinfo{person}{Kimberlee Cordova}, \bibinfo{person}{Michael Specter}, {and} \bibinfo{person}{Kevin~M Esvelt}.} \bibinfo{year}{2023}\natexlab{}.
\newblock \showarticletitle{Can large language models democratize access to dual-use biotechnology?}
\newblock \bibinfo{journal}{\emph{arXiv preprint arXiv:2306.03809}} (\bibinfo{year}{2023}).
\newblock


\bibitem[Solaiman(2023)]%
        {solaiman2023gradient}
\bibfield{author}{\bibinfo{person}{Irene Solaiman}.} \bibinfo{year}{2023}\natexlab{}.
\newblock \showarticletitle{The gradient of generative AI release: Methods and considerations}. In \bibinfo{booktitle}{\emph{Proceedings of the 2023 ACM Conference on Fairness, Accountability, and Transparency}}. \bibinfo{pages}{111--122}.
\newblock


\bibitem[Solaiman et~al\mbox{.}(2023)]%
        {solaiman2023evaluating}
\bibfield{author}{\bibinfo{person}{Irene Solaiman}, \bibinfo{person}{Zeerak Talat}, \bibinfo{person}{William Agnew}, \bibinfo{person}{Lama Ahmad}, \bibinfo{person}{Dylan Baker}, \bibinfo{person}{Su~Lin Blodgett}, \bibinfo{person}{Hal Daum{\'e}~III}, \bibinfo{person}{Jesse Dodge}, \bibinfo{person}{Ellie Evans}, \bibinfo{person}{Sara Hooker}, {et~al\mbox{.}}} \bibinfo{year}{2023}\natexlab{}.
\newblock \showarticletitle{Evaluating the Social Impact of Generative AI Systems in Systems and Society}.
\newblock \bibinfo{journal}{\emph{arXiv preprint arXiv:2306.05949}} (\bibinfo{year}{2023}).
\newblock


\bibitem[Song et~al\mbox{.}(2020)]%
        {song2020universal}
\bibfield{author}{\bibinfo{person}{Liwei Song}, \bibinfo{person}{Xinwei Yu}, \bibinfo{person}{Hsuan-Tung Peng}, {and} \bibinfo{person}{Karthik Narasimhan}.} \bibinfo{year}{2020}\natexlab{}.
\newblock \showarticletitle{Universal adversarial attacks with natural triggers for text classification}.
\newblock \bibinfo{journal}{\emph{arXiv preprint arXiv:2005.00174}} (\bibinfo{year}{2020}).
\newblock


\bibitem[Sorensen et~al\mbox{.}(2024)]%
        {sorensen2024roadmap}
\bibfield{author}{\bibinfo{person}{Taylor Sorensen}, \bibinfo{person}{Jared Moore}, \bibinfo{person}{Jillian Fisher}, \bibinfo{person}{Mitchell Gordon}, \bibinfo{person}{Niloofar Mireshghallah}, \bibinfo{person}{Christopher~Michael Rytting}, \bibinfo{person}{Andre Ye}, \bibinfo{person}{Liwei Jiang}, \bibinfo{person}{Ximing Lu}, \bibinfo{person}{Nouha Dziri}, {et~al\mbox{.}}} \bibinfo{year}{2024}\natexlab{}.
\newblock \showarticletitle{A Roadmap to Pluralistic Alignment}.
\newblock \bibinfo{journal}{\emph{arXiv preprint arXiv:2402.05070}} (\bibinfo{year}{2024}).
\newblock


\bibitem[Srivastava et~al\mbox{.}(2022)]%
        {srivastava2022beyond}
\bibfield{author}{\bibinfo{person}{Aarohi Srivastava}, \bibinfo{person}{Abhinav Rastogi}, \bibinfo{person}{Abhishek Rao}, \bibinfo{person}{Abu Awal~Md Shoeb}, \bibinfo{person}{Abubakar Abid}, \bibinfo{person}{Adam Fisch}, \bibinfo{person}{Adam~R Brown}, \bibinfo{person}{Adam Santoro}, \bibinfo{person}{Aditya Gupta}, \bibinfo{person}{Adri{\`a} Garriga-Alonso}, {et~al\mbox{.}}} \bibinfo{year}{2022}\natexlab{}.
\newblock \showarticletitle{Beyond the imitation game: Quantifying and extrapolating the capabilities of language models}.
\newblock \bibinfo{journal}{\emph{arXiv preprint arXiv:2206.04615}} (\bibinfo{year}{2022}).
\newblock


\bibitem[Sun et~al\mbox{.}(2023)]%
        {sun2023aligning}
\bibfield{author}{\bibinfo{person}{Huaman Sun}, \bibinfo{person}{Jiaxin Pei}, \bibinfo{person}{Minje Choi}, {and} \bibinfo{person}{David Jurgens}.} \bibinfo{year}{2023}\natexlab{}.
\newblock \showarticletitle{Aligning with Whom? Large Language Models Have Gender and Racial Biases in Subjective NLP Tasks}.
\newblock  (\bibinfo{year}{2023}).
\newblock
\showeprint[arxiv]{2311.09730}~[cs.CL]


\bibitem[Sun et~al\mbox{.}(2024)]%
        {sun2024trustllm}
\bibfield{author}{\bibinfo{person}{Lichao Sun}, \bibinfo{person}{Yue Huang}, \bibinfo{person}{Haoran Wang}, \bibinfo{person}{Siyuan Wu}, \bibinfo{person}{Qihui Zhang}, \bibinfo{person}{Chujie Gao}, \bibinfo{person}{Yixin Huang}, \bibinfo{person}{Wenhan Lyu}, \bibinfo{person}{Yixuan Zhang}, \bibinfo{person}{Xiner Li}, \bibinfo{person}{Zhengliang Liu}, \bibinfo{person}{Yixin Liu}, \bibinfo{person}{Yijue Wang}, \bibinfo{person}{Zhikun Zhang}, \bibinfo{person}{Bhavya Kailkhura}, \bibinfo{person}{Caiming Xiong}, \bibinfo{person}{Chao Zhang}, \bibinfo{person}{Chaowei Xiao}, \bibinfo{person}{Chunyuan Li}, \bibinfo{person}{Eric Xing}, \bibinfo{person}{Furong Huang}, \bibinfo{person}{Hao Liu}, \bibinfo{person}{Heng Ji}, \bibinfo{person}{Hongyi Wang}, \bibinfo{person}{Huan Zhang}, \bibinfo{person}{Huaxiu Yao}, \bibinfo{person}{Manolis Kellis}, \bibinfo{person}{Marinka Zitnik}, \bibinfo{person}{Meng Jiang}, \bibinfo{person}{Mohit Bansal}, \bibinfo{person}{James Zou}, \bibinfo{person}{Jian Pei}, \bibinfo{person}{Jian
  Liu}, \bibinfo{person}{Jianfeng Gao}, \bibinfo{person}{Jiawei Han}, \bibinfo{person}{Jieyu Zhao}, \bibinfo{person}{Jiliang Tang}, \bibinfo{person}{Jindong Wang}, \bibinfo{person}{John Mitchell}, \bibinfo{person}{Kai Shu}, \bibinfo{person}{Kaidi Xu}, \bibinfo{person}{Kai-Wei Chang}, \bibinfo{person}{Lifang He}, \bibinfo{person}{Lifu Huang}, \bibinfo{person}{Michael Backes}, \bibinfo{person}{Neil~Zhenqiang Gong}, \bibinfo{person}{Philip~S. Yu}, \bibinfo{person}{Pin-Yu Chen}, \bibinfo{person}{Quanquan Gu}, \bibinfo{person}{Ran Xu}, \bibinfo{person}{Rex Ying}, \bibinfo{person}{Shuiwang Ji}, \bibinfo{person}{Suman Jana}, \bibinfo{person}{Tianlong Chen}, \bibinfo{person}{Tianming Liu}, \bibinfo{person}{Tianyi Zhou}, \bibinfo{person}{Willian Wang}, \bibinfo{person}{Xiang Li}, \bibinfo{person}{Xiangliang Zhang}, \bibinfo{person}{Xiao Wang}, \bibinfo{person}{Xing Xie}, \bibinfo{person}{Xun Chen}, \bibinfo{person}{Xuyu Wang}, \bibinfo{person}{Yan Liu}, \bibinfo{person}{Yanfang Ye}, \bibinfo{person}{Yinzhi Cao}, {and}
  \bibinfo{person}{Yue Zhao}.} \bibinfo{year}{2024}\natexlab{}.
\newblock \bibinfo{title}{TrustLLM: Trustworthiness in Large Language Models}.
\newblock
\newblock
\showeprint[arxiv]{2401.05561}~[cs.CL]


\bibitem[Suri et~al\mbox{.}(2023)]%
        {suri2023large}
\bibfield{author}{\bibinfo{person}{Gaurav Suri}, \bibinfo{person}{Lily~R Slater}, \bibinfo{person}{Ali Ziaee}, {and} \bibinfo{person}{Morgan Nguyen}.} \bibinfo{year}{2023}\natexlab{}.
\newblock \showarticletitle{Do Large Language Models Show Decision Heuristics Similar to Humans? A Case Study Using GPT-3.5}.
\newblock \bibinfo{journal}{\emph{arXiv preprint arXiv:2305.04400}} (\bibinfo{year}{2023}).
\newblock


\bibitem[Tann et~al\mbox{.}(2023)]%
        {tann2023using}
\bibfield{author}{\bibinfo{person}{Wesley Tann}, \bibinfo{person}{Yuancheng Liu}, \bibinfo{person}{Jun~Heng Sim}, \bibinfo{person}{Choon~Meng Seah}, {and} \bibinfo{person}{Ee-Chien Chang}.} \bibinfo{year}{2023}\natexlab{}.
\newblock \showarticletitle{Using Large Language Models for Cybersecurity Capture-The-Flag Challenges and Certification Questions}.
\newblock \bibinfo{journal}{\emph{arXiv preprint arXiv:2308.10443}} (\bibinfo{year}{2023}).
\newblock


\bibitem[Tao et~al\mbox{.}(2023)]%
        {tao2023auditing}
\bibfield{author}{\bibinfo{person}{Yan Tao}, \bibinfo{person}{Olga Viberg}, \bibinfo{person}{Ryan~S. Baker}, {and} \bibinfo{person}{Rene~F. Kizilcec}.} \bibinfo{year}{2023}\natexlab{}.
\newblock \showarticletitle{Auditing and Mitigating Cultural Bias in LLMs}.
\newblock  (\bibinfo{year}{2023}).
\newblock
\showeprint[arxiv]{2311.14096}~[cs.CL]


\bibitem[Tegmark and Omohundro(2023)]%
        {tegmark_provably_2023}
\bibfield{author}{\bibinfo{person}{Max Tegmark} {and} \bibinfo{person}{Steve Omohundro}.} \bibinfo{year}{2023}\natexlab{}.
\newblock \showarticletitle{Provably safe systems: the only path to controllable {AGI}}.
\newblock  (\bibinfo{date}{Sept.} \bibinfo{year}{2023}).
\newblock
\urldef\tempurl%
\url{https://doi.org/10.48550/arXiv.2309.01933}
\showDOI{\tempurl}
\newblock
\shownote{arXiv:2309.01933 [cs]}.


\bibitem[Thiel(2023)]%
        {thiel2023identifying}
\bibfield{author}{\bibinfo{person}{David Thiel}.} \bibinfo{year}{2023}\natexlab{}.
\newblock \showarticletitle{Identifying and Eliminating CSAM in Generative ML Training Data and Models}.
\newblock  (\bibinfo{year}{2023}).
\newblock


\bibitem[Thiel et~al\mbox{.}(2023)]%
        {thiel2023generative}
\bibfield{author}{\bibinfo{person}{David Thiel}, \bibinfo{person}{Melissa Stroebel}, {and} \bibinfo{person}{Rebecca Portnoff}.} \bibinfo{year}{2023}\natexlab{}.
\newblock \showarticletitle{Generative ML and CSAM: Implications and Mitigations}.
\newblock  (\bibinfo{year}{2023}).
\newblock


\bibitem[Touvron et~al\mbox{.}(2023)]%
        {touvron2023llama}
\bibfield{author}{\bibinfo{person}{Hugo Touvron}, \bibinfo{person}{Louis Martin}, \bibinfo{person}{Kevin Stone}, \bibinfo{person}{Peter Albert}, \bibinfo{person}{Amjad Almahairi}, \bibinfo{person}{Yasmine Babaei}, \bibinfo{person}{Nikolay Bashlykov}, \bibinfo{person}{Soumya Batra}, \bibinfo{person}{Prajjwal Bhargava}, \bibinfo{person}{Shruti Bhosale}, \bibinfo{person}{Dan Bikel}, \bibinfo{person}{Lukas Blecher}, \bibinfo{person}{Cristian~Canton Ferrer}, \bibinfo{person}{Moya Chen}, \bibinfo{person}{Guillem Cucurull}, \bibinfo{person}{David Esiobu}, \bibinfo{person}{Jude Fernandes}, \bibinfo{person}{Jeremy Fu}, \bibinfo{person}{Wenyin Fu}, \bibinfo{person}{Brian Fuller}, \bibinfo{person}{Cynthia Gao}, \bibinfo{person}{Vedanuj Goswami}, \bibinfo{person}{Naman Goyal}, \bibinfo{person}{Anthony Hartshorn}, \bibinfo{person}{Saghar Hosseini}, \bibinfo{person}{Rui Hou}, \bibinfo{person}{Hakan Inan}, \bibinfo{person}{Marcin Kardas}, \bibinfo{person}{Viktor Kerkez}, \bibinfo{person}{Madian Khabsa},
  \bibinfo{person}{Isabel Kloumann}, \bibinfo{person}{Artem Korenev}, \bibinfo{person}{Punit~Singh Koura}, \bibinfo{person}{Marie-Anne Lachaux}, \bibinfo{person}{Thibaut Lavril}, \bibinfo{person}{Jenya Lee}, \bibinfo{person}{Diana Liskovich}, \bibinfo{person}{Yinghai Lu}, \bibinfo{person}{Yuning Mao}, \bibinfo{person}{Xavier Martinet}, \bibinfo{person}{Todor Mihaylov}, \bibinfo{person}{Pushkar Mishra}, \bibinfo{person}{Igor Molybog}, \bibinfo{person}{Yixin Nie}, \bibinfo{person}{Andrew Poulton}, \bibinfo{person}{Jeremy Reizenstein}, \bibinfo{person}{Rashi Rungta}, \bibinfo{person}{Kalyan Saladi}, \bibinfo{person}{Alan Schelten}, \bibinfo{person}{Ruan Silva}, \bibinfo{person}{Eric~Michael Smith}, \bibinfo{person}{Ranjan Subramanian}, \bibinfo{person}{Xiaoqing~Ellen Tan}, \bibinfo{person}{Binh Tang}, \bibinfo{person}{Ross Taylor}, \bibinfo{person}{Adina Williams}, \bibinfo{person}{Jian~Xiang Kuan}, \bibinfo{person}{Puxin Xu}, \bibinfo{person}{Zheng Yan}, \bibinfo{person}{Iliyan Zarov}, \bibinfo{person}{Yuchen
  Zhang}, \bibinfo{person}{Angela Fan}, \bibinfo{person}{Melanie Kambadur}, \bibinfo{person}{Sharan Narang}, \bibinfo{person}{Aurelien Rodriguez}, \bibinfo{person}{Robert Stojnic}, \bibinfo{person}{Sergey Edunov}, {and} \bibinfo{person}{Thomas Scialom}.} \bibinfo{year}{2023}\natexlab{}.
\newblock \showarticletitle{Llama 2: Open Foundation and Fine-Tuned Chat Models}.
\newblock  (\bibinfo{year}{2023}).
\newblock
\showeprint[arxiv]{2307.09288}~[cs.CL]


\bibitem[Trager et~al\mbox{.}(2023)]%
        {trager2023international}
\bibfield{author}{\bibinfo{person}{Robert Trager}, \bibinfo{person}{Ben Harack}, \bibinfo{person}{Anka Reuel}, \bibinfo{person}{Allison Carnegie}, \bibinfo{person}{Lennart Heim}, \bibinfo{person}{Lewis Ho}, \bibinfo{person}{Sarah Kreps}, \bibinfo{person}{Ranjit Lall}, \bibinfo{person}{Owen Larter}, \bibinfo{person}{Se{\'a}n~{\'O} h{\'E}igeartaigh}, {et~al\mbox{.}}} \bibinfo{year}{2023}\natexlab{}.
\newblock \showarticletitle{International governance of civilian AI: A jurisdictional certification approach}.
\newblock \bibinfo{journal}{\emph{arXiv preprint arXiv:2308.15514}} (\bibinfo{year}{2023}).
\newblock


\bibitem[Tsai et~al\mbox{.}(2023)]%
        {tsai2023ring}
\bibfield{author}{\bibinfo{person}{Yu-Lin Tsai}, \bibinfo{person}{Chia-Yi Hsu}, \bibinfo{person}{Chulin Xie}, \bibinfo{person}{Chih-Hsun Lin}, \bibinfo{person}{Jia-You Chen}, \bibinfo{person}{Bo Li}, \bibinfo{person}{Pin-Yu Chen}, \bibinfo{person}{Chia-Mu Yu}, {and} \bibinfo{person}{Chun-Ying Huang}.} \bibinfo{year}{2023}\natexlab{}.
\newblock \showarticletitle{Ring-A-Bell! How Reliable are Concept Removal Methods for Diffusion Models?}
\newblock \bibinfo{journal}{\emph{arXiv preprint arXiv:2310.10012}} (\bibinfo{year}{2023}).
\newblock


\bibitem[Turner et~al\mbox{.}(2023)]%
        {turner2023activation}
\bibfield{author}{\bibinfo{person}{Alex Turner}, \bibinfo{person}{Lisa Thiergart}, \bibinfo{person}{David Udell}, \bibinfo{person}{Gavin Leech}, \bibinfo{person}{Ulisse Mini}, {and} \bibinfo{person}{Monte MacDiarmid}.} \bibinfo{year}{2023}\natexlab{}.
\newblock \showarticletitle{Activation addition: Steering language models without optimization}.
\newblock \bibinfo{journal}{\emph{arXiv preprint arXiv:2308.10248}} (\bibinfo{year}{2023}).
\newblock


\bibitem[Turpin et~al\mbox{.}(2023)]%
        {turpin2023language}
\bibfield{author}{\bibinfo{person}{Miles Turpin}, \bibinfo{person}{Julian Michael}, \bibinfo{person}{Ethan Perez}, {and} \bibinfo{person}{Samuel~R. Bowman}.} \bibinfo{year}{2023}\natexlab{}.
\newblock \showarticletitle{Language Models Don't Always Say What They Think: Unfaithful Explanations in Chain-of-Thought Prompting}.
\newblock  (\bibinfo{year}{2023}).
\newblock
\showeprint[arxiv]{2305.04388}~[cs.CL]


\bibitem[{UK Department for Science, Innovation \& Technology}(2023)]%
        {dsit2023}
\bibfield{author}{\bibinfo{person}{{UK Department for Science, Innovation \& Technology}}.} \bibinfo{year}{2023}\natexlab{}.
\newblock \bibinfo{booktitle}{\emph{A pro-innovation approach to {AI} regulation}}.
\newblock \bibinfo{type}{{T}echnical {R}eport}.
\newblock
\urldef\tempurl%
\url{https://www.gov.uk/government/publications/ai-regulation-a-pro-innovation-approach/white-paper}
\showURL{%
\tempurl}


\bibitem[{United Nations}(2022)]%
        {un2022}
\bibfield{author}{\bibinfo{person}{{United Nations}}.} \bibinfo{year}{2022}\natexlab{}.
\newblock \bibinfo{title}{Principles for the ethical use of artificial intelligence in the {United} {Nations} system}.
\newblock
\newblock
\urldef\tempurl%
\url{https://unsceb.org/sites/default/files/2023-03/CEB_2022_2_Add.1%20%28AI%20ethics%20principles%29.pdf}
\showURL{%
\tempurl}


\bibitem[{United States National Science Foundation}(2023)]%
        {nsf2023national}
\bibfield{author}{\bibinfo{person}{{United States National Science Foundation}}.} \bibinfo{year}{2023}\natexlab{}.
\newblock \showarticletitle{National Deep Inference Facility for Very Large Language Models (NDIF)}.
\newblock  (\bibinfo{year}{2023}).
\newblock


\bibitem[{U.S. Department of Commerce} and {National Institute of Standards and Technology}(2023)]%
        {airmf2023}
\bibfield{author}{\bibinfo{person}{{U.S. Department of Commerce}} {and} \bibinfo{person}{{National Institute of Standards and Technology}}.} \bibinfo{year}{2023}\natexlab{}.
\newblock \bibinfo{title}{{AI} {Risk} {Management} {Framework}: {AI} {RMF} (1.0)}.
\newblock
\newblock
\urldef\tempurl%
\url{https://doi.org/10.6028/NIST.AI.100-1}
\showDOI{\tempurl}


\bibitem[van~den Brom(2022)]%
        {Brom2022OnsiteIA}
\bibfield{author}{\bibinfo{person}{H.~E. van~den Brom}.} \bibinfo{year}{2022}\natexlab{}.
\newblock \showarticletitle{On-site Inspection and Legal Certainty}.
\newblock \bibinfo{journal}{\emph{SSRN Electronic Journal}} (\bibinfo{year}{2022}).
\newblock
\urldef\tempurl%
\url{https://api.semanticscholar.org/CorpusID:249326468}
\showURL{%
\tempurl}


\bibitem[Wagner and Dittmar(2006)]%
        {wagner_unexpected_2006}
\bibfield{author}{\bibinfo{person}{Stephen Wagner} {and} \bibinfo{person}{Lee Dittmar}.} \bibinfo{year}{2006}\natexlab{}.
\newblock \showarticletitle{The unexpected benefits of {Sarbanes}-{Oxley}}.
\newblock \bibinfo{journal}{\emph{Harvard Business Review}} \bibinfo{volume}{84}, \bibinfo{number}{4} (\bibinfo{date}{April} \bibinfo{year}{2006}), \bibinfo{pages}{133--140; 150}.
\newblock
\showISSN{0017-8012}


\bibitem[Waldman(2019)]%
        {waldman2019privacy}
\bibfield{author}{\bibinfo{person}{Ari~Ezra Waldman}.} \bibinfo{year}{2019}\natexlab{}.
\newblock \showarticletitle{Privacy {Law}'s {False} {Promise}}.
\newblock \bibinfo{journal}{\emph{SSRN Electronic Journal}} (\bibinfo{year}{2019}).
\newblock
\showISSN{1556-5068}
\urldef\tempurl%
\url{https://doi.org/10.2139/ssrn.3339372}
\showDOI{\tempurl}


\bibitem[Wallace et~al\mbox{.}(2019)]%
        {wallace2019universal}
\bibfield{author}{\bibinfo{person}{Eric Wallace}, \bibinfo{person}{Shi Feng}, \bibinfo{person}{Nikhil Kandpal}, \bibinfo{person}{Matt Gardner}, {and} \bibinfo{person}{Sameer Singh}.} \bibinfo{year}{2019}\natexlab{}.
\newblock \showarticletitle{Universal adversarial triggers for attacking and analyzing NLP}.
\newblock \bibinfo{journal}{\emph{arXiv preprint arXiv:1908.07125}} (\bibinfo{year}{2019}).
\newblock


\bibitem[Wan et~al\mbox{.}(2023)]%
        {wan2023poisoning}
\bibfield{author}{\bibinfo{person}{Alexander Wan}, \bibinfo{person}{Eric Wallace}, \bibinfo{person}{Sheng Shen}, {and} \bibinfo{person}{Dan Klein}.} \bibinfo{year}{2023}\natexlab{}.
\newblock \showarticletitle{Poisoning Language Models During Instruction Tuning}.
\newblock  (\bibinfo{year}{2023}).
\newblock
\showeprint[arxiv]{2305.00944}~[cs.CL]


\bibitem[Wang et~al\mbox{.}(2018)]%
        {wang2018glue}
\bibfield{author}{\bibinfo{person}{Alex Wang}, \bibinfo{person}{Amanpreet Singh}, \bibinfo{person}{Julian Michael}, \bibinfo{person}{Felix Hill}, \bibinfo{person}{Omer Levy}, {and} \bibinfo{person}{Samuel~R Bowman}.} \bibinfo{year}{2018}\natexlab{}.
\newblock \showarticletitle{GLUE: A multi-task benchmark and analysis platform for natural language understanding}.
\newblock \bibinfo{journal}{\emph{arXiv preprint arXiv:1804.07461}} (\bibinfo{year}{2018}).
\newblock


\bibitem[Wang et~al\mbox{.}(2023b)]%
        {wang2023exploitability}
\bibfield{author}{\bibinfo{person}{Jiongxiao Wang}, \bibinfo{person}{Junlin Wu}, \bibinfo{person}{Muhao Chen}, \bibinfo{person}{Yevgeniy Vorobeychik}, {and} \bibinfo{person}{Chaowei Xiao}.} \bibinfo{year}{2023}\natexlab{b}.
\newblock \showarticletitle{On the Exploitability of Reinforcement Learning with Human Feedback for Large Language Models}.
\newblock  (\bibinfo{year}{2023}).
\newblock
\showeprint[arxiv]{2311.09641}~[cs.AI]


\bibitem[Wang et~al\mbox{.}(2023c)]%
        {wang2023knowledge}
\bibfield{author}{\bibinfo{person}{Song Wang}, \bibinfo{person}{Yaochen Zhu}, \bibinfo{person}{Haochen Liu}, \bibinfo{person}{Zaiyi Zheng}, \bibinfo{person}{Chen Chen}, {and} \bibinfo{person}{Jundong Li}.} \bibinfo{year}{2023}\natexlab{c}.
\newblock \showarticletitle{Knowledge Editing for Large Language Models: A Survey}.
\newblock  (\bibinfo{year}{2023}).
\newblock
\showeprint[arxiv]{2310.16218}~[cs.CL]


\bibitem[Wang et~al\mbox{.}(2023a)]%
        {wang2023adversarial}
\bibfield{author}{\bibinfo{person}{Tony~T. Wang}, \bibinfo{person}{Adam Gleave}, \bibinfo{person}{Tom Tseng}, \bibinfo{person}{Kellin Pelrine}, \bibinfo{person}{Nora Belrose}, \bibinfo{person}{Joseph Miller}, \bibinfo{person}{Michael~D. Dennis}, \bibinfo{person}{Yawen Duan}, \bibinfo{person}{Viktor Pogrebniak}, \bibinfo{person}{Sergey Levine}, {and} \bibinfo{person}{Stuart Russell}.} \bibinfo{year}{2023}\natexlab{a}.
\newblock \showarticletitle{Adversarial Policies Beat Superhuman Go AIs}.
\newblock  (\bibinfo{year}{2023}).
\newblock
\showeprint[arxiv]{2211.00241}~[cs.LG]


\bibitem[Watkins et~al\mbox{.}(2021)]%
        {watkins2021governing}
\bibfield{author}{\bibinfo{person}{Elizabeth~Anne Watkins}, \bibinfo{person}{Emanuel Moss}, \bibinfo{person}{Jacob Metcalf}, \bibinfo{person}{Ranjit Singh}, {and} \bibinfo{person}{Madeleine~Clare Elish}.} \bibinfo{year}{2021}\natexlab{}.
\newblock \showarticletitle{Governing algorithmic systems with impact assessments: Six observations}. In \bibinfo{booktitle}{\emph{Proceedings of the 2021 AAAI/ACM Conference on AI, Ethics, and Society}}. \bibinfo{pages}{1010--1022}.
\newblock


\bibitem[Wei et~al\mbox{.}(2023)]%
        {wei2023jailbroken}
\bibfield{author}{\bibinfo{person}{Alexander Wei}, \bibinfo{person}{Nika Haghtalab}, {and} \bibinfo{person}{Jacob Steinhardt}.} \bibinfo{year}{2023}\natexlab{}.
\newblock \showarticletitle{Jailbroken: How does llm safety training fail?}
\newblock \bibinfo{journal}{\emph{arXiv preprint arXiv:2307.02483}} (\bibinfo{year}{2023}).
\newblock


\bibitem[Wei et~al\mbox{.}(2022)]%
        {wei2022chain}
\bibfield{author}{\bibinfo{person}{Jason Wei}, \bibinfo{person}{Xuezhi Wang}, \bibinfo{person}{Dale Schuurmans}, \bibinfo{person}{Maarten Bosma}, \bibinfo{person}{Fei Xia}, \bibinfo{person}{Ed Chi}, \bibinfo{person}{Quoc~V Le}, \bibinfo{person}{Denny Zhou}, {et~al\mbox{.}}} \bibinfo{year}{2022}\natexlab{}.
\newblock \showarticletitle{Chain-of-thought prompting elicits reasoning in large language models}.
\newblock \bibinfo{journal}{\emph{Advances in Neural Information Processing Systems}}  \bibinfo{volume}{35} (\bibinfo{year}{2022}), \bibinfo{pages}{24824--24837}.
\newblock


\bibitem[Weidinger et~al\mbox{.}(2021)]%
        {weidinger2021ethical}
\bibfield{author}{\bibinfo{person}{Laura Weidinger}, \bibinfo{person}{John Mellor}, \bibinfo{person}{Maribeth Rauh}, \bibinfo{person}{Conor Griffin}, \bibinfo{person}{Jonathan Uesato}, \bibinfo{person}{Po-Sen Huang}, \bibinfo{person}{Myra Cheng}, \bibinfo{person}{Mia Glaese}, \bibinfo{person}{Borja Balle}, \bibinfo{person}{Atoosa Kasirzadeh}, {et~al\mbox{.}}} \bibinfo{year}{2021}\natexlab{}.
\newblock \showarticletitle{Ethical and social risks of harm from language models}.
\newblock \bibinfo{journal}{\emph{arXiv preprint arXiv:2112.04359}} (\bibinfo{year}{2021}).
\newblock


\bibitem[Weidinger et~al\mbox{.}(2023)]%
        {weidinger2023sociotechnical}
\bibfield{author}{\bibinfo{person}{Laura Weidinger}, \bibinfo{person}{Maribeth Rauh}, \bibinfo{person}{Nahema Marchal}, \bibinfo{person}{Arianna Manzini}, \bibinfo{person}{Lisa~Anne Hendricks}, \bibinfo{person}{Juan Mateos-Garcia}, \bibinfo{person}{Stevie Bergman}, \bibinfo{person}{Jackie Kay}, \bibinfo{person}{Conor Griffin}, \bibinfo{person}{Ben Bariach}, \bibinfo{person}{Iason Gabriel}, \bibinfo{person}{Verena Rieser}, {and} \bibinfo{person}{William Isaac}.} \bibinfo{year}{2023}\natexlab{}.
\newblock \showarticletitle{Sociotechnical {Safety} {Evaluation} of {Generative} {AI} {Systems}}.
\newblock  (\bibinfo{date}{Oct.} \bibinfo{year}{2023}).
\newblock
\urldef\tempurl%
\url{http://arxiv.org/abs/2310.11986}
\showURL{%
\tempurl}
\newblock
\shownote{arXiv:2310.11986 [cs]}.


\bibitem[Wen et~al\mbox{.}(2023)]%
        {wen2023hard}
\bibfield{author}{\bibinfo{person}{Yuxin Wen}, \bibinfo{person}{Neel Jain}, \bibinfo{person}{John Kirchenbauer}, \bibinfo{person}{Micah Goldblum}, \bibinfo{person}{Jonas Geiping}, {and} \bibinfo{person}{Tom Goldstein}.} \bibinfo{year}{2023}\natexlab{}.
\newblock \showarticletitle{Hard prompts made easy: Gradient-based discrete optimization for prompt tuning and discovery}.
\newblock \bibinfo{journal}{\emph{arXiv preprint arXiv:2302.03668}} (\bibinfo{year}{2023}).
\newblock


\bibitem[Westra(2021)]%
        {westra2021virtue}
\bibfield{author}{\bibinfo{person}{Evan Westra}.} \bibinfo{year}{2021}\natexlab{}.
\newblock \showarticletitle{Virtue {Signaling} and {Moral} {Progress}}.
\newblock \bibinfo{journal}{\emph{Philosophy \& Public Affairs}} \bibinfo{volume}{49}, \bibinfo{number}{2} (\bibinfo{year}{2021}), \bibinfo{pages}{156--178}.
\newblock
\showISSN{1088-4963}
\urldef\tempurl%
\url{https://doi.org/10.1111/papa.12187}
\showDOI{\tempurl}
\newblock
\shownote{\_eprint: https://onlinelibrary.wiley.com/doi/pdf/10.1111/papa.12187}.


\bibitem[White(2010)]%
        {white2010markets}
\bibfield{author}{\bibinfo{person}{Lawrence~J. White}.} \bibinfo{year}{2010}\natexlab{}.
\newblock \showarticletitle{Markets: {The} {Credit} {Rating} {Agencies}}.
\newblock \bibinfo{journal}{\emph{Journal of Economic Perspectives}} \bibinfo{volume}{24}, \bibinfo{number}{2} (\bibinfo{date}{June} \bibinfo{year}{2010}), \bibinfo{pages}{211--226}.
\newblock
\showISSN{0895-3309}
\urldef\tempurl%
\url{https://doi.org/10.1257/jep.24.2.211}
\showDOI{\tempurl}


\bibitem[Wieringa(2020)]%
        {wieringa2020account}
\bibfield{author}{\bibinfo{person}{Maranke Wieringa}.} \bibinfo{year}{2020}\natexlab{}.
\newblock \showarticletitle{What to account for when accounting for algorithms: a systematic literature review on algorithmic accountability}. In \bibinfo{booktitle}{\emph{Proceedings of the 2020 conference on fairness, accountability, and transparency}}. \bibinfo{pages}{1--18}.
\newblock


\bibitem[Wilkinson et~al\mbox{.}(2023)]%
        {wilkinson2023accountability}
\bibfield{author}{\bibinfo{person}{Daricia Wilkinson}, \bibinfo{person}{Kate Crawford}, \bibinfo{person}{Hanna Wallach}, \bibinfo{person}{Deborah Raji}, \bibinfo{person}{Bogdana Rakova}, \bibinfo{person}{Ranjit Singh}, \bibinfo{person}{Angelika Strohmayer}, {and} \bibinfo{person}{Ethan Zuckerman}.} \bibinfo{year}{2023}\natexlab{}.
\newblock \showarticletitle{Accountability in Algorithmic Systems: From Principles to Practice}. In \bibinfo{booktitle}{\emph{Extended Abstracts of the 2023 CHI Conference on Human Factors in Computing Systems}}. \bibinfo{pages}{1--4}.
\newblock


\bibitem[Wu et~al\mbox{.}(2022)]%
        {wu2022backdoorbench}
\bibfield{author}{\bibinfo{person}{Baoyuan Wu}, \bibinfo{person}{Hongrui Chen}, \bibinfo{person}{Mingda Zhang}, \bibinfo{person}{Zihao Zhu}, \bibinfo{person}{Shaokui Wei}, \bibinfo{person}{Danni Yuan}, \bibinfo{person}{Chao Shen}, {and} \bibinfo{person}{Hongyuan Zha}.} \bibinfo{year}{2022}\natexlab{}.
\newblock \showarticletitle{BackdoorBench: A Comprehensive Benchmark of Backdoor Learning}.
\newblock \bibinfo{journal}{\emph{arXiv preprint arXiv:2206.12654}} (\bibinfo{year}{2022}).
\newblock


\bibitem[Wu et~al\mbox{.}(2023)]%
        {wu2023depn}
\bibfield{author}{\bibinfo{person}{Xinwei Wu}, \bibinfo{person}{Junzhuo Li}, \bibinfo{person}{Minghui Xu}, \bibinfo{person}{Weilong Dong}, \bibinfo{person}{Shuangzhi Wu}, \bibinfo{person}{Chao Bian}, {and} \bibinfo{person}{Deyi Xiong}.} \bibinfo{year}{2023}\natexlab{}.
\newblock \showarticletitle{DEPN: Detecting and Editing Privacy Neurons in Pretrained Language Models}.
\newblock  (\bibinfo{year}{2023}).
\newblock
\showeprint[arxiv]{2310.20138}~[cs.CR]


\bibitem[Yang et~al\mbox{.}(2023)]%
        {yang2023shadow}
\bibfield{author}{\bibinfo{person}{Xianjun Yang}, \bibinfo{person}{Xiao Wang}, \bibinfo{person}{Qi Zhang}, \bibinfo{person}{Linda Petzold}, \bibinfo{person}{William~Yang Wang}, \bibinfo{person}{Xun Zhao}, {and} \bibinfo{person}{Dahua Lin}.} \bibinfo{year}{2023}\natexlab{}.
\newblock \showarticletitle{Shadow Alignment: The Ease of Subverting Safely-Aligned Language Models}.
\newblock  (\bibinfo{year}{2023}).
\newblock
\showeprint[arxiv]{2310.02949}~[cs.CL]


\bibitem[Yao et~al\mbox{.}(2023)]%
        {yao2023survey}
\bibfield{author}{\bibinfo{person}{Yifan Yao}, \bibinfo{person}{Jinhao Duan}, \bibinfo{person}{Kaidi Xu}, \bibinfo{person}{Yuanfang Cai}, \bibinfo{person}{Eric Sun}, {and} \bibinfo{person}{Yue Zhang}.} \bibinfo{year}{2023}\natexlab{}.
\newblock \showarticletitle{A Survey on Large Language Model (LLM) Security and Privacy: The Good, the Bad, and the Ugly}.
\newblock \bibinfo{journal}{\emph{arXiv preprint arXiv:2312.02003}} (\bibinfo{year}{2023}).
\newblock


\bibitem[Yew and Hadfield-Menell(2022)]%
        {yew2022penalty}
\bibfield{author}{\bibinfo{person}{Rui-Jie Yew} {and} \bibinfo{person}{Dylan Hadfield-Menell}.} \bibinfo{year}{2022}\natexlab{}.
\newblock \showarticletitle{A Penalty Default Approach to Preemptive Harm Disclosure and Mitigation for AI Systems}. In \bibinfo{booktitle}{\emph{Proceedings of the 2022 AAAI/ACM Conference on AI, Ethics, and Society}}. \bibinfo{pages}{823--830}.
\newblock


\bibitem[Yong et~al\mbox{.}(2023)]%
        {yong2023lowresource}
\bibfield{author}{\bibinfo{person}{Zheng-Xin Yong}, \bibinfo{person}{Cristina Menghini}, {and} \bibinfo{person}{Stephen~H. Bach}.} \bibinfo{year}{2023}\natexlab{}.
\newblock \showarticletitle{Low-Resource Languages Jailbreak GPT-4}.
\newblock  (\bibinfo{year}{2023}).
\newblock
\showeprint[arxiv]{2310.02446}~[cs.CL]


\bibitem[Yu et~al\mbox{.}(2023)]%
        {yu2023gptfuzzer}
\bibfield{author}{\bibinfo{person}{Jiahao Yu}, \bibinfo{person}{Xingwei Lin}, {and} \bibinfo{person}{Xinyu Xing}.} \bibinfo{year}{2023}\natexlab{}.
\newblock \showarticletitle{GPTFUZZER: Red Teaming Large Language Models with Auto-Generated Jailbreak Prompts}.
\newblock \bibinfo{journal}{\emph{arXiv preprint arXiv:2309.10253}} (\bibinfo{year}{2023}).
\newblock


\bibitem[Yuksekgonul et~al\mbox{.}(2022)]%
        {yuksekgonul2022post}
\bibfield{author}{\bibinfo{person}{Mert Yuksekgonul}, \bibinfo{person}{Maggie Wang}, {and} \bibinfo{person}{James Zou}.} \bibinfo{year}{2022}\natexlab{}.
\newblock \showarticletitle{Post-hoc concept bottleneck models}.
\newblock \bibinfo{journal}{\emph{arXiv preprint arXiv:2205.15480}} (\bibinfo{year}{2022}).
\newblock


\bibitem[Zhan et~al\mbox{.}(2023)]%
        {zhan2023removing}
\bibfield{author}{\bibinfo{person}{Qiusi Zhan}, \bibinfo{person}{Richard Fang}, \bibinfo{person}{Rohan Bindu}, \bibinfo{person}{Akul Gupta}, \bibinfo{person}{Tatsunori Hashimoto}, {and} \bibinfo{person}{Daniel Kang}.} \bibinfo{year}{2023}\natexlab{}.
\newblock \showarticletitle{Removing RLHF Protections in GPT-4 via Fine-Tuning}.
\newblock  (\bibinfo{year}{2023}).
\newblock
\showeprint[arxiv]{2311.05553}~[cs.CL]


\bibitem[Zhang et~al\mbox{.}(2022)]%
        {zhang2022explainable}
\bibfield{author}{\bibinfo{person}{Chanyuan~Abigail Zhang}, \bibinfo{person}{Soohyun Cho}, {and} \bibinfo{person}{Miklos Vasarhelyi}.} \bibinfo{year}{2022}\natexlab{}.
\newblock \showarticletitle{Explainable artificial intelligence (xai) in auditing}.
\newblock \bibinfo{journal}{\emph{International Journal of Accounting Information Systems}}  \bibinfo{volume}{46} (\bibinfo{year}{2022}), \bibinfo{pages}{100572}.
\newblock


\bibitem[Zhang et~al\mbox{.}(2023)]%
        {zhang2023adversarial}
\bibfield{author}{\bibinfo{person}{Milin Zhang}, \bibinfo{person}{Mohammad Abdi}, {and} \bibinfo{person}{Francesco Restuccia}.} \bibinfo{year}{2023}\natexlab{}.
\newblock \showarticletitle{Adversarial Machine Learning in Latent Representations of Neural Networks}.
\newblock \bibinfo{journal}{\emph{arXiv preprint arXiv:2309.17401}} (\bibinfo{year}{2023}).
\newblock


\bibitem[Zhang et~al\mbox{.}(2019)]%
        {Zhang2019AdversarialAO}
\bibfield{author}{\bibinfo{person}{W. Zhang}, \bibinfo{person}{Quan.Z Sheng}, \bibinfo{person}{Ahoud Abdulrahmn~F. Alhazmi}, {and} \bibinfo{person}{Chenliang Li}.} \bibinfo{year}{2019}\natexlab{}.
\newblock \showarticletitle{Adversarial Attacks on Deep Learning Models in Natural Language Processing: A Survey}.
\newblock \bibinfo{journal}{\emph{arXiv: Computation and Language}} (\bibinfo{year}{2019}).
\newblock
\urldef\tempurl%
\url{https://api.semanticscholar.org/CorpusID:260428188}
\showURL{%
\tempurl}


\bibitem[Zhang et~al\mbox{.}(2020)]%
        {zhang2020adversarial}
\bibfield{author}{\bibinfo{person}{Wei~Emma Zhang}, \bibinfo{person}{Quan~Z Sheng}, \bibinfo{person}{Ahoud Alhazmi}, {and} \bibinfo{person}{Chenliang Li}.} \bibinfo{year}{2020}\natexlab{}.
\newblock \showarticletitle{Adversarial attacks on deep-learning models in natural language processing: A survey}.
\newblock \bibinfo{journal}{\emph{ACM Transactions on Intelligent Systems and Technology (TIST)}} \bibinfo{volume}{11}, \bibinfo{number}{3} (\bibinfo{year}{2020}), \bibinfo{pages}{1--41}.
\newblock


\bibitem[Zhao et~al\mbox{.}(2023)]%
        {zhao2023explainability}
\bibfield{author}{\bibinfo{person}{Haiyan Zhao}, \bibinfo{person}{Hanjie Chen}, \bibinfo{person}{Fan Yang}, \bibinfo{person}{Ninghao Liu}, \bibinfo{person}{Huiqi Deng}, \bibinfo{person}{Hengyi Cai}, \bibinfo{person}{Shuaiqiang Wang}, \bibinfo{person}{Dawei Yin}, {and} \bibinfo{person}{Mengnan Du}.} \bibinfo{year}{2023}\natexlab{}.
\newblock \showarticletitle{Explainability for large language models: A survey}.
\newblock \bibinfo{journal}{\emph{ACM Transactions on Intelligent Systems and Technology}} (\bibinfo{year}{2023}).
\newblock


\bibitem[Zhong et~al\mbox{.}(2023)]%
        {zhong2023clock}
\bibfield{author}{\bibinfo{person}{Ziqian Zhong}, \bibinfo{person}{Ziming Liu}, \bibinfo{person}{Max Tegmark}, {and} \bibinfo{person}{Jacob Andreas}.} \bibinfo{year}{2023}\natexlab{}.
\newblock \showarticletitle{The Clock and the Pizza: Two Stories in Mechanistic Explanation of Neural Networks}.
\newblock  (\bibinfo{year}{2023}).
\newblock
\showeprint[arxiv]{2306.17844}~[cs.LG]


\bibitem[Zhou et~al\mbox{.}(2023)]%
        {zhou_webarena_2023}
\bibfield{author}{\bibinfo{person}{Shuyan Zhou}, \bibinfo{person}{Frank~F. Xu}, \bibinfo{person}{Hao Zhu}, \bibinfo{person}{Xuhui Zhou}, \bibinfo{person}{Robert Lo}, \bibinfo{person}{Abishek Sridhar}, \bibinfo{person}{Xianyi Cheng}, \bibinfo{person}{Tianyue Ou}, \bibinfo{person}{Yonatan Bisk}, \bibinfo{person}{Daniel Fried}, \bibinfo{person}{Uri Alon}, {and} \bibinfo{person}{Graham Neubig}.} \bibinfo{year}{2023}\natexlab{}.
\newblock \showarticletitle{{WebArena}: {A} {Realistic} {Web} {Environment} for {Building} {Autonomous} {Agents}}.
\newblock  (\bibinfo{date}{Oct.} \bibinfo{year}{2023}).
\newblock
\urldef\tempurl%
\url{https://doi.org/10.48550/arXiv.2307.13854}
\showDOI{\tempurl}
\newblock
\shownote{arXiv:2307.13854 [cs]}.


\bibitem[Zhou et~al\mbox{.}(2018)]%
        {zhou2018transferable}
\bibfield{author}{\bibinfo{person}{Wen Zhou}, \bibinfo{person}{Xin Hou}, \bibinfo{person}{Yongjun Chen}, \bibinfo{person}{Mengyun Tang}, \bibinfo{person}{Xiangqi Huang}, \bibinfo{person}{Xiang Gan}, {and} \bibinfo{person}{Yong Yang}.} \bibinfo{year}{2018}\natexlab{}.
\newblock \showarticletitle{Transferable adversarial perturbations}. In \bibinfo{booktitle}{\emph{Proceedings of the European Conference on Computer Vision (ECCV)}}. \bibinfo{pages}{452--467}.
\newblock


\bibitem[Zhou et~al\mbox{.}(2019)]%
        {zhou2019latent}
\bibfield{author}{\bibinfo{person}{Xiaowei Zhou}, \bibinfo{person}{Ivor~W Tsang}, {and} \bibinfo{person}{Jie Yin}.} \bibinfo{year}{2019}\natexlab{}.
\newblock \showarticletitle{Latent adversarial defence with boundary-guided generation}.
\newblock \bibinfo{journal}{\emph{arXiv preprint arXiv:1907.07001}} (\bibinfo{year}{2019}).
\newblock


\bibitem[Zhu et~al\mbox{.}(2019)]%
        {zhu2019freelb}
\bibfield{author}{\bibinfo{person}{Chen Zhu}, \bibinfo{person}{Yu Cheng}, \bibinfo{person}{Zhe Gan}, \bibinfo{person}{Siqi Sun}, \bibinfo{person}{Tom Goldstein}, {and} \bibinfo{person}{Jingjing Liu}.} \bibinfo{year}{2019}\natexlab{}.
\newblock \showarticletitle{Freelb: Enhanced adversarial training for natural language understanding}.
\newblock \bibinfo{journal}{\emph{arXiv preprint arXiv:1909.11764}} (\bibinfo{year}{2019}).
\newblock


\bibitem[Ziegler et~al\mbox{.}(2022)]%
        {ziegler2022adversarial}
\bibfield{author}{\bibinfo{person}{Daniel~M. Ziegler}, \bibinfo{person}{Seraphina Nix}, \bibinfo{person}{Lawrence Chan}, \bibinfo{person}{Tim Bauman}, \bibinfo{person}{Peter Schmidt-Nielsen}, \bibinfo{person}{Tao Lin}, \bibinfo{person}{Adam Scherlis}, \bibinfo{person}{Noa Nabeshima}, \bibinfo{person}{Ben Weinstein-Raun}, \bibinfo{person}{Daniel de Haas}, \bibinfo{person}{Buck Shlegeris}, {and} \bibinfo{person}{Nate Thomas}.} \bibinfo{year}{2022}\natexlab{}.
\newblock \showarticletitle{Adversarial Training for High-Stakes Reliability}.
\newblock  (\bibinfo{year}{2022}).
\newblock
\showeprint[arxiv]{2205.01663}~[cs.LG]


\bibitem[Zou et~al\mbox{.}(2023a)]%
        {zou2023representation}
\bibfield{author}{\bibinfo{person}{Andy Zou}, \bibinfo{person}{Long Phan}, \bibinfo{person}{Sarah Chen}, \bibinfo{person}{James Campbell}, \bibinfo{person}{Phillip Guo}, \bibinfo{person}{Richard Ren}, \bibinfo{person}{Alexander Pan}, \bibinfo{person}{Xuwang Yin}, \bibinfo{person}{Mantas Mazeika}, \bibinfo{person}{Ann-Kathrin Dombrowski}, {et~al\mbox{.}}} \bibinfo{year}{2023}\natexlab{a}.
\newblock \showarticletitle{Representation engineering: A top-down approach to ai transparency}.
\newblock \bibinfo{journal}{\emph{arXiv preprint arXiv:2310.01405}} (\bibinfo{year}{2023}).
\newblock


\bibitem[Zou et~al\mbox{.}(2023b)]%
        {zou2023universal}
\bibfield{author}{\bibinfo{person}{Andy Zou}, \bibinfo{person}{Zifan Wang}, \bibinfo{person}{J.~Zico Kolter}, {and} \bibinfo{person}{Matt Fredrikson}.} \bibinfo{year}{2023}\natexlab{b}.
\newblock \showarticletitle{Universal and {Transferable} {Adversarial} {Attacks} on {Aligned} {Language} {Models}}.
\newblock  (\bibinfo{date}{July} \bibinfo{year}{2023}).
\newblock
\urldef\tempurl%
\url{https://doi.org/10.48550/arXiv.2307.15043}
\showDOI{\tempurl}
\newblock
\shownote{arXiv:2307.15043 [cs]}.


\end{thebibliography}

\newpage
\appendix

\section{Motivations for External Audits} \label{app:goals}

Audits involve formally evaluating systems to assess risks, compliance with standards and regulations, and other desiderata of interest to stakeholders. 
High-quality audits from independent, external auditors have been motivated in multiple ways \citep{sharkey2024causal}:

\begin{itemize}
    \item \textbf{Identifying problems:} The most direct purpose of audits is to identify risks from unsound systems or practices. 
    \item \textbf{Incentivizing responsible development:} 
    When individual components of the development process are insufficiently documented, information necessary to contextually assess risks is lost \citep{khlaaf_how_2023}. Audits can assess the sufficiency of internal controls, risk assessment, and documentation \citep{raji2020closing, brundage_toward_2020,koessler_risk_2023, schuett_agi_2023}. 
    Greater accountability for internal practices incentivizes auditees to spend more effort on risk mitigation and documentation \citep{wagner_unexpected_2006}, especially when facing penalties or public scrutiny \citep{goh_disciplining_2013, yew2022penalty}.
    \item \textbf{Increasing transparency:} Publicly shared information from audits can help regulators and the scientific community develop a better understanding of system behaviors and limitations. % To the extent that certain findings are reported, this can also enhance public trust and allow for more democratic participation \citep{whittlestone2021governments}.
    \item \textbf{Enabling fixes to technical problems:} When problems are found during an audit, developers can then work to address them \citep{raji2019actionable}. 
    External audits can also identify risk factors that might merit further guardrails on deployment, closer monitoring of deployed systems, or follow-up studies of user impacts.
    \item \textbf{Balancing transparency and security:} Keeping systems entirely secret is maximally secure but prevents external scrutiny. Open-sourcing them allows for maximal scrutiny but can proliferate proprietary or misusable systems \citep{seger2023open}. Audits offer a middle ground that allows for some transparency and independent risk assessment with high security. 
    \item \textbf{Providing greater credibility to responsible developers:} Passing audits increases trust in developers and their systems. Hence, the public can better calibrate their trust in developers and systems.
\end{itemize}

\section{Technical Assistance as a form of Outside-the-Box Access} \label{app:technical}

In \Cref{sec:outside}, we discuss how outside-the-box access to information can help auditors conduct audits more effectively. However, for similar reasons, access to technical assistance can also be useful. For example, one resource that auditors will often need, especially for large language models, is computing infrastructure \citep{anderljung2023publicly, ojewale2024towards}. 
Further, additional technical assistance from the developers' engineers may also help because they have unique practical knowledge of working effectively with their models. This may include assistance with fine-tuning, developing realistic test cases (e.g., \citep{zhou_webarena_2023}), or integrating models with external tools that enhance capabilities to resemble real-world usage \citep{sharkey2023auditing, park_generative_2023, davidson_ai_2023, naihin_testing_2023}. Past experience with AI audits has highlighted the value of technical assistance from developers.
\begin{quote}
    \textit{After seeing the final audit report, we realized that we could have helped [METR, (formerly ARC Evals)] be more successful in identifying concerning behavior if we had known more details about their (clever and well-designed) audit approach. This is because getting models to perform near the limits of their capabilities is a fundamentally difficult research endeavor. Prompt engineering and fine-tuning language models are active research areas, with most expertise residing within AI companies. With more collaboration, we could have leveraged our deep technical knowledge of our models to help [METR] execute the evaluation more effectively.}
    
    \hspace{10pt} --Anthropic on their audit by METR (formerly ARC Evals) \citep{anthropic2023challenges}
\end{quote}
Allowing developers to arbitrarily influence audits undermines their independence, so incorporating requirements for developers to provide technical assistance into legal auditing frameworks may be difficult and is beyond the scope of this paper. 
However, auditors may find it helpful if specific requests for technical assistance are answered in good faith by auditees.

\section{Innovation on Auditing Tools} \label{app:supporting}

White-box tools for studying AI systems have long been a topic of technical interest, but research on methods often struggles to keep up with the scale and capabilities of leading AI systems \citep{bengio2023managing}.
There are gaps between the capabilities of white-box evaluation tools and what auditors may need from them. 
More progress on both foundational research and practical tools will be useful for auditors, especially for state-of-the-art large language models because of their unique versatility and complexity.

\textbf{Basic research:} Current methods have provided useful insights. However, developing a detailed mechanistic understanding is not yet possible in state-of-the-art models. 
More progress in the basic science of neural networks and efforts to study their inner workings will help further research on evaluation techniques.  
This will require progress on both developing more intrinsically understandable systems and techniques to interpret trained ones \citep{carvalho2019machine, rauker2023toward}.

\textbf{Practical tools:} The goal of research on evaluation techniques is to produce methods that can be effectively used off-the-shelf by auditors.
In the adversarial attack literature, benchmarks have largely focused on fooling networks with small perturbations to inputs instead of eliciting harm via more real-world features \citep{Zhang2019AdversarialAO, Roth2021TokenModificationAA, hendrycks2021natural, kaufmann2023testing}.
In the interpretability literature, few benchmarks connected to practical tasks exist, with it being common to judge techniques based on researcher intuition \citep{dong2017towards, miller2019explanation, krishnan2020against, adebayo2020debugging, rauker2023toward, casper2023red1}. 
Given the increasing scale and complexity of modern AI systems, developing more effective evaluation tools poses a challenge.
Fortunately, open-source and API access to advanced AI systems has enabled progress on evaluation tools.
However, no technique for benchmarking evaluation tools is more directly informative than applications on real systems.
Partnerships between researchers and developers can facilitate these.

\textbf{Secure auditing infrastructure:} As discussed in \Cref{sec:security}, granting auditors white-box access to systems via application programming interfaces or secure research environments can reduce the risk of leaks. 
However, because norms for AI audits have not yet been established, there is little infrastructure for conducting audits securely. 
For example, efforts like the US National Deep Inference Facility project \citep{nsf2023national} could make more resources available to auditors. Establishing better tools and protocols is another priority \citep{Lucaj2023AIRI}.
At the same time, it will be key to establish norms and a regulatory framework around AI audits, as has been done in other industries with audits.

\section{Beyond Access: Other Aspects of Rigorous Audits.} \label{app:beyond}

White- and outside-the-box access is necessary but not sufficient for rigorous audits. Many factors can undermine or degrade the quality of audits. We overview challenges here.

\textbf{Poorly-resourced audits:} Working with state-of-the-art AI systems and effectively evaluating them requires compute and technical expertise. While developing and commercializing advanced AI systems can be lucrative, searching for problems with them might not be profitable or financially sustainable. Existing audits have largely relied on private funding (e.g., \citep{metr, kinniment2023evaluating, openai2023gpt4, anthropic2023challenges}), rather than public funding or other more sustainable, reliable, and diversified sources of funding.

\textbf{Limitations with technical tools:} As discussed in \Cref{app:supporting}, there is a gap between existing technical tools for evaluations and the kind of tooling needed to reliably assess the safety and trustworthiness of advanced systems. Until this gap is closed, audits will be limited in identifying risks. %THIS PARAGRAPH ISN'T SUFFICIENTLY CLEAR

\textbf{Narrowly-scoped audits:} Audits may omit important evaluations. For example, early audits of GPT-4 have focused on risk-related capabilities \citep{metr, kinniment2023evaluating, openai2023gpt4} but did not appear to include external evaluation regarding other concerns such as robustness to adversarial attacks; potential for misinformation; demographic representation; or impacts on societal welfare, democracy, discrimination, and equality. 
Another way in which auditing can be narrow in scope is if it only occurs pre-deployment. A ``black cloud'' system with ever-changing components is even more difficult to evaluate than a black box. 
% Post-deployment audits to re-evaluate risks in light of new updates or findings about systems enables better risk assessment. PERHAPS CITE RE BLACK CLOUD How Is ChatGPT's Behavior Changing over Time?, BUT THAT PAPER HAS BEEN CRITICIZED

\textbf{Conflicts of interest:} Auditors may face pressure to refrain from insisting on sufficient access or conducting sufficiently rigorous audits. Auditor conflicts of interest, including collusion with auditees \citep{white2010markets, bolton2012credit} are well-known and long-standing problems \citep{goldman1974auditor-firm, moore2006conflicts}. They stem in part from the typical payment structure of auditors: auditors that produce more favorable evaluations – including due to receiving inadequate or incomplete information from audit targets – are often preferred over other auditors, leading companies to ``opinion-shop'' \citep{lennox2000companies} for comparatively lax evaluations. This can trigger a race to the bottom in which audits become progressively less rigorous and less informative \citep{baghai2020reputations, anderljung2023publicly}. This type of dynamic could emerge in the absence of adequate regulatory structures. For example, recent audits of state-of-the-art language models from OpenAI \citep{openai2023gpt4} and Anthropic \citep{anthropic2023challenges} were conducted on a voluntary basis by the Model Evaluation and Threat Research organization (METR, formerly named ARC Evals) \citep{metr}, which maintains a close relationship with both companies the details of which are not publicly disclosed.

\textbf{Exclusion of under-represented viewpoints:} Not all people agree on what behaviors from AI systems are harmful. As a result, audits can exclude under-represented groups if they are designed in a way that fails to take a wide range of interests into account \citep{birhane2022values, lazar2023ai, feffer_red-teaming_2024}.
Improving meaningful participation \citep{shur2023multiplicity, sorensen2024roadmap} and dialogue \citep{dobbe2019hard} among diverse stakeholders plays an indispensable role in improving fairness and representation.

\textbf{Cosmetic compliance:} Absent clear legal requirements, companies have an incentive to prioritize cosmetic compliance with good practices \citep{krawiec2003cosmetic}, a form of cheap talk \citep{farrell1996cheap} or virtue signaling \citep{westra2021virtue} in which audit targets create a superficial (yet misleading) appearance of good faith cooperation.

\textbf{Regulatory capture:} While governance regimes that bolster auditing standards and procedures may appear promising, they, too, can be undermined. Studies in the field of organizational science demonstrate that companies respond strategically to interventions, employing a variety of operational, political, and legal tactics \citep{oliver1991strategic} including supporting biased research \citep{abdalla2021grey}. In its simplest form, companies may selectively disclose audit-relevant information \citep{marquis2016scrutiny, hess2019transparency}, enabling them to game outcomes, including in AI audits \citep{raji2022outsider, birhane2024ai}. Meanwhile, more sophisticated and well-resourced companies can shape the underlying audit criteria, metrics, and institutions, including by selecting which auditors have privileged access to information and which do not. Legal sociologists describe this symbiotic relationship between regulators and regulated entities as “legal endogeneity”: it is precisely the actors that law seeks to control that end up controlling the law \citep{edelman1992legal, edelman2016working, waldman2019privacy, abdalla2021grey}. AI audits are especially susceptible to these dynamics because the relevant standards are currently unclear \citep{costanza-chock2022who} and audit tools are bespoke and applied inconsistently across different developers and domains \citep{costanza-chock2022who}.

\end{document}